\documentclass[11pt]{article}

\usepackage[round,authoryear]{natbib}
\usepackage[margin=1in]{geometry}
\usepackage{amsmath,amssymb,amsfonts,amsthm}
\usepackage{array}
\usepackage{cite}
\usepackage{mathtools}
\usepackage{graphicx, bm, bbm, tikz}
\usepackage{enumitem,booktabs, comment}
\usepackage{algorithm} \usepackage{subfigure}
\usepackage{algpseudocode}
\usepackage{hyperref}
\usepackage[export]{adjustbox}
\hypersetup{colorlinks=true, linkcolor=red, citecolor=blue}
\usepackage{tikz}
\usetikzlibrary{calc,arrows.meta,positioning}
\usetikzlibrary{decorations.pathreplacing}
\usepackage{bbding}
\newcommand{\E}{\mathbb{E}}
\newcommand{\Prb}{\mathbb{P}}

\newtheorem{theorem}{Theorem}[section] 
\newtheorem{lemma}[theorem]{Lemma}
\newtheorem{proposition}[theorem]{Proposition}

\theoremstyle{definition}
\newtheorem{definition}[theorem]{Definition}

\theoremstyle{remark}

\newtheoremstyle{boldremark}
  {3pt}{3pt}        
  {\normalfont}       
  {}                
  {\bfseries}       
  {.}               
  {.5em}            
  {}                

\theoremstyle{boldremark}
\newtheorem{remark}[theorem]{Remark}

\newcolumntype{C}[1]{>{\centering\arraybackslash}p{#1}}

\newcommand{\footremember}[2]{%
  \footnote{#2}%
  \newcounter{#1}%
  \setcounter{#1}{\value{footnote}}%
}

\title{Learning and Steering in Incomplete-Information Games}
\author{%
Asrın Efe Yorulmaz
\footremember{uiucay20}{University of Illinois Urbana--Champaign; e-mail: \texttt{ay20@illinois.edu}}%
\and
Tamer Başar\footremember{uiucb1} {University of Illinois Urbana--Champaign; e-mail: \texttt{basar1@illinois.edu}}%
}

\date{}

\begin{document}

\maketitle

\begin{abstract}
In many modern multi-agent systems, a mediator seeks to influence strategic decision dynamics using only outcome-level feedback and without knowing agents' utilities. In this work, we study how a mediator can learn an \emph{unknown} game and steer \emph{adaptive} agents toward outcomes that maximize her utility using only trajectory feedback. The mediator knows the game form and her own objective, but not the agents' utilities. Moreover, she commits to nonnegative payments and provides recommendations and contexts, but receives only a single realized terminal outcome rather than full-information feedback. We develop halving-type algorithms that learn \emph{utility differences} in normal-form games and \emph{conditional continuation-value differences} in extensive-form games, achieving a sample complexity of $\widetilde O(\varepsilon^{-2})$ that matches worst-case lower bounds up to logarithmic factors. We then combine these learning procedures with a \emph{Lagrangian formulation} in an online \emph{Learn--Compute--Steer} algorithm that interleaves learning, equilibrium computation, and steering toward mediator-optimal extensive-form correlated equilibria (EFCE). With high probability, the \emph{Learn--Compute--Steer} algorithm achieves a \emph{directness gap} and an \emph{optimality gap} of $\widetilde O(T^{-1/4})$, while \emph{time-average payments} are of $\widetilde O(T^{-1/8})$.
\end{abstract}

\section{Introduction}

Classical game theory is primarily concerned with a \emph{forward} problem: given agents’ preferences and the rules of interaction, how will strategic agents behave?
This perspective underlies equilibrium analysis and its many refinements, which characterize stable patterns of play assuming agents optimize against one another.
In many modern applications, however, a designer, platform, or principal does not know agents’ utilities a priori. Although much of the literature on the \emph{inverse} problem of learning the game itself from observations assumes that observed behavior corresponds to an equilibrium outcome \citep{KuleshovSchrijvers2015InverseGT}, this assumption is often untenable in practice. First, equilibrium play may be computationally expensive to obtain in large games, as computing Nash equilibria is PPAD-hard in general \citep{PPAD-hard}. Second, even when equilibria are well-defined, real-world agents typically adapt online under limited information and changing conditions, so their behavior need not settle at, or even approximate, any fixed equilibrium profile within the time horizon of interest. This raises a complementary \emph{inverse} and \emph{control-oriented} question: given only observed behavior from adaptive agents, how can a principal both \emph{learn} about the strategic environment and \emph{shape} long-run outcomes toward objectives of its own?

\vspace{6pt}
\noindent
A natural behavioral model in such settings is one where agents employ no-regret learning algorithms rather than explicitly computing equilibria, as no-regret dynamics provide a principled foundation for predicting long-run behavior in multi-agent interactions \citep{EconometricsforLearningAgents}. Moreover, when each player runs a no-regret algorithm, the induced time-averaged play approaches equilibria notions such as correlated equilibrium (CE) and coarse correlated equilibrium (CCE) in general-sum normal-form games (NFGs) \citep{hart2000simple, cesa2006prediction}. Furthermore, regret-minimization methods and their equilibria convergence guarantees extend to extensive-form games (EFGs)  \citep{CFR2007, MCCFR, CFR+}.

\vspace{6pt}
\noindent
In many socio-technical settings, however, there is also a \emph{principal} or \emph{mediator} with an objective function distinct from those of the agents. Such a mediator can intervene by providing information, recommendations, or incentives that influence agents’ learning trajectories.
In extensive-form games, a natural solution concept for this mediated interaction is \emph{extensive-form correlated equilibrium} (EFCE), which unifies information design and incentive constraints into a single implementability object \citep{VonStengelForges2008EFCE}.

\vspace{6pt}
\noindent
A central obstacle in multi-stage settings is the mediator's limited observability.
We distinguish between two feedback regimes. In the first setting, \emph{full-information feedback}, the mediator observes sufficient information to evaluate or estimate utilities for deviations beyond the realized play, e.g., full-information payoff vectors in normal-form games or counterfactual value information in EFGs. In the second setting, \emph{trajectory feedback}, in each iteration the mediator observes a single realized terminal outcome together with the information sets visited by each player along their trajectories, and can base payments only on the realized outcome. In this work, we focus on designing mediator algorithms that work in the trajectory-feedback regime. Formally, we ask the question:

\begin{center}
\emph{Without knowing the players' utilities and observing only realized trajectories, can a mediator steer no-regret learners to optimal EFCE, while her payments vanish on average?}
\end{center}
\noindent
We answer this question in the affirmative by making the following contributions.

\begin{itemize}
  \item 
  In Section~\ref{sec:learning}, we develop algorithms that learn utility differences in normal-form games (Algorithm~\ref{alg:gbs-ma}) and \emph{conditional continuation-value differences} in extensive-form games (Algorithm~\ref{alg:gbs-efg-all}), both with sample complexity $\widetilde O(\varepsilon^{-2})$. 
  This dependence matches the normal-form lower bounds of \citet{Payingagents2025} and extensive-form lower bounds in Proposition~\ref{prop:efg-inherited-lower-bounds} up to logarithmic factors. Moreover, we show that learning \emph{conditional continuation-value differences} suffices to evaluate the EFCE incentive constraints, which, as we later establish, enables the mediator to steer players toward a desired EFCE.
  
  \item
  In Section~\ref{sec:steering}, we present online
  \emph{Learn--Compute--Steer} algorithm
  (Algorithm~\ref{alg:online-efce}) that combines learning,
  equilibrium computation, and trajectory-feedback steering. We show that, 
  the directness gap and the optimality
  gap relative to the mediator's optimal EFCE value diminish with the rate of
  $\widetilde O(T^{-1/4})$, while time-average
  payments vanish with the rate of $\widetilde O(T^{-1/8})$, with high probability. Finally, in Section~\ref{sec:experiments}, we complement the analysis
  with experiments in a normal-form Stag Hunt game and
  extensive-form Kuhn Poker instances.
\end{itemize}

\subsection{Related Work}

A broad literature studies \emph{inverse} problems in games---inferring agents' preferences from observed behavior---under names such as learning from revealed preferences and inverse game theory.
Classical revealed-preference models focus on identifying utility parameters from choices under rationality assumptions \citep{BeigmanVohra2006RevealedPref, ZadimoghaddamRoth2012RevealedPref}, while inverse game-theoretic work develops identification and estimation methods for multi-agent utilities, often positing that the observed play corresponds to an equilibrium or a noised observation of it \citep{KuleshovSchrijvers2015InverseGT, WuXuYao2022InverseStackelberg, GoktasGreenwald2024InverseMAL}.
Our setting departs from these equilibrium-based formulations as we assume that the principal observes the agents’ strategies over repeated play, potentially far from equilibrium, and seeks to learn and intervene the process.

\vspace{6pt}
\noindent
Our perspective is closely aligned with recent work that studies how a mediator can simultaneously learn an unknown game and influence agents’ behavior under learning-based behavioral assumptions.
Most closely related to our learning procedure (Section~\ref{sec:learning}), \citep{Payingagents2025} show that a principal can learn an unknown game with sample complexity $\tilde{O}(\varepsilon^{-2})$ and establish a matching information-theoretic lower bound of $\tilde{\Omega}(\varepsilon^{-2})$.
A key limitation, however, is that their guarantees are developed for normal-form settings and do not extend to EFGs; moreover, their upper bounds do not match the lower bound with respect to dependence on the players’ action-space sizes, which can be prohibitively large in practice.
We address these issues in the present work, and provide a comparison of our results in Table \ref{tab:comp_nf}. Importantly, similar to \citep{Payingagents2025}, \citep{Yorulmaz2025}, in this work we assume that the players are no-regret within each mediator-provided \emph{context}.
Related themes also appear in the literature on principal--agent interaction with learning agents and on contract design under uncertainty, where the principal must reason about an agent that adapts over time \citep{GeneralizedPrincipalAgentLearningICLR25, ContractingWithLearningAgent2024, ComplexityOfContracts2021, OnlineIncentiveCompatibleLearning2020}.

\vspace{6pt}
\noindent
Building on this viewpoint, we pursue a different downstream objective. Rather than merely estimating utilities, we integrate \emph{learning the underlying game}, which constitutes the \emph{Learn} component of our framework, with \emph{equilibrium-oriented steering}, which constitutes the \emph{Compute–Steer} component, in multi-stage games under trajectory feedback. A complementary line of work studies how to steer adaptive agents by shaping rewards under partial knowledge of agents' dynamics, often using control-theoretic viewpoints \citep{HuangEtAl2025MarkovianSteering, SteeringGameDynamicsMPC2025, EvoGameDynamicsEnvFeedback2022}.
These approaches typically model specific learning dynamics, whereas, for the steering problem, we assume only regret guarantees within each mediator-provided context, as in the \emph{Learn} component of our framework, and focus on steering agents toward a mediator-optimal EFCE.

\vspace{6pt}
\noindent
Most directly related to our \emph{Learn-Compute-Steer} algorithm, Algorithm \ref{alg:epochal-online-bce-corrected}, \citep{SteeringNoRegret24} introduce a trajectory-feedback steering framework that drives no-regret agents toward a desired equilibrium while maintaining vanishing average payments.
They also ask whether there exists a fully online trajectory-feedback algorithm that attains the mediator-optimal equilibrium, and we answer this question in the affirmative in the present work assuming that players have no regret within each mediator-provided context for EFCE. Our result in this work, compared to earlier ones can be found in Table \ref{tab:comp_mg}, where offline steering refers to steering toward a known equilibrium, whereas online steering refers to simultaneously computing an equilibrium and steering the players toward it. All results hold for EFGs. Earlier work on implementation via payments is also conceptually aligned with our use of nonnegative, outcome-dependent transfers to enforce obedience \citep{kImplementation2003}.

\vspace{6pt}
\noindent
On the other hand, our mediated interaction model is closely related to the Bayesian persuasion framework and information design problems, where a sender commits to an information structure to shape receivers' behavior \citep{KamenicaGentzkow2011BayesianPersuasion, BergemannMorris2019InformationDesignJEL}.
Moreover, BCE provides a unifying implementability object that captures both Bayes plausibility and incentive constraints, while it subsumes classical correlation/communication concepts in games with information \citep{BergemannMorris2016BCE, Forges1986CommEq, Myerson1986MultistageComm}.
In extensive-form settings, related implementability notions include extensive-form correlated equilibrium (EFCE) and its computational treatments \citep{VonStengelForges2008EFCE}. Related to our mediated steering model, \citep{Yorulmaz2025} provide results characterizing what can be achieved under information-design constraints when the objective is to steer agents’ behavior. 

\begin{table}[t]
\centering
\caption{Comparison of Sample Complexity Rates for $\varepsilon$-Learning the Games}
\label{tab:comp_nf}
\begin{tabular}{C{2.5cm} C{2.3cm} C{3.3cm} C{7.5cm}}
\toprule
\small \textbf{Problem Type}
& \small \textbf{\citet{Payingagents2025}}
& \small \textbf{Our Results}
& \small \textbf{Worst-case Lower Bounds }  \\
\midrule

\small \textbf{Single-Agent NFGs}
& \small$\mathcal{O}\big(\frac{M^3}{\varepsilon^2}\big)$
& \small$\mathcal{O}\big(\frac{M}{\varepsilon^2}\big)$
  Thm.~\ref{thm:main}
& \small\( 
\Omega\big(\max\big\{\frac{nM\log(1/\varepsilon)}{\log M},\frac{C^2}{\varepsilon^2}\big\}\big) 
\) \citet{Payingagents2025}\\

\small \textbf{Multi-Agent NFGs}
& \!\!\!\!\!\small$\mathcal{O}\big(\frac{nm^3M^5+m^2M^4n^4}{\varepsilon^2}\big)$
& \small$\mathcal{O}\big(\frac{Mn^3}{\varepsilon^2}\big)$
  Thm.~\ref{thm:multi-main}
& \small \( 
\Omega\big(\max\big\{\frac{nM\log(1/\varepsilon)}{\log M},\frac{C^2}{\varepsilon^2}\big\}\big) 
\) \citet{Payingagents2025} \\

\small \textbf{EFGs}
& $\times$
& \small $\tilde{\mathcal{O}}\big(\frac{\mathcal J n^2}{\rho_{min}^2\varepsilon^2}\big)$
  Thm.~\ref{thm:learn-all-efg-allx}\!\!\!\!
& \small
\!\!\( \Omega\big(
\max\big\{
\frac{\mathcal J\log(1/\varepsilon)}{\log\mathcal J},
\frac{C^2}{\varepsilon^2},
\frac{\log(1/(2\delta))}{\rho_{\min}}
\big\}\big)
\) Prop. \ref{prop:efg-inherited-lower-bounds}\\
\bottomrule
\end{tabular}
\vspace{4pt}

\begin{minipage}{\textwidth}
\footnotesize
Here $m = \max_i m_i$, and $M=\prod_i m_i$ is the
normal-form joint-action count, and $C=\max_i C_i\ge1$.
For EFGs,
$\mathcal J\le M_{\mathrm{EF}}nI_{\max}a_{\max}$,
where $M_{\mathrm{EF}}=\prod_i\prod_{I\in\mathcal I_i}|A_I|$,
$I_{\max}=\max_i|\mathcal I_i|$, and
$a_{\max}=\max_{i,I}|A_I|$.
\end{minipage}
\end{table}

\begin{table}[t]
\centering
\caption{Comparison of directness-gap bounds in EFGs under regret of order
$\mathcal O(\sqrt T)$.}
\label{tab:comp_mg}
\small
\begin{tabular}{
  C{3.0cm}
  C{3.2cm}
  C{2.5cm}
  C{2.7cm}
  C{2.7cm}
}
\toprule
\textbf{Reference}
& \textbf{Feedback}
& \textbf{Equilibrium}
& \textbf{Offline steering}
& \textbf{Online steering} \\
\midrule

\citet{SteeringNoRegret24}
& Full feedback
& BCE
& $\mathcal O(T^{-1/4})$
& $\mathcal O(T^{-1/6})$ \\

\citet{SteeringNoRegret24}
& Trajectory feedback
& BCE
& $\mathcal O(T^{-1/4})$
& --- \\

Theorem~\ref{thm:main-corrected}
& Trajectory feedback
& EFCE
& ---
& $\widetilde{\mathcal O}(T^{-1/4})$ \\

\bottomrule
\end{tabular}

\end{table}

\vspace{6pt}
\noindent
Finally, our setting is related to work on \emph{strategizing against} or \emph{exploiting} agents, as well as to Stackelberg and commitment perspectives.
Literature shows that a principal can extract value against no-regret learners, motivating equilibrium-free analyses in repeated interaction \citep{DengEtAl2019Strategizing, StrategizingBayesianGamesCOLT22, SellingToNoRegretBuyer2017, AuctionsBetweenRegretMinimizers2022}.
In Stackelberg settings, there are works on learning optimal commitments and dynamic Stackelberg equilibria under various models of follower response \citep{ConitzerSandholm2006Commitment, LetchfordConitzerMunagala2009CommitmentLearning, HaghtalabEtAl2022StackelbergNonMyopic, LaufferEtAl2022DynamicStackelberg}.
In a related direction, \citep{Mazumdar2025} provide impossibility results and identifies what information must be known in order to steer a learning follower through the leader’s actions. Comparably, we focus on a mediator who chooses contexts, recommendations, and payments to induce mediator-optimal outcomes.

\section{Preliminaries}\label{sec:prelims}

For an integer $n\ge 1$, we write $[n]:=\{1,2,\dots,n\}$.
For a statement $p$, let $\mathbf 1\{p\}\in\{0,1\}$ denote its indicator.
For a vector $v\in \mathbb R^m$, $v[i]$ denotes its $i$-th coordinate, and
$\langle u,v\rangle$ denotes the Euclidean inner product.
We use $\widetilde{O}(\cdot)$ and $\widetilde{\Omega}(\cdot)$ to hide factors logarithmic in their argument(s). Unless stated otherwise, all utilities take values in $[0,1]$.

\subsection{Normal-Form Games}\label{subsec:nfg}
A \emph{normal-form game} is a tuple $\Gamma=(n,\{A_i\}_{i=1}^n,\{u_i\}_{i=1}^n)$, where $n$ is the number of players; each player $i$ has a finite action set $A_i$ with $|A_i|=m_i\ge 2$; and each player $i$ has a utility function $u_i:A_1\times\cdots\times A_n\to[0,1]$. Let $A:=A_1\times\cdots\times A_n$ denote the set of action profiles. As standard, we write $-i$ to refer to all players except $i$.
Define $m:=\max_i m_i$, $M:=\prod_{i=1}^n m_i$, and $M_{-i}:=\prod_{j\in[n]\setminus\{i\}}m_j$.
A (mixed) strategy of player $i$ is a distribution $x_i\in X_i:=\Delta(A_i)$.
For a mixed profile $x=(x_1,\dots,x_n)$, we write $a\sim x$ to mean $a_i\sim x_i$ independently,
and define \( u_i(x):=\E_{a\sim x}\big[u_i[a]\big].\)

\subsection{Extensive-form games}\label{subsec:efg}
We use the standard finite extensive-form game as follows.

\begin{definition}\label{def:efg}
An \emph{extensive-form game} $\Gamma$ with $n$ players has:
\begin{enumerate}
\item a player set $K$, in which we refer to a specific player with $i$ and to remaining players with $-i$;
\item a finite directed tree $H$ of \emph{histories}, \emph{nodes}, with root $\varnothing$. The edges are labeled by actions; for $h\in H$, $A_h$ denotes the set of legal actions at $h$. The leaves are \emph{terminal} histories with terminal set $Z\subset H$;
\item a partition of non-terminal histories, given as
\( 
H\setminus Z \;=\; H_C \sqcup H_1 \sqcup \cdots \sqcup H_n,
\) 
where $H_i$ is the set of nodes at which player $i$ acts, and $H_C$ is the set of chance nodes;
\item for each player $i$, a partition $\mathcal I_i$ of $H_i$ into information sets;
      all nodes $I\in\mathcal I_i$ share the same action set $A_I$;
\item bounded utilities $u_i:Z\to[0,1]$ for each player $i$;
\item for each chance node $h\in H_C$, a fixed distribution $c(\cdot\mid h)$ over $A_h$.
\end{enumerate}
\end{definition}

\vspace{6pt}
\noindent
For a history $h\in H$, let $\sigma_i(h)$ denote the sequence
of player $i$'s information-set--action pairs along the path
from the root to $h$, excluding any action at $h$ itself.
Player $i$ has \emph{perfect recall} if
$\sigma_i(h)=\sigma_i(h')$ whenever $h$ and $h'$ belong to the
same information set $I\in\mathcal I_i$.
We assume perfect recall throughout and denote the common
sequence at $I$ by $\sigma_i(I)$.
Writing $\sigma_i(I)a$ for the sequence obtained by
appending action $a\in A_I$ at $I$, we define
\( 
\Sigma_i:=\{\varnothing\}\cup
\{\sigma_i(I)a:I\in\mathcal I_i,\ a\in A_I\}.
\) Furthermore, we write $I\preceq Z$ whenever information set $I \in I_i$ precedes $Z \in I_i$ under such ordering.

\vspace{6pt}
\noindent
A pure strategy of player $i$ specifies an action at every
information set $I\in\mathcal I_i$, while a behavioral
strategy $\beta_i$ assigns a distribution
$\beta_i(\cdot\mid I)\in\Delta(A_I)$ at each such set.
The \emph{realization plan} $x_i\in\mathbb R_+^{\Sigma_i}$ records the probabilities of player $i$'s action sequences, by
\( 
x_i[\sigma_i(I)a]
=x_i[\sigma_i(I)]\beta_i(a\mid I), x_i[\varnothing]=1.
\) 
Consequently, the set of feasible realization plans is
the sequence-form polytope
\( 
X_i:=\left\{
x_i\in\mathbb R_+^{\Sigma_i}:
x_i[\varnothing]=1,
\sum_{a\in A_I}x_i[\sigma_i(I)a]=x_i[\sigma_i(I)],
\ \forall I\in\mathcal I_i
\right\}.
\) 

\vspace{6pt}
\noindent
We let $X:=X_1\times\cdots\times X_n$.
For a strategy profile $x=(x_1,\ldots,x_n)\in X$ and
a terminal history $z\in Z$, write
$x_i[z]:=x_i[\sigma_i(z)]$.
For any subset $N\subseteq[n]$, define the terminal
reach vector $\hat x_N\in\mathbb R^Z$ by
\( 
\hat x_N[z]:=\prod_{j\in N}x_j[z],
\) 
and write $\hat x:=\hat x_{[n]}$.
Thus, $\hat x[z]$ represents the joint contribution of
the players' strategies to reaching $z$, excluding chance.
Let $\pi_C(z)\in[0,1]$ denote the product of chance
probabilities along the path to $z$.
The probability of reaching $z$ is then
$\pi_C(z)\hat x[z]$, and player $i$'s expected utility is
\( 
u_i(x)=\E_{z\sim x}[u_i(z)]
=\sum_{z\in Z}u_i(z)\pi_C(z)\hat x[z].
\) 
In particular, holding $x_{-i}$ fixed, the map
$x_i\mapsto u_i(x_i,x_{-i})$ is linear in $x_i$.

\subsection{Mediator-Augmented Games and Contexts}
\label{subsec:mediator-augment}

In the considered mediator-augmented games, we simply add a mediator, denoted by player $0$, to the original
game $\Gamma$ and denote the augmented game with $\hat \Gamma$. We assume that the mediator knows the game form and her
utility $u_0:Z\to[0,1]$, but does not know the players'
utilities. In a normal-form game, terminal outcomes are simply
identified with joint action profiles.

\vspace{6pt}
\noindent
Moreover, for each player $i$, we let $\mathcal C_i$ be a \emph{context} space.
Before play in round $t$, the mediator chooses
$\kappa^{(t)}\in\prod_i\mathcal C_i$, and player $i$
observes $\kappa_i^{(t)}$. Contexts identify the \emph{learning
environments} used by the players. Crucially, we assume that each player maintains a
separate learning state for each \emph{context} and updates it
only on rounds carrying that \emph{context}. Returning to a
\emph{context} resumes its existing state; using a new \emph{context}
initializes a previously unused state. \emph{Context} selection
may depend on the preceding interaction, but is made before
the current round's play randomization and Nature's moves.
All \emph{contexts} concern the same underlying game; the
mediator's payments and recommendation protocol determine
the round-wise augmented interaction.

\vspace{6pt}
\noindent
Within a context, the mediator may also provide action
recommendations. We distinguish the original strategy
space $X_i$ from the response-strategy space $Y_i$ in the
recommendation-augmented game $\widehat\Gamma$. A strategy
in $Y_i$ specifies actions using the player's original
information and recommendations received so far. We use space $Y_i$ for all commitments under the fixed
recommendation protocol, and write $Y:=\prod_iY_i$.

\vspace{6pt}
\noindent
A commitment $\mu\in X_0$ specifies a recommendation law;
fixing it induces a game $\Gamma_\mu$ among the players.
For $x\in Y$, let $\pi(\cdot\mid\mu,x)$ be the induced
law on original terminal outcomes, and we define
\( 
u_j(\mu,x):=
\mathbb E_{z\sim\pi(\cdot\mid\mu,x)}[u_j(z)], \text{ for }
 j\in\{0,1,\ldots,n\}.
\) 
When no recommendations are sent, players use strategies
in $X_i$, and we write $\pi(\cdot\mid x)$ and $u_j(x)$.
An original-game strategy can also be viewed as a response
strategy that ignores recommendations.

\vspace{6pt}
\noindent
As mentioned, the mediator also has the ability of providing nonnegative payments. In the
normal-form learning procedures these are vectors
$p_i^{(t)}\in[0,P_{\max}]^{A_i}$, depending on each player's
chosen action. In extensive-form games they are terminal
payment rules $q_i^{(t)}:Z\to[0,P_{\max}]$. We refer to the total utility experienced by players, including payments by the mediator, as \emph{augmented payoff} or \emph{augmented utility} of a player. We denote the realized augmented payoffs as
\[
r_i^{(t)}:=
\begin{cases}
u_i(a^{(t)})+p_i^{(t)}[a_i^{(t)}],&\text{in NFGs},\\
u_i(z^{(t)})+q_i^{(t)}(z^{(t)}),&\text{in EFGs}.
\end{cases}
\]
Furthermore, for a selected strategy profile, we denote its expected augmented
utility as
\( 
v_i^{(t)}(\mu,x):=
\mathbb E\bigl[r_i^{(t)}\mid\mu,x\bigr]
\), where we condition on the past strategies and the selected payment rules and recommendations. Importantly, the expectation is taken before current recommendations,
action samples, and Nature's moves are generated. When no
recommendations are sent, the argument $\mu$ is omitted. Holding the actual opponents' selected strategies and
the mediator's choices fixed, we write $v_i^{(t)}(x_i^\star)$
for the expected augmented payoff using only
player $i$'s strategy by $x_i^\star$. 

\subsection{Online Optimization and No-Regret Algorithms}
\label{subsec:noregret}

We measure regret separately within each mediator-provided
context. For player $i$, context $c\in\mathcal C_i$, and horizon $T$, we define
\( 
\mathcal T_i(T,c):=\{t\le T:\kappa_i^{(t)}=c\},
N_i(T,c):=|\mathcal T_i(T,c)|.
\) 
For brewity in this section, let $t_{i,c}(1)<t_{i,c}(2)<\cdots<t_{i,c}(m)$ be the rounds carrying
context $c$. We let $\mathcal F_{i,c}$ contain the history
and parameters fixed before this context's first use.

\vspace{6pt}
\noindent
\textbf{Normal-form games.}
On a learning round $t$, player $i$ selects
$x_i^{(t)}\in\Delta(A_i)$ and samples
$a_i^{(t)}\sim x_i^{(t)}$. The payment vector
$p_i^{(t)}$ is chosen before the current action samples,
and the augmented utility is
\( 
u_i^{(t)}(a):=u_i(a)+p_i^{(t)}[a_i].
\) 
For $x_i\in\Delta(A_i)$ and $a_{-i}\in A_{-i}$, write
\( 
u_i^{(t)}(x_i,a_{-i})
:=
\sum_{b\in A_i}x_i[b]
\bigl(u_i(b,a_{-i})+p_i^{(t)}[b]\bigr).
\) 
The regret against the opponents' realized
actions is
\( 
R_{i,c}^{\mathrm{NF}}(m)
:=
\max_{b\in A_i}\sum_{\ell=1}^{m}
\big[
u_i^{(t_\ell)}(b,a_{-i}^{(t_\ell)})
-u_i^{(t_\ell)}(x_i^{(t_\ell)},a_{-i}^{(t_\ell)})
\big].
\) 
Following \citet{Yorulmaz2025,Payingagents2025}, we
distinguish the mixed strategies selected by an online
algorithm from their sampled actions. The corresponding
realized-action regret is
\( 
\widehat R_{i,c}^{\mathrm{NF}}(m)
:=
\max_{b\in A_i}\sum_{\ell=1}^{m}
\big[
u_i^{(t_\ell)}(b,a_{-i}^{(t_\ell)})
-u_i^{(t_\ell)}(a^{(t_\ell)})
\big],
\) 
where $u_i^{(t)}(a^{(t)})=r_i^{(t)}$. Then, we have the following result.

\begin{proposition}\label{prop:realized-regret}
Fix a context $c$ and a context horizon $m$.
Suppose that an online algorithm $\mathcal A$ satisfies
$R_{i,c}^{\mathrm{NF}}(m)\le B_{i,c}(m)$ almost surely.
Assume that the action samples are conditionally
independent given the history and strategies, and augmented utilities lie in
$[0,L]$. Then,
\( 
\Prb\big\{
\widehat R_{i,c}^{\mathrm{NF}}(m)
\le B_{i,c}(m)+L\sqrt{2m\log(1/\beta)}
\,|\,\mathcal F_{i,c}
\big\}\ge1-\beta,
\) for every $\beta\in(0,1)$.
\end{proposition}

\begin{proof}
See Appendix~\ref{app:sampling-regret}.
\end{proof}

\vspace{6pt}
\noindent
\textbf{Extensive-form games.}
On a learning round $t$, player $i$ selects a realization
plan $x_i^{(t)}\in X_i$. A terminal history records
only the actions along the realized path. The observed terminal node does not
specify the opponents' behavior at information sets
that a different strategy of player $i$ might reach.
We therefore evaluate strategies against the fixed strategies in expectation. The expected regret is
\( 
R_{i,c}^{\mathrm{EF}}(m)
:=
\max_{x_i^\star\in X_i}\sum_{\ell=1}^{m}
\big[
v_i^{(t_\ell)}(x_i^\star,x_{-i}^{(t_\ell)})
-v_i^{(t_\ell)}(x^{(t_\ell)})
\big].
\) 
Executing the selected plans produces a terminal history
$z^{(t)}$ and realized augmented payoff
$r_i^{(t)}=u_i(z^{(t)})+q_i^{(t)}(z^{(t)})$.
The realized-payoff regret is
\( 
\widehat R_{i,c}^{\mathrm{EF}}(m)
:=
\max_{x_i^\star\in X_i}\sum_{\ell=1}^{m}
\big[
v_i^{(t_\ell)}(x_i^\star,x_{-i}^{(t_\ell)})
-r_i^{(t_\ell)}
\big].
\) 
\noindent
Importantly, the same definitions apply during steering in the
recommendation-augmented game. We replace $X_i$ by $Y_i$
and include the current commitment $\mu^{(t_\ell)}$
as the first argument of $v_i^{(t_\ell)}$.
The strategy chosen in the maximization is fixed across
the context's rounds, while its actions may depend on
the recommendations received. Then, we have the following result. 

\begin{proposition}\label{prop:efg-sampling-regret}
Fix a context $c$ and a context horizon $m$.
Suppose that each episode is sampled from the distribution
induced by the strategies and mediator choices. If augmented payoffs lie in $[0,L]$, then,
\( 
\Prb\big\{
\big|
\widehat R_{i,c}^{\mathrm{EF}}(m)
-R_{i,c}^{\mathrm{EF}}(m)
\big|
\le L\sqrt{2m\log(2/\beta)}
\,|\,\mathcal F_{i,c}
\big\}\ge1-\beta,
\) for every $\beta\in(0,1)$.
\end{proposition}

\begin{proof}
See Appendix~\ref{app:sampling-regret}.
\end{proof}

\noindent
\textbf{Regret guarantees.}
When the game and interaction protocol are clear,
we suppress the superscripts $\mathrm{NF}$ and
$\mathrm{EF}$ and write $R_{i,c}(m)$ and
$\widehat R_{i,c}(m)$. We say a learner is context-wise no-regret
if
\( 
\limsup_{m\to\infty}\frac{R_{i,c}(m)}m\le0
\) 
for every context $c$ that occurs infinitely often. We assume that each player is context-wise no-regret. If $P_c$ bounds its payments,
we use realized-payoff bounds of the form
\begin{equation}\label{eq:context-regret-confidence}
\Prb\!\left\{
\widehat R_{i,c}(m)>
(1+P_c)\mathcal R_{i,c}(m,\beta)
\,\middle|\,\mathcal F_{i,c}
\right\}\le\beta,
\end{equation}
where $\mathcal R_{i,c}(m,\beta)=o(m)$ for fixed $\beta$ due to Propositions~\ref{prop:realized-regret}
and~\ref{prop:efg-sampling-regret}.
For context-wise square-root regret guarantees, we take
\( 
\mathcal R_{i,c}(m,\beta)
=C_{i,c}(\beta)\sqrt m,
\) 
with $C_{i,c}(\beta)$ independent of $m$. 

\begin{remark}[Regret bounds and sampling]
\label{rem:efg-regret-sampling}
The learning and steering analyses below use
deterministic regret inequalities. For randomized
learners, these inequalities may instead be
understood on a high-probability event. In particular, Propositions~\ref{prop:realized-regret}
and~\ref{prop:efg-sampling-regret} convert
selected-strategy regret bounds into
realized-payoff regret bounds with an additional
square-root concentration term.

\vspace{6pt}
\noindent
For a fixed horizon $T$, allocating failure
probability $\delta/(3nT)$ to each player
and learning epoch ensures that all
realized-payoff regret inequalities used by
the learning procedure hold simultaneously
with probability at least $1-\delta/3$.
The resulting regret coefficients increase
by at most a logarithmic factor. The subsequent proofs use these inequalities and account separately for
the concentration events associated with
counterfactual trajectories, reachability,
and realized outcomes.
\end{remark}

\subsection{Extensive-Form Correlated Equilibrium}
\label{subsec:efce}

We recall the extensive-form correlated equilibrium (EFCE)
of \citet{VonStengelForges2008EFCE}, using the
trigger-deviation characterization of
\citet{CelliEtAl2020EFCE}. For each player $i$, let
\( 
S_i:=\prod_{I\in\mathcal I_i}A_I,
\mathcal S:=\prod_{i=1}^n S_i
\) 
denote the sets of pure contingency plans and joint pure
profiles. Each $s_i\in S_i$ specifies an action at every
information set of player $i$, including those not reached
during play. Its sequence-form realization plan is denoted
by $\mathrm{Seq}_i(s_i)\in\mathrm{ext}(X_i)$.
For a pure profile $s\in\mathcal S$, define
\( 
U_i(s):=
\sum_{z\in Z}
u_i(z)\pi(z\mid\mathrm{Seq}(s)).
\)

\vspace{6pt}
\noindent
In the ex-ante EFCE recommendation device, the mediator
samples a joint pure profile $\bar s\sim\mu$, where
$\mu\in\Delta(\mathcal S)$, independently of Nature's
moves. Whenever player $i$ reaches an information set
$I$, the mediator privately recommends $\bar s_i(I)$.
A player receives no further recommendations after
disobeying, while recommendations to the other players
continue. The mediator's distribution may correlate
recommendations across players and information sets.

\paragraph{Trigger deviations.}
For player $i$, information set $I\in\mathcal I_i$,
and recommended action $a\in A_I$, we
write $\sigma=(I,a)$ for the corresponding sequence
$\sigma_i(I)a\in\Sigma_i$.
Let $S_i(I)\subseteq S_i$ consist of the pure plans
whose preceding actions are consistent with reaching
$I$, and define
\( 
S_i(I,a):=
\{s_i\in S_i(I):s_i(I)=a\}.
\) 
An $(I,a)$-trigger deviation follows all recommendations
until player $i$ reaches $I$ and receives recommendation
$a$. If this occurs, the player stops following
recommendations and instead uses a continuation plan
sampled from a fixed distribution
$\hat\mu_i\in\Delta(S_i(I))$.

\vspace{6pt}
\noindent
To express the resulting incentive constraints, let
$Z(I)\subseteq Z$ denote the terminal histories passing
through $I$, and let $Z(I,a)\subseteq Z(I)$ denote
those taking action $a$ at $I$.
For $z\in Z$, let $S_i(z)$ be the set of pure plans
whose actions agree with player $i$'s actions along $z$,
and write
\( 
\mathcal S(z):=\prod_{j=1}^n S_j(z),
S_{-i}(z):=\prod_{j\ne i}S_j(z).
\) 
The probability of terminal history $z$ under
obedience is
\( 
q_\mu(z):=
\pi_C(z)\sum_{\bar s\in\mathcal S(z)}\mu(\bar s).
\) 
For $z\in Z(I)$, the probability mass associated
with triggering at $(I,a)$ and subsequently reaching
$z$ is
\begin{equation}\label{eq:efce-trigger-terminal-prob}
\begin{aligned}
p^{(I,a)}_{\mu,\hat\mu_i}(z)
:={}&\pi_C(z)
\left(
\sum_{\substack{
s_i\in S_i(I,a)\\
s_{-i}\in S_{-i}(z)}}
\mu(s_i,s_{-i})
\right)\cdot
\left(
\sum_{\hat s_i\in S_i(z)}
\hat\mu_i(\hat s_i)
\right).
\end{aligned}
\end{equation}
Here $\hat\mu_i$ is supported on $S_i(I)$.
The first factor accounts for the sampled
recommendations and the opponents' obedient play;
the second accounts for the continuation chosen
after triggering.

\begin{definition}[Extensive-form correlated equilibrium]
\label{def:efce}
A distribution $\mu\in\Delta(\mathcal S)$ is an
\emph{extensive-form correlated equilibrium} if,
for every player $i$, information set
$I\in\mathcal I_i$, recommended action $a\in A_I$,
and continuation distribution
$\hat\mu_i\in\Delta(S_i(I))$,
\begin{equation}\label{eq:efce-ic}
\sum_{z\in Z(I,a)}q_\mu(z)u_i(z)
\ge
\sum_{z\in Z(I)}
p^{(I,a)}_{\mu,\hat\mu_i}(z)u_i(z).
\end{equation}
Equivalently, we define the maximum trigger-deviation gain
\begin{equation}\label{eq:efce-trigger-gap}
\begin{aligned}
\delta(\mu)
:=\max_{i\in[n]}
\max_{\substack{I\in\mathcal I_i\\a\in A_I}}
\max_{\hat\mu_i\in\Delta(S_i(I))}
\Bigg\{
&\sum_{z\in Z(I)}
p^{(I,a)}_{\mu,\hat\mu_i}(z)u_i(z)-\sum_{z\in Z(I,a)}
q_\mu(z)u_i(z)
\Bigg\}.
\end{aligned}
\end{equation}
Then $\mu$ is an EFCE if $\delta(\mu)\le0$,
and an $\varepsilon$-EFCE if $\delta(\mu)\le\varepsilon$.
\end{definition}

\vspace{6pt}
\noindent
Since the objective in \eqref{eq:efce-trigger-gap}
is linear in $\hat\mu_i$, its maximum is attained
by a pure continuation plan $\hat s_i\in S_i(I)$.
Thus, checking every player, every pair $(I,a)$,
and every pure continuation plan suffices to
characterize the set of EFCEs.

\section{Learning Games with No-Regret Players}
\label{sec:learning}

In this section, the mediator learns the players' unknown
utility differences through contexts and nonnegative
payments. Importantly, we proceed with our learning task in \emph{epochs}, which lets us to learn the utility differences separately for each $(a_i,a_{-i})$ pair. No action \emph{recommendations} are sent during a
\emph{learning} round. Instead, each epoch receives a new
context for every player. The interaction on a learning round proceeds as follows:
\begin{enumerate}
\item The mediator provides the players with the current epoch's contexts
and fixes its payment rules: $p_i^{(t)}$ in a normal-form
game and $q_i^{(t)}$ in an extensive-form game.
\item Each player $i$ uses its provided context to select and execute $x_i^{(t)}\in X_i$.
\item Player $i$ receives its realized augmented payoff
$r_i^{(t)}$. The mediator observes the joint action profile
in an NFG or the terminal history in an EFG.
\item Each player updates its strategy using current context's
learning state. The mediator updates the testing counts within the latest epoch.
\end{enumerate}
In Sections \ref{sec:single-agent-binary-search} and \ref{sec:multi-agent-guarded}, we use the
following context-wise no-regret property for all players and any horizon $T$:
\begin{equation}\label{eq:signal-wise-nfg-regret}
\widehat R_{i,c}^{\mathrm{NF}}(N_i(T,c))
=\max_{b\in A_i}
\sum_{t\in\mathcal T_i(T,c)}
\left[
u_i(b,a_{-i}^{(t)})+p_i^{(t)}[b]-r_i^{(t)}
\right]
\le C_i\sqrt{N_i(T,c)}.
\end{equation}
\noindent 
As stated, we aim to learn the players' utility differences by
choosing contexts and payments. Stated naively, this goal is impossible to achieve due to an inherent identifiability issue. In specific no-regret algorithms such as FTRL \citep{abernethy2008competing} and OMD \citep{nemirovski1983problem}, an agent's incentives depend only on \emph{utility differences} between actions.
Consequently, for any collection of functions $W_i : A_{-i} \to \mathbb{R}$ that do not depend on $a_i$,
the shifted utilities
\( 
u_i'(a_i,a_{-i}) := u_i(a_i,a_{-i}) + W_i(a_{-i})
\) 
induce the same comparisons for agent $i$ at every $(a_i,a_{-i})$.
Hence, based on partial feedback alone, one cannot distinguish $\Gamma$ from the game $\Gamma'$
obtained by applying such shifts. Therefore, utilities are only learnable up to additive terms $W_i(a_{-i})$. Formally, given a game $\Gamma$ and precision $\varepsilon>0$, we say that the principal \emph{$\varepsilon$-learns} $\Gamma$
if it outputs estimates $\hat u_i : A \to \mathbb{R}$ for each agent $i$ such that there exist functions
$W_i : A_{-i} \to \mathbb{R}$ satisfying
\( 
\bigl|u_i(a_i,a_{-i}) - \hat u_i(a_i,a_{-i}) - W_i(a_{-i})\bigr|
\;\le\; \varepsilon,
 \forall i\in[n],\ \forall (a_i,a_{-i})\in A.
\) 

\subsection{Learning Utilities in Single-Agent Normal-Form Games}
\label{sec:single-agent-binary-search}

To illustrate the underlying idea of our algorithm, we first provide our approach in a single-agent normal-form game
with action set $A=\{1,\ldots,m\}$ and utility
vector $u\in[0,1]^m$. The mediator observes the agent's realized
actions but not its utilities, and can offer a payment $p^{(t)}\in[0,2]^m$ each round. We use the
realized-action regret bound within each learning context,
\eqref{eq:signal-wise-nfg-regret}, with $n=1$ and $C_1=C\ge1$.
Since adding an action-independent constant to $u$ does not change
the agent's incentives, we seek to recover $u$ only up to such a
constant. To do so, for each $a\in A$, we define a suitable \emph{transformation}, referred to as the \emph{gap to the best action}:
\begin{equation}\label{eq:sa-threshold}
\theta_a:=\max_{b\in A}\bigl(u[b]-u[a]\bigr)
       =u_{\max}-u[a]\in[0,1],
\qquad u_{\max}:=\max_{b\in A}u[b].
\end{equation}
Trivially, learning all $\theta_a$ suffices to recover $u$ up to such a
constant. First, we state the following property regarding the design of the payment structure used to learn \emph{gap to the best
action}.  

\begin{lemma}\label{lem:feasible}
For $a\in A$ and $\Delta\in[0,2]$, let
$p^{(\Delta,a)}[a]=\Delta$ and
$p^{(\Delta,a)}[b]=0$ for $b\ne a$.
Then, for every $b\ne a$,
\( 
\bigl(u+p^{(\Delta,a)}\bigr)[a]
-\bigl(u+p^{(\Delta,a)}\bigr)[b]
=u[a]-u[b]+\Delta.
\) 
\end{lemma}

\begin{proof}
The payment $p^{(\Delta,a)}$ is nonnegative and bounded by $2$.
For every $b\neq a$,
\( 
\bigl(u+p^{(\Delta,a)}\bigr)[a]
-\bigl(u+p^{(\Delta,a)}\bigr)[b]
=u[a]+\Delta-u[b],
\) 
which is the desired identity.
\end{proof}
\noindent 
Thus, paying $\Delta>\theta_a$ makes $a$ strictly optimal,
whereas $\Delta<\theta_a$ leaves some alternative strictly better.
We exploit this threshold behavior using two payment tests per step.
For a fixed target action $a$, we maintain a bracket
$[L_k,U_k]$ containing $\theta_a$ at $k$-th epoch. We define its width, midpoint,
and \emph{payment level margin} by
\( 
w_k:=U_k-L_k, m_k:=\frac{L_k+U_k}{2},
\gamma_k:=\frac{w_k}{4},
\) 
and two \emph{payment levels} as
\( 
\Delta_k^-:=m_k-\gamma_k,
\Delta_k^+:=m_k+\gamma_k.
\) 
Each payment level uses the corresponding fixed payment
$p^{(\Delta_k^\pm,a)}$ for
\begin{equation}\label{eq:sa-epoch-length}
N_k:=\left\lceil\frac{16C^2}{\gamma_k^2}\right\rceil
\end{equation}
rounds under its own context $\kappa_{a,k}^\pm$.
The two payment levels use different contexts.
Let $X(\Delta)$ be the number of times $a$ is played in the
$N_k$-round epoch at payment level $\Delta$. We label the corresponding epoch as 
\[
\begin{array}{ll}
\textit{BEST} \quad \text{if }X(\Delta)\ge\frac34N_k,\qquad 
\textit{NOTBEST} \quad\text{if }X(\Delta)\le\frac14N_k, \qquad 
\textit{INCONCL} &\text{otherwise.}
\end{array}
\]
We say that a test is \emph{certified} if its label is not \textit{INCONCL}. In the next two lemmas, we establish two key properties of our two-level payment test. Specifically, Lemma~\ref{lem:one-sided-bounds-32prime} shows that if either payment test is \emph{certified}, we can \emph{partially} locate \(\theta_a\). Furthermore, Lemma~\ref{lem:onecert} guarantees that at least one of the two tests is \emph{certified}.

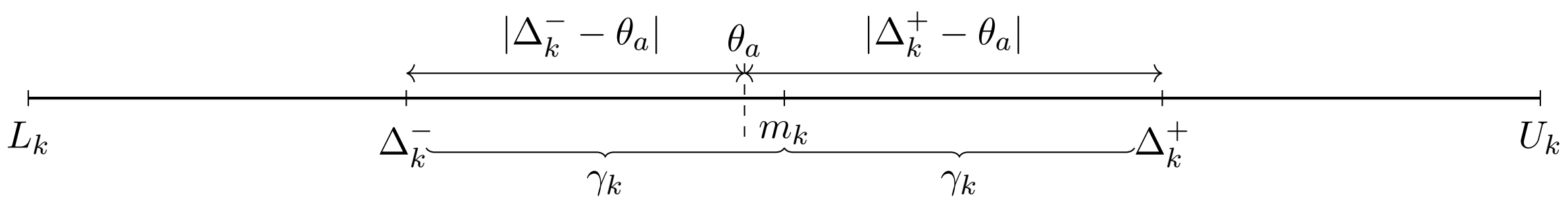
\begin{figure}[t]
\centering
\begin{tikzpicture}[x=8cm,y=1cm]

  \draw[thick] (-0.7,0) -- (1.2,0);

  \foreach \x in {-0.7,-0.225,0.25,0.725,1.2} {
    \draw (\x,0.08) -- (\x,-0.08);
  }

  \node[below] at (-0.7,-0.1) {$L_k$};
  \node[below] at (-0.225,-0.1) {$\Delta_k^-$};
  \node[below] at (0.25,-0.1) {$m_k$};
  \node[below] at (0.725,-0.1) {$\Delta_k^+$};
  \node[below] at (1.2,-0.1) {$U_k$};

  \draw[decorate,decoration={brace,mirror,raise=0.15cm}] (-0.2,-0.35) -- (0.25,-0.35);
  \node[below] at (0.025,-0.6) {$\gamma_k$};

  \draw[decorate,decoration={brace,mirror,raise=0.15cm}] (0.25,-0.35) -- (0.69,-0.35);
  \node[below] at (0.47,-0.6) {$\gamma_k$};

  \draw[dashed] (0.20,0.35) -- (0.20,-0.4);
  \node[above] at (0.2,0.28) {$\theta_a$};

  \draw[<->] (0.20,0.25) -- (-0.225,0.25);
  \node[above] at (-0.005,0.28) {$|\Delta_k^- - \theta_a|$};

  \draw[<->] (0.20,0.25) -- (0.725,0.25);
  \node[above] at (0.45,0.28) {$|\Delta_k^+ - \theta_a|$};

\end{tikzpicture}
\caption{The two payment payment levels $\Delta_k^\pm=m_k\pm\gamma_k$ around the current bracket midpoint. At least one payment level is separated from the unknown threshold $\theta_a$ by at least $\gamma_k$.}
\label{fig:onecert-geometry}
\end{figure}

\begin{lemma}
\label{lem:one-sided-bounds-32prime}
Set $\beta_k:=4C/(3\sqrt{N_k})$. For an epoch at level $\Delta$,
\begin{enumerate}
\item If its label is \textit{BEST}, then $\theta_a\le\Delta+\beta_k$.
\item If its label is \textit{NOTBEST}, then $\theta_a\ge\Delta-\beta_k$.
\item If its label is \textit{INCONCL}, then
$|\Delta-\theta_a|<\gamma_k$.
\end{enumerate}
\end{lemma}

\begin{proof}
See
Appendix~\ref{app:single-agent-proofs}.
\end{proof}

\noindent
By \eqref{eq:sa-epoch-length}, $\beta_k\le\gamma_k/3$.
Moreover, as depicted in Figure \ref{fig:onecert-geometry} the payment levels lie on opposite sides of the midpoint, and thus at
least one is separated from $\theta_a$ by $\gamma_k$.

\begin{lemma}\label{lem:onecert}
For $\Delta_k^\pm=m_k\pm\gamma_k$,
\( 
\max\bigl\{|\Delta_k^--\theta_a|,
            |\Delta_k^+-\theta_a|\bigr\}\ge\gamma_k.
\) 
Consequently, at least one of the two payment levels is certified.
\end{lemma}
\begin{proof}
See
Appendix~\ref{app:single-agent-proofs}.
\end{proof}

\noindent 
Let $\ell_k^\sigma\in\{\textit{BEST},\textit{NOTBEST},
\textit{INCONCL}\}$ be the label at $\Delta_k^\sigma$ for
$\sigma\in\{-,+\}$. Then, we update the bracket for the \(\theta_a\) by intersecting it with
the one-sided bounds supplied by all certified payment levels:
\begin{equation}\label{eq:sa-bracket-update}
\begin{aligned}
L_{k+1}&:=\max\!\left(\{L_k\}\cup
   \{\Delta_k^\sigma-\beta_k:\ell_k^\sigma=\textit{NOTBEST}\}\right),\\
U_{k+1}&:=\min\!\left(\{U_k\}\cup
   \{\Delta_k^\sigma+\beta_k:\ell_k^\sigma=\textit{BEST}\}\right).
\end{aligned}
\end{equation}
Then, we provide the following lemma regarding the contraction of our updates given by \eqref{eq:sa-bracket-update}.

\begin{lemma}\label{lem:bracket}
The update \eqref{eq:sa-bracket-update} preserves
$\theta_a\in[L_{k+1},U_{k+1}]$ and satisfies
\( 
U_{k+1}-L_{k+1}\le\frac56(U_k-L_k).
\) 
\end{lemma}
\begin{proof}
See
Appendix~\ref{app:single-agent-proofs}.
\end{proof}
\noindent 
Algorithm~\ref{alg:gbs} applies this procedure independently to
each target action and stops testing an action as soon as its bracket
has width at most $\varepsilon$. We provide our sample complexity result for Algorithm~\ref{alg:gbs} in Theorem \ref{thm:main}. 

\begin{algorithm}[H]
\caption{BinarySearch-SA}\label{alg:gbs}
\begin{algorithmic}[1]
\Require \( \varepsilon \in(0,1)\)
\For{$a\in A$}
  \State Set $k\gets0$ and $[L_0^{(a)},U_0^{(a)}]\gets[0,1]$.
  \While{$U_k^{(a)}-L_k^{(a)}>\varepsilon$}
    \State $m_k\gets (L_k^{(a)}+U_k^{(a)})/2$, 
      $\gamma_k\gets(U_k^{(a)}-L_k^{(a)})/4$,
      $N_k\gets\lceil16C^2/\gamma_k^2\rceil$,
      $\Delta_k^\pm\gets m_k\pm\gamma_k$.
    \For{$\sigma\in\{-,+\}$}
      \State Provide context $\kappa_{a,k}^{\sigma}$.
      \State Run $N_k$ rounds with $p^{(\Delta_k^\sigma,a)}$,
        retaining this context throughout the epoch.
      \State Record $X(\Delta_k^\sigma)$ and its label $\ell_k^\sigma$.
    \EndFor
    \State Compute $[L_{k+1}^{(a)},U_{k+1}^{(a)}]$
      using \eqref{eq:sa-bracket-update}; set $k\gets k+1$.
  \EndWhile
  \State $\widehat\theta_a\gets(L_k^{(a)}+U_k^{(a)})/2$.
\EndFor
\State Set $c\gets m^{-1}\sum_{b\in A}\widehat\theta_b$ and
  \textbf{return} $\widehat u[a]\gets-\widehat\theta_a+c$ for all $a\in A$.
\end{algorithmic}
\end{algorithm}

\begin{theorem}\label{thm:main}
For any $\varepsilon\in(0,1)$, Algorithm~\ref{alg:gbs}
outputs $\widehat u$ such that, with
$W:=c-u_{\max}$,
\( 
\|u+W\mathbf1-\widehat u\|_\infty\le\varepsilon.
\) 
The total number of rounds is
$T=O(mC^2/\varepsilon^2)$, and the algorithm uses
$O(m\log(1/\varepsilon))$ distinct contexts.
\end{theorem}
\begin{proof}
See
Appendix~\ref{app:single-agent-proofs}.
\end{proof}

\subsection{Learning Utilities in Multi-Agent Normal-Form Games}
\label{sec:multi-agent-guarded}

We now extend the single-agent procedure to normal-form games with $n$ players.
The additional difficulty is that, while learning player $i$'s utility
at a fixed opponents' action profile $y\in A_{-i}$, the opponents
continue to learn and need not play $y$ in every round. We therefore
combine the two-level payment search of Section~\ref{sec:single-agent-binary-search}
with \emph{pinning payments} that keep the opponents close to $y$.
The mediator observes the realized joint action $a^{(t)}$ and uses the
realized-action regret bound~\eqref{eq:signal-wise-nfg-regret} within each testing context.
Without loss of generality, we treat $C_j\ge1$. Fix $i\in[n]$, $y\in A_{-i}$, and a target action $a\in A_i$.
As in the single-agent case, we define the \emph{gap to the best action}
\begin{equation}\label{eq:ma-threshold}
\theta_{i,y}(a)
:=\max_{b\in A_i}\bigl(u_i(b,y)-u_i(a,y)\bigr)
=u_i^{\max}(y)-u_i(a,y)\in[0,1],
\qquad u_i^{\max}(y):=\max_{b\in A_i}u_i(b,y).
\end{equation}
Estimating all such gaps recovers each utility
$u_i(\cdot,y)$ up to an additive constant depending on $y$.
Now, we first state the following property regarding the design of the payments made to the learn chosen player's \emph{gap to the best action}.  

\begin{lemma}\label{lem:feasible-multi}
For $a\in A_i$ and $\Delta\in[0,2]$, let
$p_i^{(\Delta,a)}[a]=\Delta$ and
$p_i^{(\Delta,a)}[b]=0$ for $b\ne a$.
Then, for every $b\ne a$ and $y\in A_{-i}$,
\( 
\bigl(u_i(a,y)+p_i^{(\Delta,a)}[a]\bigr)
-\bigl(u_i(b,y)+p_i^{(\Delta,a)}[b]\bigr)
=u_i(a,y)-u_i(b,y)+\Delta.
\) 
\end{lemma}
\begin{proof}
For any $b\ne a$, the payment difference is
$p_i^{(\Delta,a)}[a]-p_i^{(\Delta,a)}[b]=\Delta$.
Adding this to the \(u_i(a,y)-u_i(b,y)\) proves the identity.
The payment belongs to $[0,2]^{A_i}$.
\end{proof}

\noindent
To hold the opponents near a chosen profile $y$, we pay each
$j\ne i$ according to $p_j^{\mathrm{pin}}:=p_j^{(2,y_j)}$.
Since utility differences are at most $1$, playing $y_j$
has augmented-payoff advantage at least $1$ over every other action,
regardless of the other players' actions. Consequently, the
regret guarantee for the active context bounds the number of rounds when $j$
does not play $y_j$. Summing over the opponents, an $N$-round
epoch with a new context for every relevant player satisfies
\begin{equation}\label{eq:ma-leak-bound-main}
L:=\#\{t\in[N]:a_{-i}^{(t)}\ne y\}
\le C_{-i}^{\mathrm{pin}}\sqrt N,
\qquad C_{-i}^{\mathrm{pin}}:=\sum_{j\ne i}C_j.
\end{equation}
We call these $L$ rounds \emph{leakage rounds}. This is the only
additional source of error relative to the single-agent problem:
on the other $N-L$ rounds, the target player faces the fixed
opponents' action profile $y$. For each target $(i,y,a)$, we again maintain a bracket $[L_k,U_k]$
containing $\theta_{i,y}(a)$, initially $[0,1]$. Setting
\( 
w_k:=U_k-L_k,
m_k:=\frac{L_k+U_k}{2},
\gamma_k:=\frac{w_k}{4},
\Delta_k^\pm:=m_k\pm\gamma_k, 
\) similarly each payment level $\Delta_k^\pm$ is tested in a separate epochs with the length of
\begin{equation}\label{eq:ma-epoch-length}
N_k:=\left\lceil
\left(\frac{8(C_i+2C_{-i}^{\mathrm{pin}})}{\gamma_k}\right)^2
\right\rceil
\end{equation}
rounds. In respective epochs, player $i$ receives $p_i^{(\Delta_k^\pm,a)}$
and every opponent $j\ne i$ receives $p_j^{\mathrm{pin}}$.
Let \(X_{i,y,a}(\Delta)\) denote the number of rounds in which the \emph{joint} action \((a,y)\) is observed within an epoch at level \(\Delta\). We then label the epoch as follows:
\[
\textit{BEST}:\ X_{i,y,a}(\Delta)\ge\tfrac34 N_k,\qquad
\textit{NOTBEST}:\ X_{i,y,a}(\Delta)\le\tfrac14 N_k,\qquad
\textit{INCONCL}:\ \text{otherwise}.
\]

\noindent
Analogous to Section~\ref{sec:single-agent-binary-search}, in the next two lemmas we establish two key properties of our two-level payment test. Specifically, Lemma~\ref{lem:one-sided-bounds-ma} shows that if either payment test is \emph{certified}, we can \emph{partially} locate \(\theta_a\). Furthermore, Lemma~\ref{lem:one-certified-ma} guarantees that at least one of the two tests is \emph{certified}.

\begin{lemma}[Information from a label]\label{lem:one-sided-bounds-ma}
Define
\( 
\beta_k^{\mathrm{MA}}
:=\frac85\frac{C_i+2C_{-i}^{\mathrm{pin}}}{\sqrt{N_k}}.
\) 
For an epoch at payment level $\Delta$:
\begin{enumerate}
\item If its label is \textit{BEST}, then
$\theta_{i,y}(a)\le\Delta+\beta_k^{\mathrm{MA}}$.
\item If its label is \textit{NOTBEST}, then
$\theta_{i,y}(a)\ge\Delta-\beta_k^{\mathrm{MA}}$.
\item If its label is \textit{INCONCL}, then
$|\Delta-\theta_{i,y}(a)|<\gamma_k$.
\end{enumerate}
\end{lemma}
\begin{proof}
See Appendix~\ref{app:multi-agent-proofs}.
\end{proof}

\begin{lemma}\label{lem:one-certified-ma}
For $\Delta_k^\pm=m_k\pm\gamma_k$, at least one epoch is certified.
\end{lemma}
\begin{proof}
See Appendix~\ref{app:multi-agent-proofs}.
\end{proof}

\noindent
By~\eqref{eq:ma-epoch-length}, we have
$\beta_k^{\mathrm{MA}}\le\gamma_k/5$.
Thus we may use the bracket update~\eqref{eq:sa-bracket-update},
with $\beta_k$ replaced by $\beta_k^{\mathrm{MA}}$, which yields the following contraction condition on the maintained brackets.

\begin{proposition}\label{prop:bracket-shrink} The update \eqref{eq:sa-bracket-update} preserves
$\theta_{i,y}(a)\in[L_{k+1},U_{k+1}]$ and satisfies
\( 
U_{k+1}-L_{k+1}\le\frac9{10}(U_k-L_k).
\) 
\end{proposition}
\begin{proof}
See Appendix~\ref{app:multi-agent-proofs}.
\end{proof}

\noindent
Finally, Algorithm~\ref{alg:gbs-ma} applies the procedure separately to every
player \(i\), opponents' action profile \(y \in A_{-i}\), and target \(a \in A_i\) . As in the single-agent
case, each search stops when its bracket has width at most
$\varepsilon$. We provide our sample complexity result for Algorithm~\ref{alg:gbs-ma} in Theorem \ref{thm:multi-main}.

\begin{algorithm}[H]
\caption{BinarySearch-MA}\label{alg:gbs-ma}
\begin{algorithmic}[1]
\Require \( \varepsilon \in(0,1)\)
\For{$i\in[n]$}
  \For{$y\in A_{-i}$}
    \For{$a\in A_i$}
      \State $k\gets0$, $[L_0^{(i,y,a)},U_0^{(i,y,a)}]\gets[0,1]$.
      \While{$U_k^{(i,y,a)}-L_k^{(i,y,a)}>\varepsilon$}
        \State $m_k\gets(L_k^{(i,y,a)}+U_k^{(i,y,a)})/2$,
          $\gamma_k\gets(U_k^{(i,y,a)}-L_k^{(i,y,a)})/4$.
        \State $\Delta_k^\pm\gets m_k\pm\gamma_k$ and
          $N_k\gets\left\lceil[8(C_i+2C_{-i}^{\mathrm{pin}})/\gamma_k]^2\right\rceil$.
        \For{$\sigma\in\{-,+\}$}
          \State Provide a new context $\kappa_{i,a,k}^{\sigma}$ for player $i$ and $\kappa_{i,a,j,y,k}^{\sigma}$ for player $j$, where $j\neq i$
          \State For $N_k$ rounds, pay $p_i^{(\Delta_k^\sigma,a)}$
            to $i$ and $p_j^{(2,y_j)}$ to $j\ne i$, retaining same contexts.
          \State Observe $X_{i,y,a}(\Delta_k^\sigma)$ and label the epoch
            $\ell_k^\sigma$.
        \EndFor
        \State Update $[L_{k+1}^{(i,y,a)},U_{k+1}^{(i,y,a)}]$
          by~\eqref{eq:sa-bracket-update}, with $\beta_k^{\mathrm{MA}}$;
          set $k\gets k+1$.
      \EndWhile
      \State $\widehat\theta_{i,y}(a)\gets
        (L_k^{(i,y,a)}+U_k^{(i,y,a)})/2$.
    \EndFor
    \State $c_{i,y}\gets m_i^{-1}\sum_{b\in A_i}\widehat\theta_{i,y}(b)$.
    \State $\widetilde u_i(a,y)\gets-\widehat\theta_{i,y}(a)+c_{i,y}$
      for all $a\in A_i$.
  \EndFor
\EndFor
\State \Return $\{\widetilde u_i\}_{i=1}^n$.
\end{algorithmic}
\end{algorithm}

\begin{theorem}\label{thm:multi-main}
For every $i\in[n]$ and $y\in A_{-i}$, Algorithm~\ref{alg:gbs-ma}
outputs estimates $\widetilde u_i(\cdot,y)$ such that, with
$W_i(y):=u_i^{\max}(y)-c_{i,y}$,
\( 
\max_{a\in A_i}
\bigl|u_i(a,y)-\widetilde u_i(a,y)-W_i(y)\bigr|
\le\varepsilon.
\) 
The total number of rounds satisfies
\( 
T=O\big(
\frac{M}{\varepsilon^2}\sum_{i=1}^n
(C_i+2C_{-i}^{\mathrm{pin}})^2\big).
\) 
In particular,
$T=O(Mn^3C^2/\varepsilon^2)$.
The number of private context identifiers used across all players is
$O(n^2M\log(1/\varepsilon))$.
\end{theorem}
\begin{proof}
See Appendix~\ref{app:multi-agent-proofs}.
\end{proof}

\subsection{Learning Utilities in Extensive-Form Games}
\label{sec:learn-efg}

We now extend our learning procedure to finite perfect-recall extensive-form games under trajectory feedback. Unlike in normal-form games, an observed terminal history in an extensive-form game reveals only the actions taken along the realized path, leaving behavior at unreached information sets unobserved. Since regret is evaluated against opponents' actual realization plans, their behavior in unobserved branches of the game may differ from the principal's prescribed strategies, even when they comply along the realized path. Thus, a unilateral deviation leading to unreached branches, may cause counterfactual payoffs that cannot be inferred from the observed trajectories. This creates an additional difficulty when learning \emph{conditional continuation values}.  Furthermore, the information set whose continuation values we seek to learn may be reached only on a subset of rounds. To address these difficulties, we extend the two-level payment procedure developed for normal-form games by incentivizing prescribed behavior outside the queried information set and introducing terminal-dependent payments that \emph{protect} the queried player's strategies against opponents' deviations, without requiring observation of the players' strategies.

\vspace{6pt}
\noindent 
We use the realized-payoff regret
$\widehat R_{i,c}^{\mathrm{EF}}$ defined in
Section \ref{subsec:noregret}, with space
$X_i$. For the standalone EFG learning results, the
epoch arguments use
\begin{equation}\label{eq:pathwise-per-signal-regret}
\widehat R_{i,c}^{\mathrm{EF}}(N_i(T,c))
=\max_{x_i^\star\in X_i}
\sum_{t\in\mathcal T_i(T,c)}
\bigl[v_i^{(t)}(x_i^\star)-r_i^{(t)}\bigr]
\le C_i\sqrt{N_i(T,c)}.
\end{equation}

\noindent
For EFGs, our learning objective is determined by the incentive constraints of the
extensive-form correlated equilibrium (EFCE) that we seek to steer players towards.
To understand our learning object, we consider the standard \emph{ex-ante recommendation} device defined in Section \ref{subsec:efce}. Under this protocol, evaluating a unilateral deviation amounts to
comparing the expected terminal utilities of two complete \emph{contingent}
strategies of the deviating player, with the other players' prescribed
\emph{contingent} strategies fixed. We therefore learn the expected utility
differences that determine these comparisons, rather than attempting to
recover the terminal utilities $u_i(z)$.

\vspace{6pt}
\noindent
Fixing a pure profile $\bar s\in\mathcal S:=\prod_i S_i,$ let
$\bar x:=\mathrm{Seq}(\bar s)$. For player $i$ and information set
$I\in\mathcal I_i$, we define the probability of reaching $I$ by
\( 
\rho_{i,I}(\bar s)
:=\Pr_{Z\sim\pi(\cdot\mid\bar x)}(I\preceq Z)
\). 
For $a\in A_I$, let $\bar s_i^{(I\to a)}$ agree with $\bar s_i$ at
every information set other than $I$, where it selects $a$, and let
$\bar x_i^{(I\to a)}:=\mathrm{Seq}_i(\bar s_i^{(I\to a)})$.
Then, for $\rho_{i,I}(\bar s)>0$, the \emph{conditional continuation value}
of action $a$ is defined as
\begin{equation}\label{eq:efg-reference-value}
v_{i,\bar s}^{(I)}(a)
:=\mathbb E_{Z\sim\pi(\cdot\mid
\bar x_i^{(I\to a)},\bar x_{-i})}
\bigl[u_i(Z)\mid I\preceq Z\bigr].
\end{equation}
Here, expectation is with respect to the terminal-history distribution
induced by the stated strategies and chance nodes' probabilities,
conditional on reaching $I$; in particular, it includes uncertainty
about the node reached within $I$ and subsequent chance outcomes.
As in the normal-form sections, we define the \emph{gap to the best action}
\begin{equation}\label{eq:efg-reference-threshold}
\theta_{i,I}^{\bar s}(a)
:=\max_{b\in A_I}
\bigl(v_{i,\bar s}^{(I)}(b)-v_{i,\bar s}^{(I)}(a)\bigr)
\in[0,1].
\end{equation}
Since
\( 
v_{i,\bar s}^{(I)}(a)
=\max_{b\in A_I}v_{i,\bar s}^{(I)}(b)
-\theta_{i,I}^{\bar s}(a),
\) 
learning these gaps determines the conditional continuation
values up to an action-independent additive constant. To relate these values to the players' incentives, we use
\( 
U_i(s)=
\sum_{z\in Z}u_i(z)\pi(z\mid\mathrm{Seq}(s))
\) 
to denote player $i$'s unconditional expected utility under
a pure-strategy profile $s \in \mathcal S$. Changing only the action
at $I$ does not affect the probability of reaching $I$
or the terminal distribution on histories avoiding $I$.
Consequently,
\begin{equation}\label{eq:efg-local-unconditional-difference}
\begin{aligned}
&U_i(\bar s_i^{(I\to a)},\bar s_{-i})-U_i(\bar s)=\rho_{i,I}(\bar s)
\Bigl[
v_{i,\bar s}^{(I)}(a)
-v_{i,\bar s}^{(I)}(\bar s_i(I))
\Bigr],
\end{aligned}
\end{equation}
where the right-hand side is defined to be zero
if $\rho_{i,I}(\bar s)=0$.
By \eqref{eq:efg-local-unconditional-difference},
the learned continuation-value differences determine
the payoff change from modifying a pure strategy at
one information set. Applying this identity successively
allows us to recover payoff differences between arbitrary
pure contingency plans. We now show that these differences
are sufficient to evaluate all the EFCE incentive
constraints of Definition~\ref{def:efce}.

\begin{proposition}
\label{prop:efg-values-suffice-efce}
Consider the ex-ante EFCE recommendation device
of Section~\ref{subsec:efce}.
For player $i$, a trigger $(I,a)$, and a pure
continuation plan $\hat s_i\in S_i(I)$, let
$F_i^{I,\hat s_i}(\bar s_i)\in S_i$ agree with
$\bar s_i$ before $I$ and with $\hat s_i$ at
$I$ and all subsequent information sets. The corresponding trigger-deviation gain is
\begin{equation}\label{eq:efg-efce-gain-from-pure-profiles}
\begin{aligned}
&G_i(\mu;I,a,\hat s_i)=
\sum_{\substack{\bar s\in\mathcal S\\
\bar s_i\in S_i(I,a)}}
\mu(\bar s)
\Bigl[
U_i(F_i^{I,\hat s_i}(\bar s_i),\bar s_{-i})
-U_i(\bar s)
\Bigr].
\end{aligned}
\end{equation}
These gains determine $\delta(\mu)$ in
\eqref{eq:efce-trigger-gap} by maximizing over
all players, trigger pairs, and pure continuation
plans. Suppose that, for every positive-reach
$(\bar s,i,I)$, the estimates
$\widetilde v_{i,\bar s}^{(I)}$ satisfy
\[
\max_{b\in A_I}
\left|
\widetilde v_{i,\bar s}^{(I)}(b)
-v_{i,\bar s}^{(I)}(b)
-c_{i,\bar s,I}
\right|\le\varepsilon
\]
for some action-independent constant
$c_{i,\bar s,I}$.
Then, these estimates, determine estimates
$\widehat D_i(t_i,s_i;s_{-i})$ such that
\begin{equation}\label{eq:efg-pure-plan-difference-accuracy}
\left|
\widehat D_i(t_i,s_i;s_{-i})
-\bigl[U_i(t_i,s_{-i})-U_i(s_i,s_{-i})\bigr]
\right|
\le 2|\mathcal I_i|\varepsilon
\end{equation}
for every $s_i,t_i\in S_i$ and $s_{-i}\in S_{-i}$.
In particular, they determine estimates
$\widehat G_i(\mu;I,a,\hat s_i)$ satisfying
\( 
\big|
\widehat G_i(\mu;I,a,\hat s_i)
-G_i(\mu;I,a,\hat s_i)
\big|
\le 2|\mathcal I_i|\varepsilon
\) 
uniformly over $\mu$, $(I,a)$, and $\hat s_i$,
including mixed continuation plans. In particular, defining $\widehat\delta(\mu)$
by maximizing the estimated trigger-deviation
gains as in \eqref{eq:efce-trigger-gap}, we have
\begin{equation}\label{eq:efg-efce-gap-accuracy}
\left|
\widehat\delta(\mu)-\delta(\mu)
\right|
\le
2\varepsilon\max_{i\in[n]}|\mathcal I_i|,
\qquad
\forall\mu\in\Delta(\mathcal S).
\end{equation}
Consequently, if $\widehat\delta(\mu)\le\eta$,
then $\mu$ is an
$\bigl(\eta+2\varepsilon
\max_i|\mathcal I_i|\bigr)$-EFCE.
\end{proposition}

\begin{proof}
See Appendix~\ref{app:efg-learning-proofs}.
\end{proof}

\vspace{6pt}
\noindent  
As in the multi-agent normal-form case,
we use payments to induce the opponents to follow their
prescribed strategies. Thus, we first describe our indicator function designs to construct payments to learn conditioned continuation values. For a terminal history $z$, let $a_{J,i}(z)$ denote the action
taken at an information set $J$ visited along the path to $z$ by player $i$.
To measure whether the players follow the pure profile
$\bar s$, we define the following indicators
\begin{equation}\label{eq:efg-compliance-def}
\begin{aligned}
\bar d_j(z)&:=\mathbf1\{a_{J,j}(z)=\bar s_j(J)
 \text{ for every visited }J\in\mathcal I_j\},\\
\bar d_{i,I}^{\mathrm{out}}(z)&:=\mathbf1\{a_{J,i}(z)=\bar s_i(J)
 \text{ for every visited }J\in\mathcal I_i\setminus\{I\}\},\\
\bar d_{-i}(z)&:=\prod_{j\ne i}\bar d_j(z),\qquad
\bar d_{i,I}(z):=\bar d_{-i}(z)\bar d_{i,I}^{\mathrm{out}}(z).
\end{aligned}
\end{equation}
In particular, $\bar d_{i,I}(z)=1$
means that all opponents follow their prescribed actions and
player $i$ follows $\bar s_i$ outside $I$, while its action at
$I$ remains unrestricted. Now, fix a learning query $(\bar s,i,I,a)$ and a payment level
$\Delta\in[0,1)$. For $B\ge2$, each opponent $j\ne i$
receives
\begin{equation}\label{eq:efg-insured-opponents}
q_j(z):=B\bar d_j(z).
\end{equation}
Since utilities belong to $[0,1]$, this payment
provides an incentive of at least $B-1$ for following
$\bar s_j$ along the realized trajectory. For the queried player $i$, we additionally need to
account for opponents' actions at information sets that
may be reached under an alternative strategy but not
during actual play. We therefore define
\begin{equation}\label{eq:efg-insured-payment}
q_i^{(I,a,\Delta)}(z)
:=\bar d_{i,I}^{\mathrm{out}}(z)
\Bigl[(B-\Delta)
+\Delta\bar d_{-i}(z)
 \mathbf1\{I\preceq z,\ a_I(z)=a\}
+(1+\Delta)(1-\bar d_{-i}(z))\Bigr].
\end{equation}
Breaking down the design of our payment, it can be seen that if player $i$ follows $\bar s_i$ outside $I$ and all
opponents comply, it receives the baseline payment $B-\Delta$
and an additional $\Delta$ when $I$ is reached and
action $a$ is selected. Moreover, if an opponent deviates,
player $i$ instead receives $B+1$, provided that it remains
compliant outside $I$. Otherwise, its payment is zero.
Thus, all payments are nonnegative, with
$0\le q_j\le B$ for $j\ne i$ and $0\le q_i\le B+1$.

\vspace{6pt}
\noindent
Here, we remark that the additional case of $B+1$ payment is essential for applying
the regret guarantees as the first two cases are rather intuitive to guarantee the compliance of the players. When evaluating an alternative
strategy of player $i$, the opponents may take actions
different from $\bar s_{-i}$ at information sets that
were not visited during actual play, therefore not accounted for. Since
utilities are bounded by one, this payment ensures that
such deviations cannot reduce the alternative strategy's
augmented payoff below its value when the opponents
follow $\bar s_{-i}$.

\vspace{6pt}
\noindent 
We now describe how the principal learns the relevant continuation values for each $(\bar s,i,I,a)$ with reach probability $r:=\rho_{i,I}(\bar s)>0$, through testing epochs. First, let $0<\rho\le r$ be a known lower bound. Such a bound is used to ensure that the information set $I$ is visited sufficiently often during the corresponding testing epoch. Since the mediator knows the game tree and the chance probabilities, it can compute the reach probability under the prescribed pure-strategy profile $\bar s$, and hence obtain a suitable positive lower bound whenever $I$ is reachable.
For an epoch of $N$ rounds with fixed payment $\Delta$,
define the following indicators to count
\begin{equation}\label{eq:efg-clean-counts}
\begin{aligned}
V&:=\sum_{t=1}^N\mathbf1\{\bar d_{i,I}(z^{(t)})=1,
 I\preceq z^{(t)}\}, \quad 
X:=\sum_{t=1}^N\mathbf1\{\bar d_{i,I}(z^{(t)})=1,
 I\preceq z^{(t)},\ a_I(z^{(t)})=a\}.
\end{aligned}
\end{equation}
Here, $V$ counts the rounds in which $I$ is reached
and all players follow their prescribed strategies
outside $I$, while $X$ counts those rounds in which
player $i$ also chooses $a$ at $I$.
Unlike in normal-form games, we consider the frequency
$X/V$ only among these rounds, rather than among
all $N$ rounds. For $V>0$, label the epoch \textit{BEST} if
$X\ge3V/4$, \textit{NOTBEST} if $X\le V/4$,
and \textit{INCONCL} otherwise. If $V=0$,
declare \textit{INCONCL}. To account for the deviations outside $I$ and
the randomness of terminal outcomes, for an epoch
failure probability $\alpha\in(0,1)$ define
\begin{equation}\label{eq:efg-epoch-constants}
\begin{aligned}
\lambda_\alpha:=\sqrt{2\log(4n/\alpha)},\quad 
Q_i(\alpha):=\frac{\sum_{j\ne i}C_j
 +(n-1)(B+1)\lambda_\alpha}{B-1}, \quad 
H_i(\alpha):=C_i+4Q_i(\alpha)+4\lambda_\alpha.
\end{aligned}
\end{equation}
Here, the term $Q_i(\alpha)$ bounds the contribution of
opponents' deviations using their regret guarantees,
while $H_i(\alpha)$ combines this contribution with
player $i$'s regret and the concentration terms.
These quantities determine the number of rounds
required for a \emph{reliable} payment test.

\vspace{6pt}
\noindent 
We now apply the two-level payment search idea from
Sections~\ref{sec:single-agent-binary-search}--
\ref{sec:multi-agent-guarded}.
For $\theta:=\theta_{i,I}^{\bar s}(a)$, we
maintain a bracket $[L_k,U_k]\subseteq[0,1]$,
initially $[0,1]$, and set
\begin{equation}\label{eq:efg-two-payment levels}
\begin{aligned}
\tau_k&:=U_k-L_k,\qquad
m_k:=\frac{L_k+U_k}{2},\qquad
\gamma_k:=\frac{\tau_k}{4},\\
\Delta_k^-&:=m_k-\gamma_k,\qquad
\Delta_k^+:=m_k+\gamma_k.
\end{aligned}
\end{equation}
Each level $\Delta_k^\sigma$, $\sigma\in\{-,+\}$,
is tested in a separate epoch of length
\begin{equation}\label{eq:efg-insured-epoch-length}
N_k:=\left\lceil
\left(\frac{16H_i(\alpha)}{\rho\gamma_k}\right)^2
\right\rceil,
\qquad 0<\rho\le\rho_{i,I}(\bar s).
\end{equation}
Here, the dependence on $\rho$ accounts for the fact that
$I$ is reached only on a subset of rounds, while
$H_i(\alpha)$ accounts for regret, opponents'
deviations, and sampling uncertainty.  As in the normal-form sections, a sufficiently
separated payment level yields a certified label,
which provides a one-sided bound on $\theta$.
The Lemma \ref{lem:certified-epoch-pathwise} establishes these properties
for extensive-form games. 

\begin{lemma}
\label{lem:certified-epoch-pathwise}
Fix a positive-reach query $(\bar s,i,I,a)$, a bracket of width
$\tau_k>0$ containing $\theta$, and either payment level
$\Delta\in\{\Delta_k^-,\Delta_k^+\}$.
Use the payments
\eqref{eq:efg-insured-opponents}--\eqref{eq:efg-insured-payment}
for $N_k$ \eqref{eq:efg-insured-epoch-length}, rounds.
Then, with probability of at least $1-\alpha$, the observed counts
satisfy $V\ge15\rho N_k/16>0$. Moreover:
\begin{enumerate}
\item If $\Delta\ge\theta+\gamma_k$, the label is \textit{BEST}.
\item If $\Delta\le\theta-\gamma_k$, the label is \textit{NOTBEST}.
\item Define
\( 
\beta_k^{\mathrm{EF}}
:=\frac{2H_i(\alpha)}{\rho\sqrt{N_k}}
\le\frac{\gamma_k}{8}.
\) 
Then, a \textit{BEST} label implies $\theta\le\Delta+\beta_k^{\mathrm{EF}}$,
while a \textit{NOTBEST} label implies
$\theta\ge\Delta-\beta_k^{\mathrm{EF}}$.
An \textit{INCONCL} label implies $|\Delta-\theta|<\gamma_k$.
\end{enumerate}
\end{lemma}

\begin{proof}
See Appendix~\ref{subsec:d2}.
\end{proof}
\noindent 
As in the normal-form setting (Lemmas~\ref{lem:onecert} and~\ref{lem:one-certified-ma}), the two payment levels straddle the unknown threshold in the sense that
\( 
\max\{|\Delta_k^--\theta|,|\Delta_k^+-\theta|\}\ge\gamma_k.
\) 
Thus, with probability of at least $1-2\alpha$, at least one of
the two epochs is certified. Let $\ell_k^\sigma$ be the label at
$\Delta_k^\sigma$ and update the bracket by intersecting it with
all certified one-sided bounds:

\begin{equation}\label{eq:efg-two-payment level-update}
\begin{aligned}
\widetilde L_{k+1}
&:=
\max\!\left(
\{L_k\}\cup
\{\Delta_k^\sigma-\beta_k^{\mathrm{EF}}:
  \ell_k^\sigma=\textit{NOTBEST}\}
\right),\\
\widetilde U_{k+1}
&:=
\min\!\left(
\{U_k\}\cup
\{\Delta_k^\sigma+\beta_k^{\mathrm{EF}}:
  \ell_k^\sigma=\textit{BEST}\}
\right),\\
[L_{k+1},U_{k+1}]
&:=
\begin{cases}
[\widetilde L_{k+1},\widetilde U_{k+1}],
&
\begin{array}{l}
\widetilde L_{k+1}\le\widetilde U_{k+1}
\text{ and at least one test is labelled},\\
\end{array}\\[4pt]
[m_k-\gamma_k,m_k+\gamma_k],
&\text{otherwise}.
\end{cases}
\end{aligned}
\end{equation}

\begin{proposition}
\label{prop:efg-bracket-shrink}
On the event that both tests satisfy
Lemma~\ref{lem:certified-epoch-pathwise}, the update
\eqref{eq:efg-two-payment level-update} preserves
$\theta\in[L_{k+1},U_{k+1}]$ and obeys
\( 
U_{k+1}-L_{k+1}\le\frac{25}{32}(U_k-L_k).
\) 
\end{proposition}
\begin{proof}
See Appendix~\ref{subsec:d2}.
\end{proof}

\begin{algorithm}[t]
\caption{\textsc{BinarySearch-Traj}}
\label{alg:gbs-trajlocal}
\begin{algorithmic}[1]
\Require Pure profile $\bar s$, target $(i,I,a)$,
  $\varepsilon\in(0,1)$, $\delta_{\mathrm{loc}}\in(0,1)$,
 $B\ge2$, and $0<\rho\le\rho_{i,I}(\bar s)$.
\State Set $K\gets\lceil\log_{32/25}(1/(2\varepsilon))\rceil$,
 $\alpha\gets\delta_{\mathrm{loc}}/(2K)$.
\State Initialize $k\gets0$ and $[L_0,U_0]\gets[0,1]$.
\While{$U_k-L_k>2\varepsilon$}
  \State Compute $m_k,\gamma_k,\Delta_k^\pm$ using
  \eqref{eq:efg-two-payment levels}, $N_k$ using
  \eqref{eq:efg-insured-epoch-length}, and
  $\beta_k^{\mathrm{EF}}$.
  \For{$\sigma\in\{-,+\}$}
    \State Provide a new context $\kappa_{i,a,I,k}^{\sigma}$ for player $i$ and $\kappa_{i,a,j,\bar s,I,k}^{\sigma}$ for player $j$, where $j\neq i$.
    \State Retain these contexts throughout its $N_k$ episodes.
    \State Run $N_k$ rounds with payments
    \eqref{eq:efg-insured-opponents}--
    \eqref{eq:efg-insured-payment}
    at $\Delta=\Delta_k^\sigma$.
    \State Compute $(V_k^\sigma,X_k^\sigma)$ using
    \eqref{eq:efg-clean-counts} and determine
    the label $\ell_k^\sigma$.
  \EndFor
  \State Update $[L_{k+1},U_{k+1}]$ using
  \eqref{eq:efg-two-payment level-update}.
  \State Set $k\gets k+1$.
\EndWhile
\State $\widehat\theta_{i,I}^{\bar s}(a)
 \gets(L_k+U_k)/2$.
\State \Return $\widehat\theta_{i,I}^{\bar s}(a)$.
\end{algorithmic}
\end{algorithm}

\vspace{6pt}
\noindent
We now use the preceding results to construct our learning
algorithm for extensive-form games given in Algorithm \ref{alg:gbs-trajlocal}. As in the normal-form
sections, we apply the two-payment search separately to each
target action. For a fixed query $(\bar s,i,I,a)$, the mediator
tests the two payment levels $\Delta_k^-$ and $\Delta_k^+$
in separate epochs, assigns their labels using the observed
counts $(V,X)$, and updates the bracket according to
\eqref{eq:efg-two-payment level-update}. By
Proposition~\ref{prop:efg-bracket-shrink}, each successful
iteration preserves the unknown threshold and reduces the
bracket width. Algorithm \ref{alg:gbs-trajlocal} terminates when the bracket width is at most
$2\varepsilon$ and returns its midpoint as the estimate of
$\theta_{i,I}^{\bar s}(a)$.
The following theorem establishes
the accuracy and sample complexity of Algorithm \ref{alg:gbs-trajlocal}, when all such tests satisfy the high-probability statements.

\begin{theorem}
\label{thm:learn-efg-traj-pathwise}
Fix a positive-reach query $(\bar s,i,I,a)$ and a known
lower bound $0<\rho\le\rho_{i,I}(\bar s)$.
Algorithm~\ref{alg:gbs-trajlocal} returns
$\widehat\theta_{i,I}^{\bar s}(a)$ such that
\( 
\Prb\big(
\big|
\widehat\theta_{i,I}^{\bar s}(a)
-\theta_{i,I}^{\bar s}(a)
\big|\le\varepsilon
\big)\ge1-\delta_{\mathrm{loc}}.
\) 
Let
\( 
K:=\left\lceil
\log_{32/25}\frac{1}{2\varepsilon}
\right\rceil,
\alpha:=\frac{\delta_{\mathrm{loc}}}{2K}.
\) 
The total number of rounds satisfies
\( 
T_{\bar s,i,I,a}
\le 2K+\frac{2097152}{399}
\frac{H_i(\alpha)^2}{\rho^2\varepsilon^2}.
\) 
\end{theorem}
\begin{proof}
See Appendix~\ref{app:efg-learning-proofs}.
\end{proof}

\begin{algorithm}[t]
\caption{\textsc{BinarySearch-EFG}:
all positive-reach queries}
\label{alg:gbs-efg-all}
\begin{algorithmic}[1]
\Require
 $\varepsilon\in(0,1)$,
 $\delta\in(0,1)$,
 $B\ge2$, and reach lower bounds \(\rho_{\bar s,i,I} \)  \( \forall \bar s \in \mathcal S, i \in [n],I\in\mathcal I_i \) .
\State Enumerate $\mathcal S=\prod_i S_i$ and
 compute $\mathcal J$ from
 \eqref{eq:efg-total-query-count}.
\ForAll{$\bar s\in\mathcal S$}
  \ForAll{$i\in[n]$ and $I\in\mathcal I_i$}
    \If{$\rho_{i,I}(\bar s)>0$}
      \State Choose
      $0<\rho_{\bar s,i,I}
       \le\rho_{i,I}(\bar s)$.
      \ForAll{$a\in A_I$}
        \State $\widehat\theta_{i,I}^{\bar s}(a)
        \gets\textsc{BinarySearch-Traj}
        (\bar s,i,I,a,\varepsilon,
        \delta/\mathcal J,B,\rho_{\bar s,i,I})$.
      \EndFor
      \State $c_{i,\bar s,I}
      \gets |A_I|^{-1}
      \sum_{b\in A_I}
      \widehat\theta_{i,I}^{\bar s}(b)$.
      \ForAll{$a\in A_I$}
        \State $\widetilde v_{i,\bar s}^{(I)}(a)
        \gets-\widehat\theta_{i,I}^{\bar s}(a)
        +c_{i,\bar s,I}$.
      \EndFor
    \EndIf
  \EndFor
\EndFor
\State \Return
$\{\widetilde v_{i,\bar s}^{(I)}(a):
\rho_{i,I}(\bar s)>0\}$.
\end{algorithmic}
\end{algorithm}

\vspace{6pt}
\noindent
We next extend the procedure to learn the conditional
continuation values for all players, prescribed pure profiles,
and information sets with positive reach probability.
Since the game form and chance probabilities are known,
the mediator can determine which information sets are
reachable under each pure profile and compute their
reach probabilities. We therefore consider the collection
of queries
\begin{equation}\label{eq:efg-total-query-count}
\begin{aligned}
\mathcal J
&:=\#\{(\bar s,i,I,a):
 \bar s\in\mathcal S,\ I\in\mathcal I_i,
 a\in A_I,\ \rho_{i,I}(\bar s)>0\}\le|\mathcal S|
 \sum_{i=1}^n\sum_{I\in\mathcal I_i}|A_I|.
\end{aligned}
\end{equation}
For each positive-reach query, Algorithm~\ref{alg:gbs-trajlocal}
estimates the corresponding gap
$\theta_{i,I}^{\bar s}(a)$ using the single-query
guarantee, with failure probability $\delta/\mathcal J$
allocated to each query. By a union bound, all estimates
are simultaneously accurate with probability of at least
$1-\delta$. Once all actions at a given $(\bar s,i,I)$ have been
tested, the mediator reconstructs the continuation-value
vector by setting
\( 
\widetilde v_{i,\bar s}^{(I)}(a)
:=-\widehat\theta_{i,I}^{\bar s}(a)
+c_{i,\bar s,I},\) where \( 
c_{i,\bar s,I}
:=\frac{1}{|A_I|}
\sum_{b\in A_I}\widehat\theta_{i,I}^{\bar s}(b).
\) 
This normalization centers the estimated continuation
values at zero while preserving all pairwise utility
differences needed to evaluate unilateral deviations.
The full procedure is presented in
Algorithm~\ref{alg:gbs-efg-all}, and its overall
learning guarantee is summarized below.

\begin{theorem}
\label{thm:learn-all-efg-allx}
Suppose $\mathcal J\ge1$, and let
$\rho_{\bar s,i,I}>0$ be the reach lower
bounds for all $(\bar s,i,I,a)$, with
\( 
\rho_{\min}:=
\min_{\rho_{i,I}(\bar s)>0}\rho_{\bar s,i,I}>0.
\) 
Define
\( 
K:=\left\lceil
\log_{32/25}\frac{1}{2\varepsilon}
\right\rceil,
\alpha:=\frac{\delta}{2\mathcal JK},
H_{\max}(\alpha):=\max_i H_i(\alpha).
\) 
Then, with probability of at least $1-\delta$,
for every $(\bar s,i,I)$,
there exists an action-independent constant
$c_{i,\bar s,I}^\star$ such that
\( 
\max_{a\in A_I}
\left|
\widetilde v_{i,\bar s}^{(I)}(a)
-v_{i,\bar s}^{(I)}(a)
-c_{i,\bar s,I}^\star
\right|\le\varepsilon.
\) 
The total number of trajectories satisfies
\( 
T_{\mathrm{tot}}
\le 2\mathcal JK+\frac{2097152}{399}
\frac{\mathcal JH_{\max}(\alpha)^2}
{\rho_{\min}^2\varepsilon^2}, 
\) 
and at most
$2n\mathcal JK$ distinct contexts
are used across the players.
\end{theorem}
\begin{proof}
See Appendix~\ref{app:efg-learning-proofs}.
\end{proof}

\vspace{6pt}
\noindent
We next establish lower bounds for learning the complete table of
conditional continuation values. The first two constructions adapt
arguments of \citet[Theorems~4.3 and~5.5]{Payingagents2025}
to our continuation-value learning target. The third construction
captures an additional difficulty specific to trajectory feedback
in extensive-form games: reaching an information set may itself
be unlikely.

\begin{proposition}
\label{prop:efg-inherited-lower-bounds}
Fix $\delta\in(0,1/2)$ and
$\varepsilon>0$. Under trajectory feedback,
learning all positive-reach
conditional continuation values to accuracy $\varepsilon$,\vspace{2pt}
up to an action-independent constant at each reference information-set triple,
takes at least
\( 
T=\Omega\big(
\max\big\{
\frac{\mathcal J\log(1/\varepsilon)}{\log\mathcal J},
\frac{C^2}{\varepsilon^2},
\frac{\log(1/(2\delta))}{\rho_{\min}}
\big\}\big) 
\) rounds. 
\end{proposition}
\begin{proof}
See Appendix~\ref{app:efg-learning-proofs}.
\end{proof}

\section{Online trajectory-feedback steering to an optimal EFCE}
\label{sec:steering}
\label{sec:learn-compute-steer}

We now use the \emph{continuation values} learned in
Section~\ref{sec:learn-efg} to \emph{steer} the players toward an EFCE that maximizes the mediator's utility through \emph{recommendations} and \emph{payments} in an online manner.
As assumed before, we assume the mediator knows the game form, the chance probabilities,
and her objective $u_0:Z\to[0,1]$, but does not know the
players' terminal utilities.
Similarly, she observes the realized terminal history in each round. 

\subsection{The steering problem}
\label{subsec:efce-steering-problem}

In this section we introduce the \emph{steering} problem. We let $\Gamma$ denote the original game and $\widehat\Gamma$
the mediator-augmented game introduced in Section \ref{subsec:mediator-augment}.
In the EFCE protocol, we define the mediator's commitment space as $X_0 := \Delta(\mathcal S)$. Moreover, the mediator samples a complete pure strategy profile $\bar s \sim \mu$ independently of Nature and reveals each recommendation only when the corresponding information set is reached. We say, fixing $\mu\in X_0$ gives
the induced game $\Gamma_\mu$ among the players, before
payments are added.

\vspace{6pt}
\noindent
We use the spaces $X_i$ and $Y_i$ introduced in
Section~\ref{subsec:mediator-augment}. \emph{Learning rounds}
use original-game realization plans in $X_i$, and the
\emph{steering rounds} use recommendation-dependent responses
in $Y_i$ as players can condition their strategies on the mediator provided recommendations. The space $Y_i$ and the steering context are
fixed while the recommendation law changes.

\vspace{6pt}
\noindent 
Let $d_i\in Y_i$ be the pure strategy that always follows
the mediator's recommendations, completed arbitrarily at
histories following its own disobedience, and put
$d:=(d_1,\ldots,d_n)$. This is one fixed rule, independent
of $\mu$. For $x\in Y$, let $\pi(\cdot\mid\mu,x)$ be the
induced distribution on the original terminal histories
$Z$, and write, for $j\in\{0,1,\ldots,n\}$,
\( 
u_j(\mu,x):=
\mathbb E_{z\sim\pi(\cdot\mid\mu,x)}[u_j(z)].
\) 

\vspace{6pt}
\noindent 
The \emph{revelation principle} concerns the equilibrium outcomes
we seek to implement; see also
\citet[Section~6.2]{SteeringNoRegret24}, and \citet{Zhang2023steer}.
In particular, the \emph{revelation principle} tells us that, for our recommendation protocol,
if $x$ is a Nash equilibrium of $\Gamma_\mu$, there exists
$\mu'\in\Delta(\mathcal S)$ such that
\( 
\pi(\cdot\mid\mu',d)=\pi(\cdot\mid\mu,x)\),
$d$ is a Nash equilibrium of $\Gamma_{\mu'}$. Therefore, it suffices to optimize over recommendation laws
under which $d$ is a Nash equilibrium of $\Gamma_{\mu'}$ without loss of generality. Then, we define the notion of \emph{equilibrium gap} \( \Delta_{\text{eq}}(\mu)\) as follows.  

\begin{definition}
\label{def:efce-equilibrium-gap}
For $\mu\in\Delta(\mathcal S)$, define
\( 
\Delta_{\mathrm{eq},i}(\mu):=
\max_{x_i\in Y_i}
\bigl[u_i(\mu,x_i,d_{-i})-u_i(\mu,d)\bigr],\) and \( 
\Delta_{\mathrm{eq}}(\mu):=
\max_{i\in[n]}\Delta_{\mathrm{eq},i}(\mu).
\) 
The mediator's optimal EFCE value is
\( 
u_0^\star:=
\max_{\substack{\mu\in\Delta(\mathcal S),
                  \Delta_{\mathrm{eq}}(\mu)=0}}
u_0(\mu,d).
\) 
\end{definition}

\vspace{6pt}
\noindent 
Thus $\Delta_{\mathrm{eq},i}(\mu)$ is the largest expected
gain that player $i$ can obtain by departing from obedience. In particular, we have the relation
\[
\begin{aligned}
\Delta_{\mathrm{eq}}(\mu)=0
&\quad\Longleftrightarrow\quad
d\text{ is a Nash equilibrium of }\Gamma_\mu \quad\Longleftrightarrow\quad
\mu\text{ is an EFCE of }\Gamma.
\end{aligned}
\]

\vspace{6pt}
\noindent 
The mediator uses learning contexts and one persistent
steering context for each player to differentiate between the environments. Let $\mathcal L_T$ be
the learning rounds through $T$ and
$\mathcal C_T:=[T]\setminus\mathcal L_T$ the steering rounds.
On a learning round, the active test has a reference
profile $\bar s^{(t)}$, sends no recommendations, and uses
the payments of Section~\ref{sec:learn-efg}.
Players select strategies as $x^{(t)}\in X$.

\vspace{6pt}
\noindent 
All steering rounds use the same context
$\kappa_i^{\mathrm{st}}$ for player $i$,
so the player's can differentiate the steering rounds from learning rounds. On a steering round, the mediator chooses
$\mu_{\mathrm{play}}^{(t)}\in\Delta(\mathcal S)$, samples
$\bar s^{(t)}\sim\mu_{\mathrm{play}}^{(t)}$ independently of
the current Nature randomization, and sends sequential
EFCE recommendations. Players select response strategies
$x^{(t)}\in Y$ before these recommendations are sampled.
The payment rule $q_i^{(t)}(\bar s,z)$ may use the mediator's
own sampled profile and the observed terminal history.

\vspace{6pt}
\noindent 
On steering rounds, we let $d_i[\bar s,z]\in\{0,1\}$ indicate
whether player $i$ follows every recommendation received
along $z$, and put
$\hat d[\bar s,z]:=\prod_i d_i[\bar s,z]$.
The strategy $d_i\in Y_i$ follows recommendations;
the indicator $d_i[\bar s,z]$ records agreement on a
realized trajectory. These indicators require no
observation at unreached information sets. We define the expected
augmented steering utility as
\( 
v_i^{(t)}(\mu,x):=
\mathbb E_{\bar s\sim\mu,\,z\mid\bar s,x}
\bigl[u_i(z)+q_i^{(t)}(\bar s,z)\bigr].
\) 
Moreover, we define the actual and reference terminal laws by
\[
\begin{aligned}
\pi_{\mathrm{act}}^{(t)}&:=
\begin{cases}
\pi(\cdot\mid x^{(t)}),&t\in\mathcal L_T,\\
\pi(\cdot\mid\mu_{\mathrm{play}}^{(t)},x^{(t)}),
&t\in\mathcal C_T,
\end{cases}\\[2mm]
\pi_{\mathrm{ref}}^{(t)}&:=
\begin{cases}
\pi(\cdot\mid\mathrm{Seq}(\bar s^{(t)})),
&t\in\mathcal L_T,\\
\pi(\cdot\mid\mu_{\mathrm{play}}^{(t)},d),
&t\in\mathcal C_T.
\end{cases}
\end{aligned}
\]
Thus a learning round is compared with its reference play, while a steering round is compared
with obedience under its recommendation law.

\begin{definition}[Directness gap, optimality gap,
and average payments]
\label{def:efce-steering-gaps}
For a horizon $T$, define
\begin{align}
DG(T)
&:=
\left\|
\frac1T\sum_{t=1}^T
\bigl(\pi_{\mathrm{act}}^{(t)}
-\pi_{\mathrm{ref}}^{(t)}\bigr)
\right\|_1,
\label{eq:efce-directness-gap}\\
OG(T)
&:=u_0^\star
-\frac1T\sum_{t=1}^T u_0(z^{(t)}),
\label{eq:efce-optimality-gap}\\
\mathrm{Pay}_i(T)
&:=\frac1T\sum_{t=1}^T
q_i^{(t)}(\bar s^{(t)},z^{(t)}).
\label{eq:efce-average-payments}
\end{align}
\end{definition}

\vspace{6pt}
\noindent 
Then, we define the steering problem as choosing contexts,
recommendation laws on steering rounds, and payment rules
using the mediator's available observations so that
\[
DG(T)\longrightarrow0,\qquad
[OG(T)]_+\longrightarrow0,\qquad
\mathrm{Pay}_i(T)\longrightarrow0
\quad\text{for every }i\in[n].
\] 
Here it is important to note that the optimality benchmark
is the original game's optimal EFCE value without
payments. Its positive shortfall is used because
intermediate play can give the mediator more than
$u_0^\star$. Moreover, vanishing average payments permits a growing total
expenditure provided that it is sublinear in $T$.

\vspace{6pt}
\noindent 
On the other hand, if the players' utilities were known, one could first
compute an optimal EFCE by linear programming and then
steer toward that fixed recommendation law. Here, the incentive constraints are initially unknown, and the explicit optimization problem can be large, making it generally more expensive to solve than performing a single step of a regret minimizer \citep{SteeringNoRegret24}. In particular, a distribution over $\mathcal S$ has one coordinate for each complete pure strategy profile, and $|\mathcal S|$ can be exponential in the size of $\Gamma$. Therefore, our approach instead updates the
recommendation law through regret minimization steps through a
Lagrangian formulation, using increasingly accurate
continuation estimates while steering proceeds.

\vspace{6pt}
\noindent 
We distinguish the regret guarantees used in the
learning and steering rounds.
Each epoch within the learning steps uses a new context for every player.
For a learning epoch of prescribed length $m$, with play
times $t_1,\ldots,t_m$, and unique context for each player, we use the realized-payoff regret
bound from Section \ref{sec:learn-efg}:
\begin{equation}\label{eq:efce-conditional-query-regret}
\max_{x_i^\star\in X_i}
\sum_{\ell=1}^m
\bigl[
v_i^{(t_\ell)}(x_i^\star)-r_i^{(t_\ell)}
\bigr]
\le C_i\sqrt m.
\end{equation}

\vspace{6pt}
\noindent
Index the steering rounds by $k=1,\ldots,N$, excluding learning rounds, and let $P$ bound the
payments. Then, for steering expected-payoff
regret in the recommendation-augmented game is,
\begin{equation}\label{eq:efce-steering-regret}
R_{i,\kappa_i^{\mathrm{st}}}(N)
=
\max_{x_i^\star\in Y_i}
\sum_{k=1}^N
\bigl[
v_i^{(k)}(\mu^{(k)},x_i^\star,x_{-i}^{(k)})
-v_i^{(k)}(\mu^{(k)},x^{(k)})
\bigr]
\le (P+1)R_i(N).
\end{equation}
Importantly, for square-root regret guarantees, we write
\( 
R_i(N):=C_i^{\mathrm{st}}\sqrt N.
\)

\subsection{The Lagrangian formulation}
\label{subsec:efce-lagrangian}

We now use the learned continuation values to compute
a recommendation law that maximizes the mediator's
utility subject to the EFCE incentive constraints.
Following \citet{Zhang2023steer}, we express these
constraints through deviation gains and optimize a
Lagrangian formulation using regret minimization.

\vspace{6pt}
\noindent
For each player $i$, let $\mathcal D_i$ be the finite
set of pure strategies in $Y_i$, and write
\( 
\mathcal D:=\{0\}\cup
\{(i,\tau_i):i\in[n],\ \tau_i\in\mathcal D_i\}.
\) 
For a sampled recommendation profile $\bar s$,
let $A_{\bar s,(i,\tau_i)}$ be player $i$'s expected
gain from using $\tau_i$ when the other players follow
their recommendations. We call these matrices, $A$, as \emph{continuation tables}. Thus, for every recommendation law $\mu$,
\begin{equation}\label{eq:efce-gain-matrix}
u_i(\mu,\tau_i,d_{-i})-u_i(\mu,d)
=
\sum_{\bar s\in\mathcal S}
\mu(\bar s)A_{\bar s,(i,\tau_i)},
\qquad A_{\bar s,0}:=0.
\end{equation}
Since a pure strategy attains the maximum deviation
gain, we can write 
\( 
\Delta_{\mathrm{eq}}(\mu)
=
\max_{y\in\Delta(\mathcal D)}\mu^\top Ay.
\) 
Here $y$ forms a weighted average of the deviation
gains. Its maximum is attained by putting all weight
on a deviation with the largest gain. The incentive constraints can first
be written as
\[
\max_{x_i\in Y_i}u_i(\mu,x_i,d_{-i})
\le u_i(\mu,d),
\qquad i\in[n].
\]
They hold for every strategy if, and only if, they hold
for every pure strategy, which is precisely
$A^\top\mu\le0$.
Writing $c[\bar s]:=U_0(\bar s)$, the mediator's
optimization problem becomes
\begin{equation}\label{eq:efce-primal-program}
u_0^\star
=
\max_{\mu\in\Delta(\mathcal S)}
\bigl\{c^\top\mu:A^\top\mu\le0\bigr\}.
\end{equation}
The vector $c$ is known because the mediator's
objective and the chance probabilities are known.
We apply the Lagrangian approach of
\citet{Zhang2023steer} to the finite
constraints $A^\top\mu\le0$.
Lemma~\ref{lem:efce-lagrangian-justification} gives
\[
u_0^\star
=
\max_{\mu\in\Delta(\mathcal S)}
\inf_{\substack{\ell\ge0\\y\in\Delta(\mathcal D)}}
\bigl[c^\top\mu-\ell\mu^\top Ay\bigr].
\]
Here $\ell$ is an optimization variable.
For computation, we replace the unrestricted
multiplier by a bounded one, $0\le\ell\le\lambda$,
where $\lambda\ge1$ is chosen before the run.
This restriction changes the optimization problem.
Proposition~\ref{prop:efce-saddle-bound} bounds
the resulting objective and equilibrium errors.
For every $\mu$, we write the new problem as,
\[
\begin{aligned}
&\min_{\substack{0\le\ell\le\lambda\\
                 y\in\Delta(\mathcal D)}}
\bigl[c^\top\mu-\ell\mu^\top Ay\bigr]
=c^\top\mu-\lambda\Delta_{\mathrm{eq}}(\mu),
\end{aligned}
\]
because $\Delta_{\mathrm{eq}}(\mu)\ge0$.
Dividing by $\lambda$ yields
\begin{equation}\label{eq:efce-lagrangian}
\max_{\mu\in\Delta(\mathcal S)}
\min_{y\in\Delta(\mathcal D)}L_\lambda(\mu,y),
\qquad
L_\lambda(\mu,y)
:=
\frac{c^\top\mu}{\lambda}-\mu^\top Ay.
\end{equation}
Thus
\[
\min_{y\in\Delta(\mathcal D)}L_\lambda(\mu,y)
=
\frac{u_0(\mu,d)}{\lambda}
-\Delta_{\mathrm{eq}}(\mu).
\]
Now, one problem about this game is that the mediator initially does not know the $A$. Therefore, the payoff differences defining $A$ are determined
by the continuation values learned in
Section~\ref{sec:learn-efg}.
By Proposition~\ref{prop:efg-values-suffice-efce}, as shown in Appendix \ref{app:efce-steering-proofs}, 
a continuation table with accuracy $\varepsilon$
yields an estimate $\widehat A$ satisfying
\begin{equation}\label{eq:efce-gain-matrix-accuracy}
\|\widehat A-A\|_\infty
\le 2\varepsilon\max_i|\mathcal I_i|,
\end{equation}
where $I_{\max}:=\max_i|\mathcal I_i|$ and
$\|M\|_\infty$ denotes the maximum absolute
entry of $M$.
Letting $\widehat A^{(t)}$ be the estimate available before
round $t$, the mediator updates two regret-minimization algorithms
using
\[
\widehat L_t(\mu,y)
:=
\frac{c^\top\mu}{\lambda}
-\mu^\top\widehat A^{(t)}y.
\]
One algorithm maximizes over recommendation laws
$\mu\in\Delta(\mathcal S)$, and the other one minimizes
over $y\in\Delta(\mathcal D)$.
As learning proceeds, the same algorithms continue
with the updated estimates of $\widehat A$.
The following result bounds the objective and
equilibrium errors of the mediator in terms of their regrets and
the error in the learned gains. Ultimately, it demonstrates that choosing a suitably growing $\lambda$ is essential. 
\begin{proposition}
\label{prop:efce-saddle-bound}
Fix $\lambda\ge1$, and let $\mu^{(t)},y^{(t)}$
be the iterates of the mediator's two algorithms.
Suppose that their cumulative regrets for the estimated
payoff sequence have nonnegative bounds $R_0(t)$ and
$R_{\min}(t)$, respectively. Suppose also that
$\|\widehat A^{(r)}-A\|_\infty\le e_r$
for every $r\le t$. Define
\( 
\bar\mu^{(t)}
:=\frac1t\sum_{r=1}^t\mu^{(r)},
\) \( 
a_t
:=
\frac{R_0(t)+R_{\min}(t)}{t}
+\frac2t\sum_{r=1}^te_r.
\) 
Then,
\begin{align}
u_0(\bar\mu^{(t)},d)
&\ge u_0^\star-\lambda a_t,
\label{eq:efce-saddle-objective}\\
\Delta_{\mathrm{eq}}(\bar\mu^{(t)})
&\le\lambda^{-1}+a_t.
\label{eq:efce-saddle-feasibility}
\end{align}
\end{proposition}

\begin{proof}
See Appendix~\ref{app:efce-steering-proofs}.
\end{proof}

\subsection{The Learn-Compute-Steer algorithm}
\label{subsec:efce-online-algorithm}

We now present our \emph{Learn-Compute-Steer} algorithm,
Algorithm~\ref{alg:online-efce}, which combines learning
continuation values, computing an optimal EFCE for the
mediator, and steering players towards it.
The rounds are split between learning and steering.
For $q\in(0,1)$, round $t$ is a learning round when
\( 
\lfloor t^q\rfloor>\lfloor(t-1)^q\rfloor.
\) 
Thus, when $\mathcal J>0$, the first $t$ rounds contain
exactly $\lfloor t^q\rfloor$ learning trajectories.
On a learning round, the players receive the payments
in \eqref{eq:efg-insured-opponents}--\eqref{eq:efg-insured-payment};
on a steering round, they receive the payments in
\eqref{eq:efce-trajectory-payment}.
The mediator's two regret-minimization algorithms
are updated on every round, using the matrix
$\widehat A^{(t)}$ obtained from the most recently
completed continuation table.

\vspace{6pt}
\noindent
For the \emph{learning} part of the Algorithm \ref{alg:online-efce}, we learn the continuation table, $A$, repeatedly, starting
with a requested accuracy $\varepsilon\in(0,1/2)$.
Each run of Algorithm~\ref{alg:gbs-efg-all} keeps this
accuracy fixed until it finishes.
Once the table is complete, we construct the corresponding
matrix $\widehat A$, halve $\varepsilon$, and begin
the next run.
While the next table is being learned, the mediator
continues using the previous matrix.
Before the first table is complete, we use
$\widehat A=0$.
A table completed at the end of round $t$ is first
used in round $t+1$.

\vspace{6pt}
\noindent
For $\mathcal J>0$, a run requesting accuracy
$\varepsilon$ uses the confidence allocation
\begin{equation}\label{eq:efce-call-confidence}
\begin{aligned}
K(\varepsilon)
:=\left\lceil
\log_{32/25}\frac{1}{2\varepsilon}
\right\rceil, \quad 
M(\varepsilon)
:=2\mathcal J K(\varepsilon),\quad 
\delta_{\mathrm{call}}(\varepsilon)
:=\frac{\delta\varepsilon}{3},\quad 
\zeta(\varepsilon)
:=\frac{\delta_{\mathrm{call}}(\varepsilon)}
{M(\varepsilon)}.
\end{aligned}
\end{equation}
There are at most $M(\varepsilon)$ testing epochs,
each with sampling-failure probability at most
$\zeta(\varepsilon)$.
Their lengths use the regret constants $C_i$
and
\( 
H(\varepsilon)
:=\max_i H_i\bigl(\zeta(\varepsilon)\bigr),
\) 
with $H_i$ defined in
\eqref{eq:efg-epoch-constants}.
Since the requested accuracy is halved after each
completed run, the sum of
$\delta_{\mathrm{call}}(\varepsilon)$ over all runs
is at most $\delta/3$.

\begin{algorithm}[t]
\caption{Learning continuation values during steering}
\label{alg:efce-learning-during-steering}
\begin{algorithmic}[1]
\Require $\varepsilon\in(0,1/2)$, $\delta$, $B\ge2$, reach lower bounds \(\rho_{\bar s,i,I} \)  \( \forall \bar s \in \mathcal S, i \in [n],I\in\mathcal I_i \) ,
and $C_i$ in
\eqref{eq:efce-conditional-query-regret}.
\State Compute $\mathcal J$ from
\eqref{eq:efg-total-query-count};
set $\widehat A\gets0$.

\Procedure{Start learning}{}
  \State Compute the confidence levels in
  \eqref{eq:efce-call-confidence}
  at the current accuracy $\varepsilon$.
  \State Start Algorithm~\ref{alg:gbs-efg-all}
  with accuracy $\varepsilon$, sampling input
  $\delta_{\mathrm{call}}(\varepsilon)$,
  and regret constants $C_i$.
\EndProcedure

\If{$\mathcal J>0$}
  \State \Call{Start learning}{}.
\EndIf

\Procedure{One learning round}{}
  \State Continue the current epoch for one
  trajectory, using its prescribed contexts and payments.
  \State Update its counts from the observed
  terminal history.
  \If{the continuation table is complete}
    \State Construct $\widehat A$ using
    \eqref{eq:efce-estimated-gain-matrix},
    for use from the next round.
    \State Set $\varepsilon\gets\varepsilon/2$.
    \State \Call{Start learning}{}.
  \EndIf
\EndProcedure
\end{algorithmic}
\end{algorithm}

\vspace{6pt}
\noindent
The learning algorithm advances only on learning rounds.
If a testing epoch is interrupted by steering rounds,
it keeps its reference profile, contexts, payment rule,
prescribed length, and accumulated counts.
The next learning round continues the same epoch.
Algorithm~\ref{alg:efce-learning-during-steering}
describes these updates.
If $\mathcal J=0$, then $A=0$ and every round is
a steering round.

\vspace{6pt}
\noindent
We next describe the \emph{computing} part of the Algorithm \ref{alg:online-efce}.
At round $t$, the two regret-minimization algorithms
select $\mu^{(t)}$ and $y^{(t)}$ and are updated
with the reward vectors
\begin{equation}\label{eq:efce-update-vectors}
g_0^{(t)}
:=\frac{c}{\lambda}-\widehat A^{(t)}y^{(t)},
\qquad
g_{\min}^{(t)}
:=(\widehat A^{(t)})^\top\mu^{(t)}.
\end{equation}
The first algorithm maximizes its reward over
$\Delta(\mathcal S)$.
The second one also maximizes its reward, over
$\Delta(\mathcal D)$; because the term involving
$y$ in $\widehat L_t$ has a negative sign,
this minimizes $\widehat L_t$.
These updates require only $c$ and $\widehat A^{(t)}$,
and thus the players do not need to report their strategies. Therefore, a no-regret algorithm like multiplicative weights (MWU) gives square-root
regret for both algorithms, with constants that
do not depend on $\lambda$
\citep{yoavschapire}.
Lemma~\ref{lem:efce-computed-regret}
states the resulting bounds.

\vspace{6pt}
\noindent
Finally, for the \emph{steering} part of Algorithm \ref{alg:online-efce}, on a round $t$, the mediator uses the
running average $\bar\mu^{(t)}$ as the recommendation
law, as required by
Proposition~\ref{prop:efce-saddle-bound}.
She samples $\bar s^{(t)}\sim\bar\mu^{(t)}$
and reveals recommendations as play proceeds,
stopping recommendations to a player after that
player first disobeys.
After observing the terminal history, she makes
the payments of \textsc{TrajectorySteer}
\citep{SteeringNoRegret24}:
\begin{equation}\label{eq:efce-trajectory-payment}
q_i(\bar s,z)
:=
\alpha\hat d[\bar s,z]
+
P d_i[\bar s,z]\bigl(1-\hat d[\bar s,z]\bigr).
\end{equation}

\begin{algorithm}[t]
\caption{Learn-Compute-Steer}
\label{alg:online-efce}
\label{alg:epochal-online-bce-corrected}
\label{alg:learn-theta-then-compute-steer}
\begin{algorithmic}[1]
\Require Horizon $T$, confidence $\delta$,
initial accuracy $\varepsilon\in(0,1/2)$,
and the inputs to
Algorithm~\ref{alg:efce-learning-during-steering}.

\State Set $q\gets3/4$, $\lambda\gets T^{1/8}$,
$\alpha\gets T^{-1/8}$, and $P\gets8nT^{1/8}$.
\State Initialize
Algorithm~\ref{alg:efce-learning-during-steering}
and regret minimizers on
$\Delta(\mathcal S)$ and $\Delta(\mathcal D)$.

\For{$t=1,\ldots,T$}
  \State Read the latest matrix $\widehat A^{(t)}$.
  \State Obtain $\mu^{(t)}$ and $y^{(t)}$
  from the two regret minimizers.
  \State Compute
  $\bar\mu^{(t)}
  \gets t^{-1}\sum_{r=1}^t\mu^{(r)}$.
  \State Update both regret minimizers using
  \eqref{eq:efce-update-vectors}.

  \If{$\mathcal J>0$ and
  $\lfloor t^q\rfloor>\lfloor(t-1)^q\rfloor$}
    \State Execute one learning round of
    Algorithm~\ref{alg:efce-learning-during-steering}.
  \Else
    \State Provide context $\kappa_i^{\mathrm{st}}$
    to each player $i$.
    \State Sample
    $\bar s^{(t)}\sim\bar\mu^{(t)}$, and reveal recommendations
    as play proceeds.
    \State Observe $z^{(t)}$ and pay each player
    $q_i(\bar s^{(t)},z^{(t)})$ from
    \eqref{eq:efce-trajectory-payment}.
  \EndIf
\EndFor
\end{algorithmic}
\end{algorithm}

\vspace{6pt}
\noindent
We now analyze the payments and outcomes of
Algorithm~\ref{alg:online-efce}.
As mentioned, on each steering round $t$, the algorithm uses
$\bar\mu^{(t)}$, whose equilibrium gap is bounded
by Proposition~\ref{prop:efce-saddle-bound}.
The following Lemma \ref{lem:efce-changing-policy-steering} relates the average of these
gaps and the players' regret to their probability
of disobeying and their expected payments.

\begin{lemma}
\label{lem:efce-changing-policy-steering}
Consider $N$ steering rounds with recommendation
laws $\mu^{(k)}$ and fixed payment parameters
$0<\alpha\le1$ and $P\ge2$, satisfying
$\alpha P\ge8n$.
Define
\[
\bar\eta_N
:=\frac1N\sum_{k=1}^N
\Delta_{\mathrm{eq}}(\mu^{(k)}),
\qquad
D^{(k)}
:=\sum_i
\Prb\bigl\{
d_i[\bar s^{(k)},z^{(k)}]=0
\bigr\}.
\]
Each probability is conditional on the history and
the strategies selected before the recommendations
and terminal history of round $k$ are sampled.
If \eqref{eq:efce-steering-regret} holds, then
\begin{equation}\label{eq:efce-changing-policy-bound}
\frac1N\sum_{k=1}^N D^{(k)}
\le
\frac2\alpha
\left[
\left(1+\frac{4n}{P}\right)
\frac{(P+1)\sum_i R_i(N)}{N}
+n\bar\eta_N
\right].
\end{equation}
Each player's average expected payment over these
rounds is at most
\( 
\alpha+\frac{P}{N}\sum_{k=1}^N D^{(k)}.
\) 
\end{lemma}

\begin{proof}
See Appendix~\ref{app:efce-steering-proofs}.
\end{proof}

\vspace{6pt}
\noindent
Finally, the following theorem combines the low-deviation guarantees for recommendations and payments established in Lemma \ref{lem:efce-changing-policy-steering} with the learning schedules of Section~\ref{sec:learn-efg} to show that the optimality gap, directness gap, and time-averaged payments all vanish under Algorithm \ref{alg:online-efce}.

\begin{theorem}
\label{thm:online-efce}
\label{thm:main-corrected}
Fix an initial accuracy $\varepsilon\in(0,1/2)$
independently of $T$ and $\delta$.
Assume that the regret inequalities
\eqref{eq:efce-conditional-query-regret}
and \eqref{eq:efce-steering-regret},
with $R_i(N)=C_i^{\mathrm{st}}\sqrt N$.
Let $T\ge64$ and $\delta\in(0,1)$.
Run Algorithm~\ref{alg:online-efce}
with the confidence allocation
\eqref{eq:efce-call-confidence}
and the internal regret minimizers of
Lemma~\ref{lem:efce-computed-regret}.
Define
\( 
a:=\frac12,\) \( 
\Lambda_T:=1+\log\!\left(
\frac{n\max\{1,\mathcal J\}(T+1)^2}{\delta}
\right).
\) 
There is a constant $C$ for all $i$, independent of $T$ and $\delta$, such that,
with probability at least $1-\delta$,
\begin{align}
\mathrm{Pay}_i(T)
\le C\Lambda_T^a T^{-1/8}, \qquad
DG(T)
\le C\Lambda_T^a T^{-1/4},\qquad
[OG(T)]_+
\le C\Lambda_T^a T^{-1/4}.
\label{eq:online-efce-gap-rates}
\end{align}
\end{theorem}

\begin{proof}
See Appendix~\ref{app:efce-steering-proofs}.
\end{proof}

\section{Experiments}
\label{sec:experiments}

We evaluate the online learn--compute--steer procedure, Algorithm \ref{alg:learn-theta-then-compute-steer} on a normal-form
Stag Hunt game and an extensive-form, two-player Kuhn poker game \citep{Kuhn1950}.
In the experiments, the mediator updates its recommendation distribution by MWU \citep{yoavschapire} and its internal deviation policies by CFR+ \citep{CFR+}. During steering, players also use CFR+ in the recommendation-augmented game.

\vspace{6pt}
\noindent
At steering round $k$, let $Z_k$ be its
sampled terminal history. With principal terminal reward $c(z)$ and optimal
equilibrium value $v^\star$, the plotted quantities are
\( 
\widehat{\mathrm{OG}}_N
=\big[v^\star-\frac1N\sum_{k=1}^N c(Z_k)\big]_+, 
\widehat{\mathrm{Pay}}_{i,N}
=\frac1N\sum_{k=1}^N q_{i,t_k},
\mathrm{DG}_N
=\big\|\frac1N\sum_{k=1}^N
\bigl(\pi_k^{\mathrm{act}}-\pi_k^{\mathrm{ref}}\bigr)\big\|_1.
\) We use 20 simulation seeds for Stag Hunt and 10 for Kuhn poker, with fixed
initialization within each seed comparison. 

\begin{figure*}[!t]
\centering
\begin{minipage}[t]{0.32\textwidth}
\centering
\includegraphics[width=\linewidth]{\detokenize{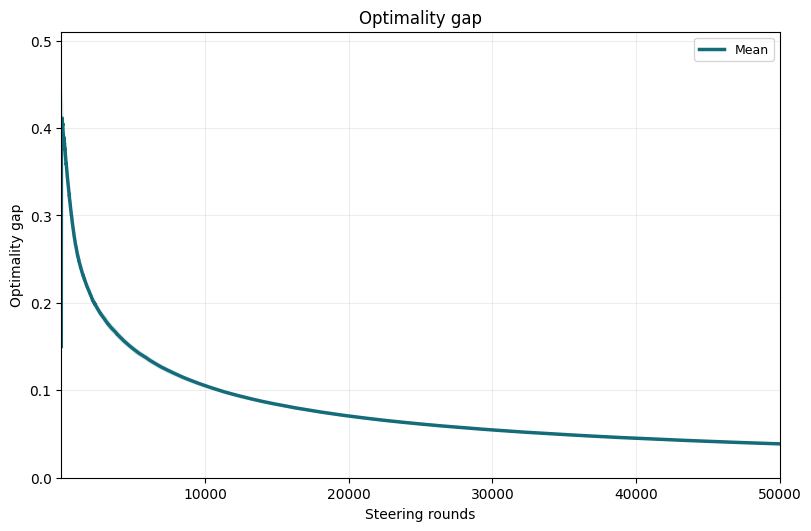}}
\par\small (a) Optimality gap.
\end{minipage}\hfill
\begin{minipage}[t]{0.32\textwidth}
\centering
\includegraphics[width=\linewidth]{\detokenize{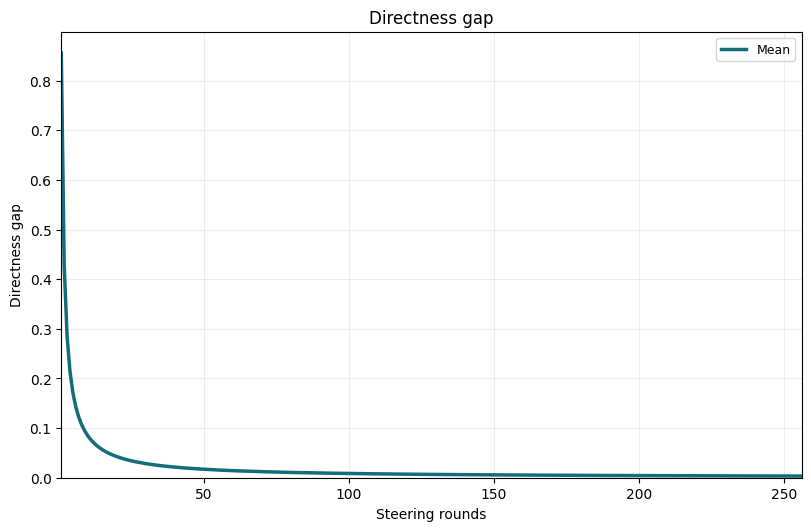}}
\par\small (b) Directness gap.
\end{minipage}\hfill
\begin{minipage}[t]{0.32\textwidth}
\centering
\includegraphics[width=\linewidth]{\detokenize{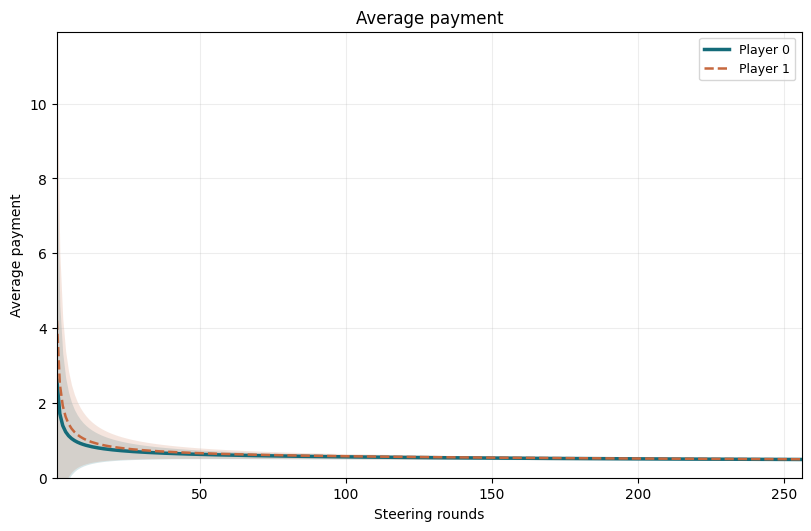}}
\par\small (c) Average realized payments.
\end{minipage}
\caption{\textbf{Stag Hunt.} Means over 20 independent runs, with shading
denoting one sample standard deviation. Panel (a) displays the first
50,000 steering rounds; panels (b)-(c) display the first 256.}
\label{fig:stag-metrics}
\end{figure*}

\subsection{Normal-form Stag Hunt}
\label{subsec:stag-hunt}

For normal-form game experiments, we consider a Stag Hunt game, where Table~\ref{tab:stag-hunt} gives the players' utilities. The principal
maximizes normalized social welfare, $(u_1+u_2)/2$, so that
$v^\star=1$, attained at $(S,S)$. This outcome is also a strict Nash
equilibrium of the original game. Thus the principal optimum is already
equilibrium-feasible, making this a useful steering example with little
conflict between objective improvement and feasibility.

\begin{table}[h]
\centering
\begin{tabular}{c|cc}
 & Stag $(S)$ & Hare $(H)$ \\
\hline
Stag $(S)$ & $(1.0,\,1.0)$ & $(0.0,\,0.7)$ \\
Hare $(H)$ & $(0.7,\,0.0)$ & $(0.6,\,0.6)$
\end{tabular}
\caption{Payoff matrix of the Stag Hunt game. Each entry shows $(u_0,u_1)$.}
\label{tab:stag-hunt}
\end{table}
\noindent
Figure~\ref{fig:stag-metrics} illustrates the different time scales of
obedience and recommendation improvement. Players follow the mediator's
recommendations after the first updates. Once both players obey, the
deviation-compensation payment is no longer incurred, and each player
receives only the compliance bonus $\alpha_t$. The payment curves thus
reflect both the dilution of early compensation payments and the subsequent
decay of $\alpha_t$. Optimality improves more gradually because the mediator
must continue improving her recommendation distribution.

\begin{figure}[!t]
\centering
\includegraphics[scale=0.1, width=\linewidth]{\detokenize{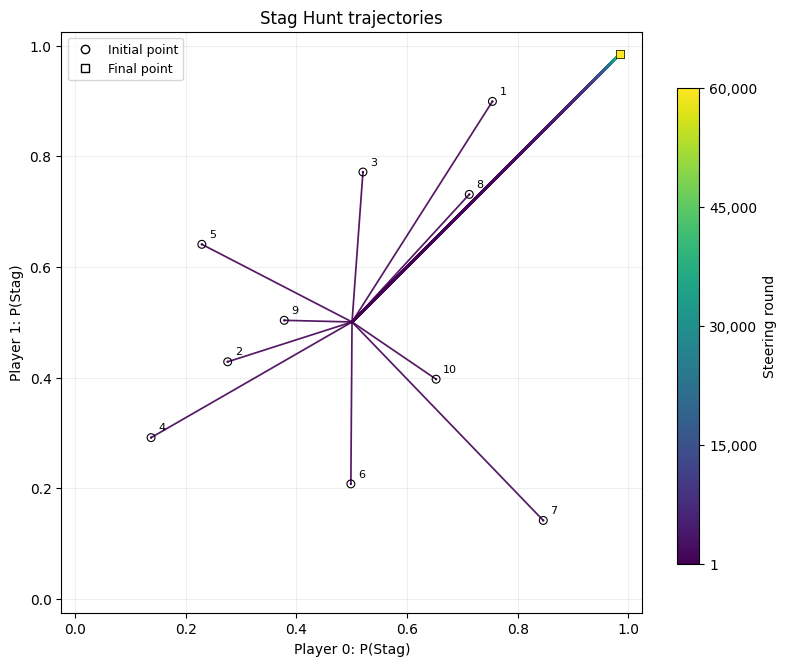}}
\caption{\textbf{Stag Hunt trajectories.} Ten initial points, displayed
through round 60,000. Displayed coordinates are current marginal action
probabilities and color indicates the steering round.}
\label{fig:stag-trajectories}
\end{figure}

\medskip
\noindent
Figure~\ref{fig:stag-trajectories} complements these metrics with trajectories
from ten stratified initial points in $[0.1,0.9]^2$. The initial movement toward $(0.5,0.5)$ reflects players learning to follow the initially near-uniform
recommendations. Once obedience is established, the trajectories track the
recommendation marginals. Their subsequent diagonal movement toward $(1,1)$
reflects the symmetry of the game and the mediator's computation progressively
concentrating recommendations on the optimal outcome $(S,S)$. Thus, the
trajectories distinguish the rapid establishment of obedience from the
slower improvement of the recommendations being followed.

\subsection{Extensive-form game: two-player Kuhn poker}
\label{subsec:kuhn}

\begin{figure*}[!t]
\centering
\begin{minipage}[t]{0.32\textwidth}
\centering
\includegraphics[width=\linewidth]{\detokenize{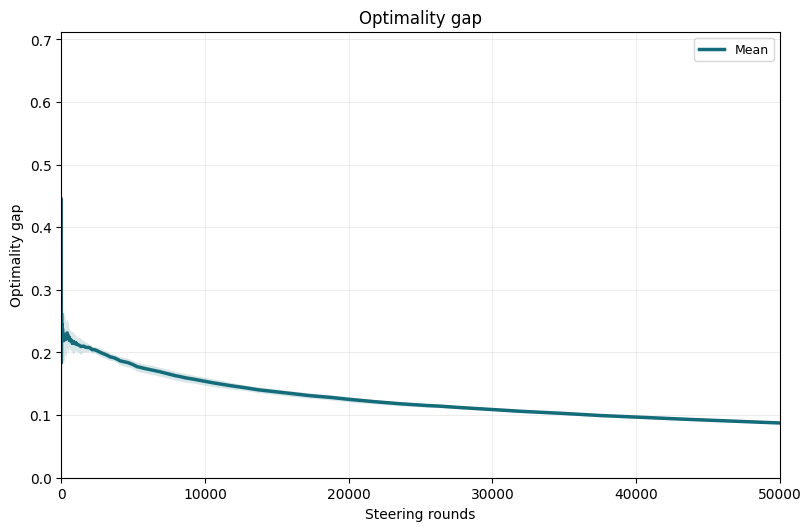}}
\par\small (a) Optimality gap.
\end{minipage}\hfill
\begin{minipage}[t]{0.32\textwidth}
\centering
\includegraphics[width=\linewidth]{\detokenize{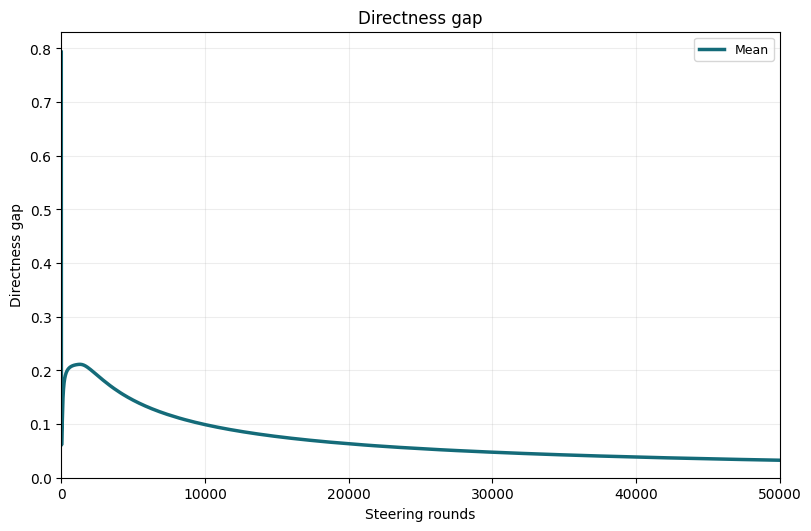}}
\par\small (b) Directness gap.
\end{minipage}\hfill
\begin{minipage}[t]{0.32\textwidth}
\centering
\includegraphics[width=\linewidth]{\detokenize{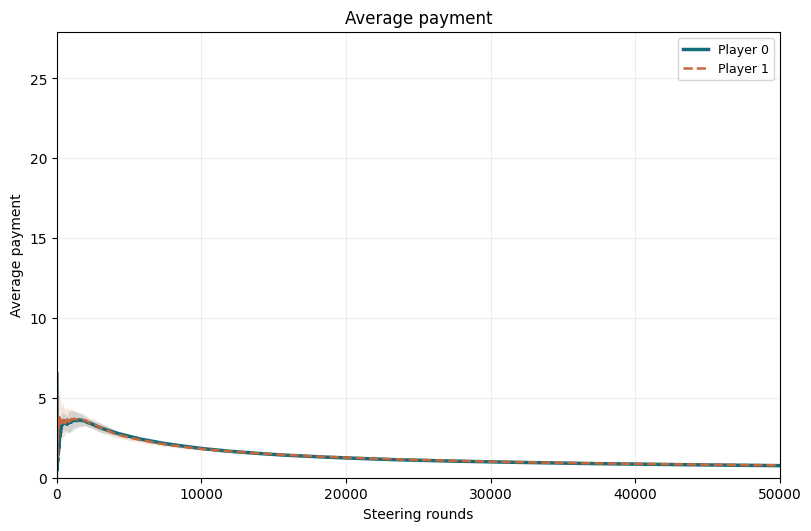}}
\par\small (c) Average realized payments.
\end{minipage}
\caption{\textbf{Two-player Kuhn poker.} Cumulative steering-round
metrics over the first 50,000 rounds. Curves show means over ten simulation seeds, with one sample
standard deviation shaded.}
\label{fig:kuhn-metrics}
\end{figure*}

For extensive-form game experiments, we consider the standard two-player Kuhn poker \citep{Kuhn1950}. We use 10 different runs for our experiments. Figure~\ref{fig:kuhn-metrics} displays the first 50,000 steering rounds. Following an initial transient, the three metrics decrease, supporting our findings in Theorem \ref{thm:main-corrected}.

\section{Conclusion}

In this paper, we have developed online learning and steering algorithms for general-sum EFGs under trajectory feedback. For game learning, our algorithm achieves a sample complexity of \(\mathcal{O}(\epsilon^{-2})\), matching the information-theoretic lower bound of \citep{Payingagents2025}. For steering, we have provided an online algorithm that steers players with no regret within each mediator-provided context toward the mediator's optimal equilibrium set at a rate of \(\mathcal{O}(T^{-1/4})\). We further have illustrated the performance of our algorithms through numerical experiments on representative EFGs and normal-form games.

\vspace{6pt} 
\noindent
Several directions remain open, including improving convergence rates of the defined gap functions, developing trajectory-implementable incentive mechanisms with lesser payments, and extending the framework to Markov games and partially observable Markov games (POMGs). Another important direction would be relaxation of the assumption of no regret within each mediator-provided context in favor of more standard learning guarantees. Finally, alternative optimization reductions and trajectory-feedback implementations of existing full-information incentive schemes may offer pathways to stronger theoretical and practical guarantees.

\section{Acknowledgments}

The authors acknowledge that LLM tools, such as ChatGPT Pro, were used in this work to assist with proofs, experiments, and proofreading. Research of the authors was supported in part by the US Army Research Office (ARO) Grant W911NF-24-1-0085.

\newpage
\bibliographystyle{plainnat} 
\bibliography{references}  

\newpage
\appendix 

\section{Proofs for Section \ref{subsec:noregret}}
\label{app:sampling-regret}

\begin{proof}[Proof of Proposition~\ref{prop:realized-regret}]
Fix player $i$, context $c$, and its prescribed local
horizon $m$, and write $t_\ell:=t_{i,c}(\ell)$.
Let $\mathcal G_\ell$ contain the full history before
play at $t_\ell$, the selected contexts, mixed strategies,
and payment rules, excluding the current action samples.
Define
\( 
D_\ell:=
u_i^{(t_\ell)}
(x_i^{(t_\ell)},a_{-i}^{(t_\ell)})
-u_i^{(t_\ell)}(a^{(t_\ell)}).
\) 
Conditional independence of the current action samples
gives
\[
\mathbb E\!\left[
u_i^{(t_\ell)}(a^{(t_\ell)})
\,\middle|\,
\mathcal G_\ell,a_{-i}^{(t_\ell)}
\right]
=
u_i^{(t_\ell)}
(x_i^{(t_\ell)},a_{-i}^{(t_\ell)}).
\]
Hence $\mathbb E[D_\ell\mid\mathcal G_\ell]=0$ and
$|D_\ell|\le L$.
The post-play history at $t_\ell$ is contained in the
pre-play history at $t_{\ell+1}$, including any intervening
rounds. Since context selection precedes the current
action samples, these increments form a martingale
difference sequence on the context's occurrences. So, we have
\( 
\widehat R_{i,c}^{\mathrm{NF}}(m)
-R_{i,c}^{\mathrm{NF}}(m)
=\sum_{\ell=1}^{m}D_\ell.
\) 
Finally, conditional Azuma--Hoeffding gives
\[
\Prb\left\{
\sum_{\ell=1}^{m}D_\ell>
L\sqrt{2m\log(1/\beta)}
\,\middle|\,\mathcal F_{i,c}
\right\}\le\beta.
\]
Combining this with the almost-sure bound
$R_{i,c}^{\mathrm{NF}}(m)\le B_{i,c}(m)$ proves the claim.
\end{proof}

\begin{proof}[Proof of Proposition~\ref{prop:efg-sampling-regret}]
Fix player $i$, context $c$, and its prescribed local
horizon $m$, and write $t_\ell:=t_{i,c}(\ell)$.
Let $\mathcal G_\ell$ contain the full pre-play history,
selected contexts, strategies, and payment rules.
During steering, it also contains the chosen commitment,
but excludes its sampled recommendations and the
current play randomization. Put
\( 
D_\ell:=
v_i^{(t_\ell)}(x^{(t_\ell)})-r_i^{(t_\ell)},
\) 
retaining the first argument $\mu^{(t_\ell)}$ in
$v_i^{(t_\ell)}$ during steering.
The sampling assumption gives
$\mathbb E[D_\ell\mid\mathcal G_\ell]=0$ and
$|D_\ell|\le L$.
As in the normal-form proof, the \emph{predictable} selection
of contexts preserves this martingale difference
property on their own occurrences. Subtracting the regret definitions gives 
\( 
\widehat R_{i,c}^{\mathrm{EF}}(m)
-R_{i,c}^{\mathrm{EF}}(m)
=\sum_{\ell=1}^{m}D_\ell.
\) 
Finally, conditional two-sided Azuma--Hoeffding yields
\[
\Prb\!\left\{
\left|\sum_{\ell=1}^{m}D_\ell\right|
>L\sqrt{2m\log(2/\beta)}
\,\middle|\,\mathcal F_{i,c}
\right\}\le\beta.
\]
This completes the proof of the result.
\end{proof}

\section{Proofs for Section \ref{sec:single-agent-binary-search}}
\label{app:single-agent-proofs}

\begin{lemma}\label{lem:majority}
For an epoch at payment level $\Delta$, if
$\Delta\ge\theta_a+\gamma_k$, then
$X(\Delta)\ge3N_k/4$; if
$\Delta\le\theta_a-\gamma_k$, then
$X(\Delta)\le N_k/4$.
\end{lemma}

\begin{proof}[Proof of Lemma~\ref{lem:majority}]
Fix an $N_k$-round testing epoch and write
$w_\Delta(b):=u[b]+p^{(\Delta,a)}[b]$.
Since the context is used only for this epoch, the
regret assumption for the active context gives, for any fixed $b\in A$,
\begin{equation}\label{eq:sa-app-regret}
\sum_{t=1}^{N_k}\bigl(w_\Delta(b)-w_\Delta(a_t)\bigr)
\le C\sqrt{N_k}.
\end{equation}
Suppose first that $\Delta\ge\theta_a+\gamma_k$.
For any $b\neq a$,
\( 
w_\Delta(a)-w_\Delta(b)
=u[a]+\Delta-u[b]\ge\Delta-\theta_a\ge\gamma_k.
\) 
Thus, with $N_k-X(\Delta)$ rounds on which $a_t\neq a$,
\[
\gamma_k\bigl(N_k-X(\Delta)\bigr)
\le\sum_{t=1}^{N_k}\bigl(w_\Delta(a)-w_\Delta(a_t)\bigr)
\le C\sqrt{N_k}.
\]
Since $N_k\ge16C^2/\gamma_k^2$, this implies
$N_k-X(\Delta)\le N_k/4$ and hence
$X(\Delta)\ge3N_k/4$.

\vspace{6pt}
\noindent 
Now suppose $\Delta\le\theta_a-\gamma_k$.
Let $b^\star\in\arg\max_b u[b]$. Then
$w_\Delta(b^\star)-w_\Delta(a)
=\theta_a-\Delta\ge\gamma_k$;
moreover, $b^\star$ maximizes $w_\Delta$ because
$\theta_a>\Delta$. Consequently, every occurrence of $a_t=a$
incurs regret at least $\gamma_k$ relative to $b^\star$,
while all other rounds have nonnegative regret. Applying
\eqref{eq:sa-app-regret} with $b=b^\star$ yields
$\gamma_k X(\Delta)\le C\sqrt{N_k}$, or
$X(\Delta)\le N_k/4$.
\end{proof}

\begin{proof}[Proof of Lemma~\ref{lem:one-sided-bounds-32prime}]
Let $w_\Delta(b):=u[b]+p^{(\Delta,a)}[b]$ and choose
$b^\star\in\arg\max_b w_\Delta(b)$. By
\eqref{eq:sa-app-regret},
\begin{equation}\label{eq:sa-app-best-regret}
\sum_{t=1}^{N_k}\bigl(w_\Delta(b^\star)-w_\Delta(a_t)\bigr)
\le C\sqrt{N_k}.
\end{equation}
If the epoch is labeled \textit{BEST}, then
$X(\Delta)\ge3N_k/4$. For every round with $a_t=a$,
\[
w_\Delta(b^\star)-w_\Delta(a_t)
=\max\{0,\theta_a-\Delta\}.
\]
All the other terms in \eqref{eq:sa-app-best-regret} are
nonnegative. Hence,
\( 
\frac34 N_k\max\{0,\theta_a-\Delta\}
\le C\sqrt{N_k},
\) 
which implies $\theta_a\le\Delta+\beta_k$.

\vspace{6pt}
\noindent
If the epoch is labeled \textit{NOTBEST}, then
$X(\Delta)\le N_k/4$. If $\theta_a\ge\Delta$, the
claimed inequality is immediate. Otherwise $a$ is an
optimal action under $w_\Delta$, and each round with
$a_t\neq a$ incurs regret at least $\Delta-\theta_a$.
Thus
\[
\frac34N_k(\Delta-\theta_a)
\le\sum_{t=1}^{N_k}\bigl(w_\Delta(a)-w_\Delta(a_t)\bigr)
\le C\sqrt{N_k},
\]
and thus $\theta_a\ge\Delta-\beta_k$. Finally, if the epoch is \textit{INCONCL}, it is neither
\textit{BEST} nor \textit{NOTBEST}. The contrapositive of
Lemma~\ref{lem:majority} implies
$|\Delta-\theta_a|<\gamma_k$.
\end{proof}

\begin{proof}[Proof of Lemma~\ref{lem:onecert}]
Put $x:=\theta_a-m_k$. The two distances are
$|\gamma_k-x|$ and $|\gamma_k+x|$, and therefore
\[
\max\{|\Delta_k^+-\theta_a|,
      |\Delta_k^--\theta_a|\}
=\gamma_k+|x|\ge\gamma_k.
\]
Lemma~\ref{lem:majority} then ensures that the
corresponding epoch is certified.
\end{proof}

\begin{proof}[Proof of Lemma~\ref{lem:bracket}]
By Lemma~\ref{lem:one-sided-bounds-32prime}, every
one-sided bound used in \eqref{eq:sa-bracket-update}
contains $\theta_a$; hence the updated bracket also contains it.
Moreover, \eqref{eq:sa-epoch-length} gives
\[
\beta_k=\frac{4C}{3\sqrt{N_k}}\le\frac{\gamma_k}{3}.
\]
By Lemma~\ref{lem:onecert}, at least one payment level is certified.
Write $\tau_k=U_k-L_k=4\gamma_k$.
If the lower-level test $\Delta_k^-$ is labeled
\textit{BEST}, then
\[
U_{k+1}-L_{k+1}\le
(\Delta_k^-+\beta_k)-L_k
=\gamma_k+\beta_k\le\frac13\tau_k.
\]
If the upper-level test $\Delta_k^+$ is labeled
\textit{NOTBEST}, the symmetric estimate gives
$U_{k+1}-L_{k+1}\le \tau_k/3$.
In the remaining possible certified cases, a
\textit{NOTBEST} label at $\Delta_k^-$ gives
\[
U_{k+1}-L_{k+1}\le
U_k-(\Delta_k^--\beta_k)
=3\gamma_k+\beta_k\le\frac56\tau_k,
\]
while a \textit{BEST} label at $\Delta_k^+$ gives the
symmetric bound. Since at least one of these cases
holds, $U_{k+1}-L_{k+1}\le5\tau_k/6$.
\end{proof}

\begin{proof}[Proof of Theorem~\ref{thm:main}]
Fix $a\in A$ and let $\tau_k^{(a)}$ be its bracket width
before the $k$th iteration, i.e. \( [L_k, U_k]\) . Lemma~\ref{lem:bracket} gives
\[
\theta_a\in[L_k^{(a)},U_k^{(a)}],\qquad
\tau_k^{(a)}\le\left(\frac56\right)^k.
\]
Hence the stopping rule is reached after at most
$K:=\lceil\log_{6/5}(1/\varepsilon)\rceil$ updates.
For the final bracket midpoint,
$|\widehat\theta_a-\theta_a|\le\varepsilon$.
Since $u[a]=u_{\max}-\theta_a$ and
$\widehat u[a]=-\widehat\theta_a+c$, choosing
$W=c-u_{\max}$ gives
\[
u[a]+W-\widehat u[a]
=\widehat\theta_a-\theta_a.
\]
Thus $\|u+W\mathbf1-\widehat u\|_\infty\le\varepsilon$. It remains to count the rounds. Let $k_a$ be the index
of the final \emph{executed} iteration for action $a$.
The algorithm executes iteration $k_a$ only if
$\tau_{k_a}^{(a)}>\varepsilon$. Since every update satisfies
$\tau_{j+1}^{(a)}\le(5/6)\tau_j^{(a)}$, for $0\le j\le k_a$,
\[
\tau_j^{(a)}\ge
\left(\frac65\right)^{k_a-j}\tau_{k_a}^{(a)}
>\varepsilon\left(\frac65\right)^{k_a-j}.
\]
As $\gamma_j=\tau_j^{(a)}/4$ and
$N_j=\lceil16C^2/\gamma_j^2\rceil$,
\begin{align*}
\sum_{j=0}^{k_a}2N_j
&\le 512C^2\sum_{j=0}^{k_a}\frac{1}{(\tau_j^{(a)})^2}
       +2(k_a+1)\\
&\le\frac{512C^2}{\varepsilon^2}
 \sum_{r=0}^{k_a}\left(\frac56\right)^{2r}
 +2(K+1)\\
&\le\frac{512}{1-(5/6)^2}\frac{C^2}{\varepsilon^2}
 +2(K+1)
 =O\!\left(\frac{C^2}{\varepsilon^2}\right),
\end{align*}
where the final bound uses $C\ge1$ and
$\varepsilon\in(0,1)$. Summing over the $m$ actions
establishes the stated round complexity. Each iteration
uses two distinct contexts, and there are at most
$K$ iterations per action, giving
$O(m\log(1/\varepsilon))$ contexts in total.
\end{proof}

\section{Proofs for Section \ref{sec:multi-agent-guarded}} 
\label{app:multi-agent-proofs}

Throughout this appendix, we fix a player $i$, a fixed opponents' action profile
$y\in A_{-i}$, and a target action $a\in A_i$. We first establish the opponent-pinning and \emph{leakage} bounds
used throughout the analysis. We define \( 
w_t(b):=u_i(b,a_{-i}^{(t)})
          +p_i^{(\Delta,a)}[b],
w_y(b):=u_i(b,y)+p_i^{(\Delta,a)}[b]
\), and let
\( 
\mathcal L
:=\{t\in[N]:a_{-i}^{(t)}\ne y\},
\mathcal C:=[N]\setminus\mathcal L,
\) to describe \emph{leakage} and \emph{compliant} rounds.  In particular, the following
lemma controls player $i$'s regret on rounds in which the
opponents play the prescribed action profile $y$.

\begin{lemma}
\label{lem:ma-pinning-leakage}
Consider an epoch of $N$ rounds in which every opponent
$j\ne i$ receives the pinning payment $p_j^{(2,y_j)}$
and player $i$ receives the fixed payment level payment
$p_i^{(\Delta,a)}$, with $\Delta\in[0,1]$.
Let, $L:=|\mathcal L|$ and
$C_{-i}^{\mathrm{pin}}:=\sum_{j\ne i}C_j$.
Then,
\( 
L\le C_{-i}^{\mathrm{pin}}\sqrt N,
\) 
and for every $b\in\arg\max_{c\in A_i}w_y(c)$,
\begin{equation}\label{eq:ma-app-compliant-regret}
0\le
\sum_{t\in\mathcal C}
\bigl(w_y(b)-w_y(a_i^{(t)})\bigr)
\le
(C_i+2C_{-i}^{\mathrm{pin}})\sqrt N.
\end{equation}
\end{lemma}

\begin{proof}
For each $j\ne i$, the pinning payment
$p_j^{(2,y_j)}$ ensures that, for every realized
$a_{-j}^{(t)}$ and every $b\ne y_j$,
\( 
\bigl(u_j(y_j,a_{-j}^{(t)})+2\bigr)
-\bigl(u_j(b,a_{-j}^{(t)})+0\bigr)\ge1.
\) 
Applying~\eqref{eq:signal-wise-nfg-regret} therefore gives
\[
\sum_{t=1}^N
\mathbf1\{a_j^{(t)}\ne y_j\}
\le C_j\sqrt N.
\]
Since an opponent-profile deviation requires at least
one opponent to deviate, summing these indicator bounds
over $j\ne i$ yields
\[
L
\le\sum_{j\ne i}\sum_{t=1}^N
\mathbf1\{a_j^{(t)}\ne y_j\}
\le C_{-i}^{\mathrm{pin}}\sqrt N.
\] 
Next, since $u_i\in[0,1]$ and the payments
are fixed throughout the epoch, for all $b,b'\in A_i$,
\begin{equation}\label{eq:ma-app-leak-difference}
\left|
[w_t(b)-w_t(b')]
-[w_y(b)-w_y(b')]
\right|
\le2\mathbf1\{t\in\mathcal L\}.
\end{equation}
Then,~\eqref{eq:signal-wise-nfg-regret} gives, for any fixed
$b\in A_i$,
\begin{equation}\label{eq:ma-app-regret}
\sum_{t=1}^N
\bigl(w_t(b)-w_t(a_i^{(t)})\bigr)
\le C_i\sqrt N.
\end{equation}
Now take $b\in\arg\max_{c\in A_i}w_y(c)$.
Every term $w_y(b)-w_y(a_i^{(t)})$ is
nonnegative. Consequently,
by~\eqref{eq:ma-app-leak-difference}
and~\eqref{eq:ma-app-regret},
\begin{align*}
0
&\le
\sum_{t\in\mathcal C}
\bigl(w_y(b)-w_y(a_i^{(t)})\bigr)\\
&\le
\sum_{t=1}^N
\bigl(w_y(b)-w_y(a_i^{(t)})\bigr)\\
&\le
\sum_{t=1}^N
\bigl(w_t(b)-w_t(a_i^{(t)})\bigr)+2L\\
&\le C_i\sqrt N+2L\\
&\le(C_i+2C_{-i}^{\mathrm{pin}})\sqrt N.
\end{align*}
This proves~\eqref{eq:ma-app-compliant-regret}.
\end{proof}

\begin{lemma}
\label{lem:majority-leak}
For an epoch at payment level $\Delta$, if
$\Delta\ge\theta_{i,y}(a)+\gamma_k$, then
$X_{i,y,a}(\Delta)\ge3N_k/4$; if
$\Delta\le\theta_{i,y}(a)-\gamma_k$, then
$X_{i,y,a}(\Delta)\le N_k/4$.
\end{lemma}

\begin{proof}[Proof of Lemma~\ref{lem:majority-leak}]
Write $\theta=\theta_{i,y}(a)$ and $N=N_k$.
By the choice~\eqref{eq:ma-epoch-length}
and Lemma~\ref{lem:ma-pinning-leakage},
\begin{equation}\label{eq:ma-app-budget}
\frac{(C_i+2C_{-i}^{\mathrm{pin}})\sqrt N}
{\gamma_k}
\le\frac N8,
\qquad
L\le C_{-i}^{\mathrm{pin}}\sqrt N
\le\frac N8.
\end{equation}
If $\Delta\ge\theta+\gamma_k$, then
$w_y(a)-w_y(b)\ge\gamma_k$ for every $b\ne a$,
so $a$ maximizes $w_y$. Let
\( 
M:=\sum_{t\in\mathcal C}
\mathbf1\{a_i^{(t)}\ne a\}.
\) 
Every such round contributes at least $\gamma_k$
to the left-hand side of
\eqref{eq:ma-app-compliant-regret}, and thus
$M\le N/8$. Therefore the number of joint plays
of $(a,y)$ is
\[
X_{i,y,a}(\Delta)
=|\mathcal C|-M
=N-L-M\ge\frac34N.
\]
If $\Delta\le\theta-\gamma_k$, let
$b^\star\in\arg\max_{b\in A_i}u_i(b,y)$.
Then, $b^\star\ne a$ and $b^\star$ maximizes $w_y$.
Every compliant round with $a_i^{(t)}=a$
contributes
\[
w_y(b^\star)-w_y(a)
=\theta-\Delta\ge\gamma_k
\]
to the left-hand side of
\eqref{eq:ma-app-compliant-regret}.
Thus the compliant number of occurrences of $a$,
which equals $X_{i,y,a}(\Delta)$, is at most
$N/8\le N/4$.
This proves both claims.
\end{proof}

\begin{proof}[Proof of Lemma~\ref{lem:one-sided-bounds-ma}]
Write $\theta=\theta_{i,y}(a)$, $N=N_k$, and
$D_i:=C_i+2C_{-i}^{\mathrm{pin}}$.
By Lemma~\ref{lem:ma-pinning-leakage},
$L\le C_{-i}^{\mathrm{pin}}\sqrt N$;
by~\eqref{eq:ma-app-budget}, $L\le N/8$.

\noindent
Suppose first that the epoch is labeled
\textit{BEST}, so
$X_{i,y,a}(\Delta)\ge3N/4$.
If $\theta\le\Delta$, there is nothing to prove.
Otherwise take
$b^\star\in\arg\max_bu_i(b,y)$.
Since $\theta>\Delta$, $b^\star$ maximizes
$w_y$, and for every compliant occurrence
of $a$ its advantage equals $\theta-\Delta$.
Because $X_{i,y,a}(\Delta)$ counts joint plays
of $(a,y)$, it already counts only compliant
occurrences of $a$.
There are therefore at least
$3N/4\ge5N/8$ such rounds. Hence,
by~\eqref{eq:ma-app-compliant-regret},
\[
\frac58N(\theta-\Delta)
\le C_i\sqrt N+2L
\le D_i\sqrt N.
\]
Thus
\[
\theta
\le\Delta+\frac{8D_i}{5\sqrt N}
=\Delta+\beta_k^{\mathrm{MA}}.
\]

\noindent
Next suppose that the epoch is labeled
\textit{NOTBEST}, so
$X_{i,y,a}(\Delta)\le N/4$.
If $\theta\ge\Delta$, the conclusion
is immediate. Otherwise $a$ maximizes
$w_y$, with advantage at least
$\Delta-\theta$ over every other action.
The number of compliant rounds not playing $a$ is
\[
M_{\mathcal C}
:=\sum_{t\in\mathcal C}
\mathbf1\{a_i^{(t)}\ne a\}
=N-L-X_{i,y,a}(\Delta)
\ge\frac58N.
\]
Using~\eqref{eq:ma-app-compliant-regret}
with comparator $a$ yields
\[
\frac58N(\Delta-\theta)
\le C_i\sqrt N+2L
\le D_i\sqrt N,
\]
and thus $\theta\ge\Delta-\beta_k^{\mathrm{MA}}$. Finally, if the label is \textit{INCONCL},
the contrapositive of
Lemma~\ref{lem:majority-leak} gives
$|\Delta-\theta|<\gamma_k$.
\end{proof}

\begin{proof}[Proof of Lemma~\ref{lem:one-certified-ma}]
Put $x:=\theta_{i,y}(a)-m_k$. Then,
\[
\max\{
|\Delta_k^--\theta_{i,y}(a)|,
|\Delta_k^+-\theta_{i,y}(a)|
\}
=\gamma_k+|x|
\ge\gamma_k.
\]
Lemma~\ref{lem:majority-leak}
certifies at least one test.
\end{proof}

\begin{proof}[Proof of Proposition~\ref{prop:bracket-shrink}] Write $\tau_k=U_k-L_k=4\gamma_k$. Then,
Lemma~\ref{lem:one-sided-bounds-ma}
ensures that every certified label yields
a one-sided bound containing
$\theta_{i,y}(a)$. Their intersection with
the old bracket, \(\tau_k\), therefore preserves containment.
The epoch length yields
\[
\beta_k^{\mathrm{MA}}
=\frac85
\frac{C_i+2C_{-i}^{\mathrm{pin}}}
{\sqrt{N_k}}
\le\frac{\gamma_k}{5}.
\]
By Lemma~\ref{lem:one-certified-ma},
a certified payment level always exists.
If $\Delta_k^-$ is labeled \textit{BEST}
or $\Delta_k^+$ is labeled \textit{NOTBEST},
the updated width is at most
$\gamma_k+\beta_k^{\mathrm{MA}}$.
In the other certified cases, it is at most
$3\gamma_k+\beta_k^{\mathrm{MA}}$.
Since $\tau_k=4\gamma_k$, in all cases
\( 
\tau_{k+1}
\le3\gamma_k+\beta_k^{\mathrm{MA}}
\le\frac9{10}\tau_k.
\) 
\end{proof}

\begin{theorem}[Sample complexity for a fixed opponents' profile]
\label{thm:per-context}
Fix a player $i$ and an opponent profile
$y\in A_{-i}$.
Algorithm~\ref{alg:gbs-ma} estimates
$\theta_{i,y}(a)$ to accuracy $\varepsilon$
for every $a\in A_i$, using at most
\( 
T_{i,y}
=O\big(
\frac{m_i(C_i+2C_{-i}^{\mathrm{pin}})^2}
{\varepsilon^2}
\big)
\) 
rounds devoted to this fixed profile.
\end{theorem}

\begin{proof}[Proof of Theorem~\ref{thm:per-context}]
Fix a target action $a$ and write
$D_i=C_i+2C_{-i}^{\mathrm{pin}}$.
Let $J_a$ be the number of completed updates
for that target.
Proposition~\ref{prop:bracket-shrink} gives
$\tau_{k+1}\le(9/10)\tau_k$ and, because the
algorithm stops at width at most $\varepsilon$,
\( 
J_a\le
K:=\left\lceil
\log_{10/9}(1/\varepsilon)
\right\rceil.
\) 
If $J_a\ge1$, the last executed update has
$\tau_{J_a-1}>\varepsilon$.
Thus, for every $0\le k\le J_a-1$,
\[
\tau_k
\ge
\left(\frac{10}{9}\right)^{J_a-1-k}
\tau_{J_a-1}
>
\varepsilon
\left(\frac{10}{9}\right)^{J_a-1-k}.
\]
Since $\gamma_k=\tau_k/4$ and
$N_k\le64D_i^2/\gamma_k^2+1$,
the number of rounds for action $a$ satisfies
\begin{align*}
\sum_{k=0}^{J_a-1}2N_k
&\le128D_i^2
\sum_{k=0}^{J_a-1}\frac1{\gamma_k^2}
+2J_a\\
&\le
\frac{2048D_i^2}{\varepsilon^2}
\sum_{r=0}^{J_a-1}
\left(\frac9{10}\right)^{2r}
+2K\\
&\le
\frac{2048}{1-(9/10)^2}
\frac{D_i^2}{\varepsilon^2}
+2K\\
&=O(D_i^2/\varepsilon^2).
\end{align*}
As each bracket update contains the searched solution \(\theta_{i,y}(a)\) the final midpoint has
\( 
|\widehat\theta_{i,y}(a)-\theta_{i,y}(a)|
\le\varepsilon.
\) 
Summing the cost over $m_i$ actions proves
the theorem.
\end{proof}

\begin{proof}[Proof of Theorem~\ref{thm:multi-main}]
By Theorem~\ref{thm:per-context},
every pair $(i,y)$ requires
$O(m_iD_i^2/\varepsilon^2)$ rounds, with
$D_i=C_i+2C_{-i}^{\mathrm{pin}}$.
Since there are $M_{-i}$ opponents' action profiles for player $i$ and $m_iM_{-i}=M$, summing gives
\[
T
=O\left(
\frac1{\varepsilon^2}
\sum_{i=1}^n M_{-i}m_iD_i^2
\right)
=O\left(
\frac M{\varepsilon^2}
\sum_{i=1}^nD_i^2
\right).
\]
If $C_i\le C$ for every player, then
$D_i\le(2n-1)C$, giving
\( 
T=O(Mn^3C^2/\varepsilon^2).
\) 

\noindent
For accuracy, let
\( 
e_{i,y}(a)
:=\widehat\theta_{i,y}(a)-\theta_{i,y}(a).
\) 
Theorem~\ref{thm:per-context} gives
$|e_{i,y}(a)|\le\varepsilon$
for all $(i,y,a)$.
Using
\( 
u_i(a,y)
=u_i^{\max}(y)-\theta_{i,y}(a),
\) 
$\widetilde u_i(a,y)
=-\widehat\theta_{i,y}(a)+c_{i,y}$,
and $W_i(y)=u_i^{\max}(y)-c_{i,y}$,
we have
\[
u_i(a,y)-\widetilde u_i(a,y)-W_i(y)
=\widehat\theta_{i,y}(a)-\theta_{i,y}(a)
=e_{i,y}(a).
\]
This proves the claimed uniform
reconstruction error. Moreover, there are $nM$ target triples $(i,y,a)$,
each using at most $2K$ epochs.
With a new context for each of $n$ players
in every epoch, the total number of private
context identifiers is
$O(n^2M\log(1/\varepsilon))$.
\end{proof}

\section{Proofs for Section \ref{sec:learn-efg}}
\label{app:efg-learning-proofs}

We prove the learning guarantees for the payments given in
\eqref{eq:efg-insured-opponents}--\eqref{eq:efg-insured-payment}.
Throughout this appendix, fix one epoch of $N$ rounds for a query
$(\bar s,i,I,a)$ with $r:=\rho_{i,I}(\bar s)>0$, fixed
payment $\Delta\in[0,1)$, and one new learning
context for every player. On the events where the epoch
inequalities \eqref{eq:pathwise-per-signal-regret} hold,
the proofs below are apart from the explicitly
stated martingale concentration events. Those events
are proved under the original filtration and intersected
with regret success afterward.
For brevity, let
\[
d^{\rm out}(z):=\bar d_{i,I}^{\rm out}(z),\quad
 d_{-i}(z):=\bar d_{-i}(z),\quad
 D_t^{\rm out}:=1-d^{\rm out}(z^{(t)}),\quad
 D_{-i,t}:=1-d_{-i}(z^{(t)}).
\]

\subsection{Opponent pinning and upper-bounding number of deviations}

\begin{lemma}
\label{lem:one-round-forcing-pathwise}
Fix an opponent $j\ne i$ and a round $t$ with payment
$q_j(z)=B\bar d_j(z)$, where $B\ge2$. Let $Z_t$ be the realized terminal history. To define
a counterfactual play, keep the same chance outcomes
and the same contingent actions of all players other
than $j$, but require player $j$ to follow the pure
reference strategy $\bar s_j$ at every information
set. Denote the resulting terminal history by
$Z_t^{(j)}$. Then
\[
u_j(Z_t^{(j)})+B
-\bigl(u_j(Z_t)+B\bar d_j(Z_t)\bigr)
\ge
(B-1)(1-\bar d_j(Z_t)).
\]
\end{lemma}

\begin{proof}
If $\bar d_j(Z_t)=1$, replacing player $j$'s
strategy by $\bar s_j$ does not change any action
along the realized path. Since chance outcomes
and the other players' contingent actions are
also unchanged, the counterfactual play follows
the same path and reaches the same terminal:
$Z_t^{(j)}=Z_t$. Both plays therefore give
player $j$ the same payoff and payment
$B$, so the displayed difference is zero. 

\vspace{6pt}
\noindent
If $\bar d_j(Z_t)=0$, player $j$ deviates from
$\bar s_j$ at some information set on the
realized path. Thus, its actual payment is
zero, whereas following $\bar s_j$ guarantees
payment $B$.
Since utilities belong to $[0,1]$,
\( 
u_j(Z_t^{(j)})-u_j(Z_t)\ge-1.
\) 
Consequently,
\( 
u_j(Z_t^{(j)})+B-u_j(Z_t)
\ge B-1.
\) 
Combining the two cases proves the claim.
\end{proof}

\begin{lemma}[Number of observed opponent deviations]
\label{lem:leak-mass-epoch-opponents}
Fix an opponent $j\ne i$ and an $N$-round testing
epoch in which player $j$ receives the payment
$q_j(z)=B\bar d_j(z)$, with $B\ge2$.
Then, for any $\zeta\in(0,1)$, with probability
at least $1-\zeta$,
\begin{equation}\label{eq:efg-opp-deviations}
\sum_{t=1}^N(1-\bar d_j(Z_t))
\le
\frac{C_j+(B+1)\sqrt{2\log(1/\zeta)}}
{B-1}\sqrt N.
\end{equation}
\end{lemma}

\begin{proof}

\vspace{6pt}
\noindent
Consider the counterfactual terminal history
$Z_t^{(j)}$ defined in
Lemma~\ref{lem:one-round-forcing-pathwise}:
player $j$ follows $\bar s_j$, while chance and
all other players use the same random outcomes
as in the actual play. Define the counterfactual
augmented payoff
\( 
Y_t:=u_j(Z_t^{(j)})+B.
\) 
Since following $\bar s_j$ guarantees payment
$B$, Lemma~\ref{lem:one-round-forcing-pathwise}
gives, on every realized round,
\begin{equation}\label{eq:efg-forcing-pathwise-sum}
Y_t-r_j^{(t)}
\ge (B-1)(1-\bar d_j(Z_t)),
\end{equation}
where
$r_j^{(t)}=u_j(Z_t)+B\bar d_j(Z_t)$
is player $j$'s actual augmented payoff. Summing \eqref{eq:efg-forcing-pathwise-sum}
over the epoch, we obtain
\begin{equation}\label{eq:efg-forcing-sum}
(B-1)\sum_{t=1}^N(1-\bar d_j(Z_t))
\le \sum_{t=1}^N(Y_t-r_j^{(t)}).
\end{equation}
The remaining step is to control the right-hand
side using player $j$'s regret guarantee.
The difficulty is that regret compares the
actual payoff with the \emph{expected} payoff
of $\bar s_j$, rather than its sampled
counterfactual payoff $Y_t$.

\vspace{6pt}
\noindent
Let $\mathcal G_t$ contain the history preceding
round $t$, together with the strategies and
payment rules selected for that round, but not
the current action or chance realizations.
Recall that $Z_t^{(j)}$ is obtained by replacing
only player $j$'s strategy with $\bar s_j$,
while preserving the underlying randomness
of Nature and the other players. Conditional on $\mathcal G_t$, this counterfactual
terminal history has the distribution induced by
the strategy profile $(\bar x_j,x_{-j}^{(t)})$,
where $\bar x_j=\mathrm{Seq}_j(\bar s_j)$.
Therefore, we have
\( 
\mathbb E[Y_t\mid\mathcal G_t]
=\sum_{z\in Z}
\pi(z\mid\bar x_j,x_{-j}^{(t)})
\bigl(u_j(z)+B\bigr)=v_j^{(t)}(\bar x_j).
\) 
Consequently, sample error
$\xi_t:=Y_t-v_j^{(t)}(\bar x_j)$
has conditional mean zero and satisfies
$|\xi_t|\le B+1$. Then, $(\xi_t)_{t=1}^N$ is a martingale
difference sequence. Moreover, since
$u_j\in[0,1]$, both $Y_t$ and its
conditional expectation lie in $[B,B+1]$,
so $|\xi_t|\le B+1$. 
We can now decompose
\begin{align*}
\sum_{t=1}^N(Y_t-r_j^{(t)})
&=
\sum_{t=1}^N
\bigl[v_j^{(t)}(\bar x_j)-r_j^{(t)}\bigr]
+\sum_{t=1}^N\xi_t\le C_j\sqrt N+\sum_{t=1}^N\xi_t,
\end{align*}
where the inequality follows from the
regret guarantee for the active context
\eqref{eq:pathwise-per-signal-regret}
applied to the fixed $\bar x_j$. Finally, Azuma--Hoeffding gives, with probability at least $1-\zeta$,
\( 
\sum_{t=1}^N\xi_t
\le (B+1)\sqrt{2N\log(1/\zeta)}.
\) 
Combining this inequality with
\eqref{eq:efg-forcing-sum} yields
\[
(B-1)\sum_{t=1}^N(1-\bar d_j(Z_t))
\le
\left[C_j+(B+1)\sqrt{2\log(1/\zeta)}\right]
\sqrt N,
\]
which proves the claim.
\end{proof}

\vspace{6pt}
\noindent
Lemma~\ref{lem:leak-mass-epoch-opponents}
shows that opponent deviations along the
observed trajectories occur on relatively
few rounds. However, this alone is not
sufficient to control the queried player's
regret: an alternative strategy may lead
to information sets that were never visited
during actual play, where the opponents'
behavior need not agree with $\bar s_{-i}$.
The following lemma addresses this difficulty by providing an upper bound on the number of mistakes made by player \(i\) at the queried information set and deviations made by player \(i\) outside that information set, in terms of the deviations of other players and regret and concentration terms.

\begin{lemma}[Mistakes and deviations outside $I$]
\label{lem:mistake-bound-decoupled}
Fix a query $(\bar s,i,I,a)$ and an $N$-round
testing epoch with payment level $\Delta\in[0,1)$.
Write $\theta:=\theta_{i,I}^{\bar s}(a)$.
For $\gamma>0$, suppose that either
$\Delta\ge\theta+\gamma$ or
$\Delta\le\theta-\gamma$, and define
\[
M:=
\begin{cases}
\displaystyle
\sum_{t=1}^N
\mathbf1\{\bar d_{i,I}(z^{(t)})=1,\,
I\preceq z^{(t)},\,
a_I(z^{(t)})\ne a\},
& \Delta\ge\theta+\gamma,\\[3mm]
\displaystyle
\sum_{t=1}^N
\mathbf1\{\bar d_{i,I}(z^{(t)})=1,\,
I\preceq z^{(t)},\,
a_I(z^{(t)})=a\},
& \Delta\le\theta-\gamma.
\end{cases}
\]
For $\gamma=0$, allow any
$\Delta\in[0,1)$ and set $M=0$. Then, for any $\zeta\in(0,1)$, with probability at least $1-\zeta$,
\begin{equation}\label{eq:efg-insured-joint-bound}
\begin{aligned}
\gamma M
+(B-1-\Delta)\sum_{t=1}^N D_t^{\rm out}
&\le C_i\sqrt N
+(2+\Delta)\sum_{t=1}^N D_{-i,t}+(1+\Delta+\gamma)
\sqrt{2N\log(1/\zeta)}.
\end{aligned}
\end{equation}
\end{lemma}

\begin{proof}
Fix the pure profile $\bar s$ specified by
the learning query. In round $t$, the players
instead select realization plans $x^{(t)}$,
which need not equal $\mathrm{Seq}(\bar s)$.
For the analysis, sample a complete pure
contingency plan for each player according
to a behavioral strategy implementing
$x^{(t)}$, together with Nature's actions
at every chance node. 

\vspace{6pt}
\noindent 
Now, let $b_t$ be player $i$'s sampled action
at $I$, whether or not $I$ is reached, according to behavioral strategy implementing
$x^{(t)}$.
Using the sampled chance outcomes, define
$\widetilde Z_t$ as the terminal generated
by the pure-strategy profile
\( 
(\bar s_i^{(I\to b_t)},\bar s_{-i}).
\) 
Thus every opponent $j\ne i$ chooses
$\bar s_j(J)$ at each information set $J$,
while player $i$ chooses $\bar s_i(J)$
at every $J\ne I$ and chooses $b_t$ at $I$.
Define
\begin{equation}\label{eq:efg-ideal-reward}
\widetilde R_t
:=u_i(\widetilde Z_t)+B-\Delta
+\Delta\mathbf1\{
I\preceq\widetilde Z_t,\ b_t=a
\}.
\end{equation}
Since $u_i\in[0,1]$, we have
$\widetilde R_t\in[B-\Delta,B+1]$. Write
\( 
w(b):=v_{i,\bar s}^{(I)}(b)
+\Delta\mathbf1\{b=a\},
b^\star\in\arg\max_{b\in A_I}w(b),
\) 
and fix the realization plan
\( 
\bar x_i^\star
:=\mathrm{Seq}_i(\bar s_i^{(I\to b^\star)}).
\) 
We first compare the expected augmented
payoff of $\bar x_i^\star$ against the
opponents' actual strategies with its payoff
when the opponents use $\bar s_{-i}$.

\vspace{6pt}
\noindent 
Consider a play in which player $i$ uses
$\bar s_i^{(I\to b^\star)}$ and the opponents
use their sampled round-$t$ contingency
plans. If every opponent chooses its
prescribed action $\bar s_j(J)$ along
this play, replacing all opponents by
$\bar s_{-i}$ leaves the terminal unchanged.
Otherwise, some opponent deviates along
the resulting terminal history $Z$, so
$\bar d_{-i}(Z)=0$. Since player $i$
follows $\bar s_i$ outside $I$,
$\bar d_{i,I}^{\rm out}(Z)=1$, and
\eqref{eq:efg-insured-payment} gives
\( 
q_i^{(I,a,\Delta)}(Z)
=(B-\Delta)+(1+\Delta)=B+1.
\) 
Its payment is therefore at
least $B+1$. In contrast, the payment obtained when all opponents use
$\bar s_{-i}$ is at most $B$.
Using the same chance outcomes in these
two plays and taking expectations yield
\begin{equation}\label{eq:efg-comparator-domination}
\begin{aligned}
v_i^{(t)}(\bar x_i^\star)
\ge
\mathbb E_{Z\sim\pi(\cdot\mid
\bar x_i^\star,\bar x_{-i})}
\bigl[
u_i(Z)+q_i^{(I,a,\Delta)}(Z)
\bigr],
\end{aligned}
\end{equation}
where
$\bar x_{-i}:=\mathrm{Seq}_{-i}(\bar s_{-i})$. Let $\mathcal F_{t^-}$ contain the selected
round-$t$ strategies and the preceding history,
but not the current action or chance realizations.
Write
\( 
r:=\rho_{i,I}(\bar s),
\pi_t(b):=\Pr(b_t=b\mid\mathcal F_{t^-}).
\) 
Since changing the action at $I$ does not affect
its reach probability or the terminal outcomes
on paths avoiding $I$, we have
\[
\begin{aligned}
&\mathbb E_{Z\sim\pi(\cdot\mid
\bar x_i^\star,\bar x_{-i})}
[u_i(Z)+q_i^{(I,a,\Delta)}(Z)]
-\mathbb E[\widetilde R_t\mid\mathcal F_{t^-}]=
r\sum_{b\in A_I}\pi_t(b)
\bigl[w(b^\star)-w(b)\bigr].
\end{aligned}
\]
Combining this identity with
\eqref{eq:efg-comparator-domination}
gives
\begin{equation}\label{eq:efg-reference-gap}
\begin{aligned}
v_i^{(t)}(\bar x_i^\star)
-\mathbb E[\widetilde R_t\mid\mathcal F_{t^-}]
&\ge
r\sum_{b\in A_I}\pi_t(b)
\bigl[w(b^\star)-w(b)\bigr]
\ge0,
\end{aligned}
\end{equation}
where the last inequality follows from
$b^\star\in\arg\max_{b\in A_I}w(b)$. Now suppose that
$\Delta\ge\theta+\gamma$.
By the definition of $\theta$,
\[
w(a)-w(b)
=v_{i,\bar s}^{(I)}(a)+\Delta
-v_{i,\bar s}^{(I)}(b)
\ge\Delta-\theta\ge\gamma
\]
for every $b\ne a$.
Thus $b^\star=a$. Alternatively, if
$\Delta\le\theta-\gamma$,
choose $b^\star\in\arg\max_b
v_{i,\bar s}^{(I)}(b)$.
Then, $b^\star\ne a$ and
\[
w(b^\star)-w(a)
=\theta-\Delta\ge\gamma.
\]
Therefore, in either case, choosing
an action counted as a \emph{mistake} loses at least $\gamma$ in the payoff $w$. Now, let $\xi_t$ indicate that
$\widetilde Z_t$ reaches $I$ and
$b_t$ is such a mistake. Since
$b_t$ is sampled independently of
the reference play's preceding chance
outcomes, \eqref{eq:efg-reference-gap}
implies
\begin{equation}\label{eq:efg-gap-mistake}
v_i^{(t)}(\bar x_i^\star)
\ge
\mathbb E[
\widetilde R_t+\gamma\xi_t
\mid\mathcal F_{t^-}].
\end{equation}
We next relate this inequality to the
actual payoff
\( 
R_t:=u_i(Z_t)
+q_i^{(I,a,\Delta)}(Z_t).
\) 
By the payment rule, we have
\begin{equation}\label{eq:efg-actual-domination}
R_t\le
\widetilde R_t
-(B-1-\Delta)D_t^{\rm out}
+(2+\Delta)D_{-i,t}.
\end{equation}
Indeed, if both deviation indicators
vanish, the actual and hypothetical
plays coincide, and thus
$R_t=\widetilde R_t$.
If $D_t^{\rm out}=1$, player $i$
receives no payment and $R_t\le1$,
whereas
\[
\widetilde R_t-(B-1-\Delta)\ge1.
\]
If $D_t^{\rm out}=0$ and
$D_{-i,t}=1$, the insured payment
gives $R_t\le B+2$, while
\[
\widetilde R_t+(2+\Delta)\ge B+2.
\]
Hence \eqref{eq:efg-actual-domination}
holds in all cases. Furthermore, whenever an observed
round is counted by $M$, all players
have followed $\bar s$ along that
trajectory outside $I$. The actual
and hypothetical plays therefore
agree up to and including $I$, and
$\xi_t=1$. It follows that
\( 
M\le\sum_{t=1}^N\xi_t.
\) 
Combining this with
\eqref{eq:efg-actual-domination} gives
\[
\begin{aligned}
&\gamma M+
(B-1-\Delta)\sum_{t=1}^N D_t^{\rm out}\le
\sum_{t=1}^N
(\widetilde R_t+\gamma\xi_t-R_t)
+(2+\Delta)\sum_{t=1}^N D_{-i,t}.
\end{aligned}
\]
By \eqref{eq:efg-gap-mistake},
\[
\begin{aligned}
&\widetilde R_t+\gamma\xi_t-R_t\le
v_i^{(t)}(\bar x_i^\star)-R_t+
\widetilde R_t+\gamma\xi_t
-\mathbb E[
\widetilde R_t+\gamma\xi_t
\mid\mathcal F_{t^-}].
\end{aligned}
\]
Summing over $t$ and applying the
regret bound for the active context
\eqref{eq:pathwise-per-signal-regret}
to the fixed plan $\bar x_i^\star$
yields
\[
\begin{aligned}
&\gamma M+
(B-1-\Delta)\sum_{t=1}^N D_t^{\rm out}\le
C_i\sqrt N
+(2+\Delta)\sum_{t=1}^N D_{-i,t}+
\sum_{t=1}^N
\Bigl[
\widetilde R_t+\gamma\xi_t
-\mathbb E[
\widetilde R_t+\gamma\xi_t
\mid\mathcal F_{t^-}]
\Bigr].
\end{aligned}
\]
The final sum is a martingale sum,
because each summand has conditional
mean zero. Moreover,
\( 
B-\Delta
\le\widetilde R_t+\gamma\xi_t
\le B+1+\gamma,
\) 
so its increments have
absolute value at most
$1+\Delta+\gamma$. Therefore, 
Azuma--Hoeffding bounds the sum by
\( 
(1+\Delta+\gamma)
\sqrt{2N\log(1/\zeta)}
\) 
with probability at least $1-\zeta$,
proving \eqref{eq:efg-insured-joint-bound}. 
\end{proof}

\subsection{Reachability, epoch labeling and bracket shrinking} \label{subsec:d2}

\begin{lemma}[Reachability under observed compliance]
\label{lem:reach-concentration}
Let $R_t^{\rm ref}$ indicate whether the
pure profile $\bar s$ reaches $I$ under
the chance outcomes sampled in round $t$,
and define
\( 
V^{\rm ref}:=\sum_{t=1}^N R_t^{\rm ref}.
\) 
Then,
\begin{equation}\label{eq:efg-reach-pathwise}
V\ge V^{\rm ref}
-\sum_{t=1}^N D_t^{\rm out}
-\sum_{t=1}^N D_{-i,t}.
\end{equation}
Moreover, for every $\zeta\in(0,1)$,
\begin{equation}\label{eq:efg-reach-martingale}
\Prb\left(
V^{\rm ref}\ge
\rho_{i,I}(\bar s)N
-\sqrt{2N\log(1/\zeta)}
\right)\ge1-\zeta.
\end{equation}
\end{lemma}

\begin{proof}
Fix a round $t$ and use the same sampled
chance outcomes to execute the actual
strategy profile and the pure profile
$\bar s$. If
$D_t^{\rm out}=D_{-i,t}=0$, every action
taken along the actual trajectory before
reaching $I$ agrees with $\bar s$.
Hence, if the trajectory under $\bar s$
reaches $I$, the actual trajectory must
also reach $I$. Since no player has
deviated outside $I$, this round is
counted by $V$. Thus, for every round,
\[
\mathbf1\{\bar d_{i,I}(Z_t)=1,\,
I\preceq Z_t\}
\ge R_t^{\rm ref}
-D_t^{\rm out}-D_{-i,t}.
\]
Summing over $t$ proves
\eqref{eq:efg-reach-pathwise}. Finally, under the fixed pure profile
$\bar s$, reaching $I$ depends only
on Nature's random choices. Therefore,
conditional on the history before round $t$,
\( 
\mathbb E[R_t^{\rm ref}\mid
\mathcal F_{t^-}]
=\rho_{i,I}(\bar s).
\) 
Since $R_t^{\rm ref}\in\{0,1\}$,
the variables
$R_t^{\rm ref}-\rho_{i,I}(\bar s)$
are martingale differences bounded
in absolute value by $1$.
The Azuma--Hoeffding inequality
then gives
\( 
\Prb\big(
\sum_{t=1}^N
(R_t^{\rm ref}-\rho_{i,I}(\bar s))
<-\sqrt{2N\log(1/\zeta)}
\big)\le\zeta,
\) 
which is equivalent to
\eqref{eq:efg-reach-martingale}.
\end{proof}

\begin{proof}[Proof of Lemma~\ref{lem:certified-epoch-pathwise}]
Fix a query $(\bar s,i,I,a)$ and one of the two
payment levels $\Delta$ at stage $k$. Write
\( 
\gamma:=\gamma_k,
\lambda:=\lambda_\alpha,
Q:=Q_i(\alpha),
A:=C_i+3Q+3\lambda,
\) 
so that $H_i(\alpha)=A+Q+\lambda$.
Since
$\Delta_k^-=L_k+\gamma$ and
$\Delta_k^+=U_k-\gamma$, both payment
levels satisfy
\begin{equation}\label{eq:efg-two-payment level-margin-app}
0\le\Delta\le1-\gamma,
\qquad
\kappa:=B-1-\Delta\ge\gamma.
\end{equation}
Apply Lemma~\ref{lem:leak-mass-epoch-opponents}
to each opponent, Lemma~\ref{lem:reach-concentration}
to the number of visits to $I$, and
Lemma~\ref{lem:mistake-bound-decoupled}
with $\gamma=0$ to player $i$'s deviations
outside $I$. When $\Delta\ne\theta$, also
apply the Lemma~\ref{lem:mistake-bound-decoupled} with the margin
$g:=|\Delta-\theta|>0$.
Assign failure probability $\alpha/(4n)$
to each condition. Since there are at
most $(n-1)+3\le4n$ events, all the
resulting bounds hold simultaneously
with probability at least $1-\alpha$.
We work on this event throughout.

\vspace{6pt}
\noindent
First, due to Lemma \ref{lem:leak-mass-epoch-opponents}, summing the opponent-deviation
bounds gives
\begin{equation}\label{eq:efg-two-payment level-opp-app}
D_{-i}:=\sum_{t=1}^N D_{-i,t}
\le Q\sqrt N.
\end{equation}
Let $D^{\rm out}:=\sum_{t=1}^N D_t^{\rm out}$.
Since $\Delta\le1$ and $g\le1$, applying \eqref{eq:efg-two-payment level-opp-app} and
Lemma~\ref{lem:mistake-bound-decoupled}
with margins $0$ and $g$, respectively,
yields
\begin{equation}\label{eq:efg-two-payment level-out-mistake-app}
\kappa D^{\rm out}\le A\sqrt N,
\qquad
gM_g\le A\sqrt N\quad(g>0),
\end{equation}
where $M_g$ counts visits to $I$
on which all players are obedient and the action is suboptimal in
the direction determined by
$\Delta-\theta$.
In particular, \eqref{eq:efg-two-payment level-margin-app}
implies
\[
D^{\rm out}\le\frac{A}{\gamma}\sqrt N.
\]
Combining these bounds with
Lemma~\ref{lem:reach-concentration}
and $r:=\rho_{i,I}(\bar s)\ge\rho$ gives
\begin{align}
V
&\ge V^{\rm ref}-D^{\rm out}-D_{-i}
\nonumber\\
&\ge \rho N-
\left(\lambda+Q+\frac{A}{\gamma}\right)
\sqrt N
\nonumber\\
&\ge \rho N-\frac{H_i(\alpha)}{\gamma}\sqrt N,
\label{eq:efg-two-payment level-reach-app}
\end{align}
where the last inequality uses $\gamma\le1$.
By the choice of epoch length,
\( 
\sqrt N\ge\frac{16H_i(\alpha)}{\rho\gamma},
\) 
and therefore
\begin{equation}\label{eq:efg-two-payment level-reach-positive-app}
V\ge\frac{15}{16}\rho N>0.
\end{equation}
Suppose first that
$\Delta\ge\theta+\gamma$.
Then $g=\Delta-\theta\ge\gamma$,
and $M_g$ counts visits to $I$
on which player $i$ chooses an action
different from $a$. By
\eqref{eq:efg-two-payment level-out-mistake-app},
\[
V-X=M_g
\le\frac{A}{g}\sqrt N
\le\frac{A}{\gamma}\sqrt N
\le\frac{\rho N}{16}.
\]
Together with
$V\ge15\rho N/16$, this gives
\[
\frac{V-X}{V}\le\frac1{15}<\frac14,
\]
so $X\ge3V/4$ and the epoch is
labeled \textit{BEST}. Similarly, if $\Delta\le\theta-\gamma$,
then $g=\theta-\Delta\ge\gamma$
and $M_g=X$. Hence
\[
X\le\frac{\rho N}{16}
\le\frac{V}{15}<\frac{V}{4},
\]
so the epoch is labeled \textit{NOTBEST}. We next establish the one-sided bounds.
Suppose that the observed label is
\textit{BEST}. If $\theta\le\Delta$,
then $\theta\le\Delta+\beta_k^{\rm EF}$
holds immediately. Otherwise,
$g:=\theta-\Delta>0$ and $M_g=X$.
By \eqref{eq:efg-two-payment level-out-mistake-app}, \( \frac{\rho N}{16} \le \frac{V}{15}\) and
$X\ge3V/4$,
\[
\begin{aligned}
\theta-\Delta
&\le\frac{A\sqrt N}{X}\\
&\le\frac{A\sqrt N}{3V/4}\\
&\le\frac{64A}{45\rho\sqrt N}
\le\frac{2H_i(\alpha)}{\rho\sqrt N}
=:\beta_k^{\rm EF}.
\end{aligned}
\]
Thus, $\theta\le\Delta+\beta_k^{\rm EF}$. If the label is \textit{NOTBEST},
the bound is immediate when
$\theta\ge\Delta$. Otherwise,
$g:=\Delta-\theta>0$ and
$M_g=V-X\ge3V/4$. The same
calculation yields
\( 
\Delta-\theta\le\beta_k^{\rm EF},
\) 
or $\theta\ge\Delta-\beta_k^{\rm EF}$. Furthermore, the epoch length gives
\[
\beta_k^{\rm EF}
=\frac{2H_i(\alpha)}{\rho\sqrt N}
\le\frac{\gamma}{8}.
\]
Finally, the two separated cases
proved above show that
$|\Delta-\theta|\ge\gamma$
always produces a certified label.
Consequently, an \textit{INCONCL}
label implies $|\Delta-\theta|<\gamma$.
All conclusions hold on the same
event of probability at least
$1-\alpha$, completing the proof.
\end{proof}

\begin{proof}[Proof of Proposition~\ref{prop:efg-bracket-shrink}]
Fix an iteration on which both test-epoch guarantees hold.
The one-sided bounds in Lemma~\ref{lem:certified-epoch-pathwise}
show that every bound added in
\eqref{eq:efg-two-payment level-update} contains the unknown threshold
$\theta$. Thus the intersection remains nonempty and contains
$\theta$.
The payment design yields
$\Delta_k^-=L_k+\tau_k/4$ and
$\Delta_k^+=L_k+3\tau_k/4$.
On the conditioned event, at least one of the payments has distance at least $\gamma_k$ from $\theta$,
so the corresponding epoch is certified.
A \textit{BEST} label at $\Delta_k^\sigma$ gives
$U_{k+1}\le\Delta_k^\sigma+\beta_k^{\mathrm{EF}}$;
a \textit{NOTBEST} label gives
$L_{k+1}\ge\Delta_k^\sigma-\beta_k^{\mathrm{EF}}$.
The least restrictive such update is either
\textit{BEST} at $\Delta_k^+$ or \textit{NOTBEST} at
$\Delta_k^-$; both leave width of at most
$3\tau_k/4+\beta_k^{\mathrm{EF}}$.
Since $\beta_k^{\mathrm{EF}}\le\gamma_k/8=\tau_k/32$,
\[
U_{k+1}-L_{k+1}\le\left(\frac34+\frac1{32}\right)\tau_k
=\frac{25}{32}\tau_k.
\]
Any additional certified bound only reduces the width.
If the candidate intersection is empty or neither epoch is
certified, the fallback interval has width $\tau_k/2$.
In every case the width therefore contracts by at least a
factor of $25/32$, while on the successful event no fallback occurs
and the threshold remains inside the updated bracket.
\end{proof}

\subsection{Termination and complexity of the two-payment level search}

\begin{proof}[Proof of Theorem~\ref{thm:learn-efg-traj-pathwise}]
For $\varepsilon\ge1/2$, the initial midpoint
$1/2$ is already $\varepsilon$-accurate, so
no testing rounds are needed. Hence, suppose that
$0<\varepsilon<1/2$ and let
\( 
q:=\frac{25}{32},
K:=\left\lceil\log_{1/q}
\frac{1}{2\varepsilon}\right\rceil,
\alpha:=\frac{\delta_{\mathrm{loc}}}{2K}.
\) 

\vspace{6pt}
\noindent 
By Proposition~\ref{prop:efg-bracket-shrink},
whenever both tests in iteration $k$ satisfy
Lemma~\ref{lem:certified-epoch-pathwise},
the updated bracket contains
$\theta_{i,I}^{\bar s}(a)$ and its width
decreases by at least a factor of $q$.
The update rule also guarantees this
contraction on unsuccessful tests through
its fallback. Thus, starting from a bracket
of width $1$, the algorithm executes at most
$K$ iterations and hence at most $2K$ tests.

\vspace{6pt}
\noindent 
Each test satisfies
Lemma~\ref{lem:certified-epoch-pathwise} with conditional probability of at least
$1-\alpha$, given the history preceding
that test. A union bound over at most
$2K$ tests therefore shows that all
executed tests satisfy the Lemma~\ref{lem:certified-epoch-pathwise}
simultaneously with probability at least
\( 
1-2K\alpha=1-\delta_{\mathrm{loc}}.
\) 

\vspace{6pt}
\noindent 
On this event, the bracket contains
$\theta_{i,I}^{\bar s}(a)$ throughout
the algorithm. At termination, its
width is at most $2\varepsilon$, i.e. \(q^K \le 2\varepsilon\) , so
the returned midpoint satisfies
\( 
\big|
\widehat\theta_{i,I}^{\bar s}(a)
-\theta_{i,I}^{\bar s}(a)
\big|\le\varepsilon.
\)

\vspace{6pt}
\noindent 
We now bound the number of trajectories.
Let $J\le K$ be the number of executed
iterations, and let $\tau_k$ denote the bracket
width before iteration $k$.
Since the algorithm executes iteration
$J-1$, we have $\tau_{J-1}>2\varepsilon$.
Moreover, $\tau_{k+1}\le q \tau_k$ implies
\[
\tau_k\ge q^{-(J-1-k)}\tau_{J-1},
\qquad 0\le k<J.
\]
Consequently,
\[
\begin{aligned}
\sum_{k=0}^{J-1}\frac1{\tau_k^2}
&\le\frac1{\tau_{J-1}^2}
\sum_{\ell=0}^{J-1}q^{2\ell}\le\frac1{4\varepsilon^2(1-q^2)}.
\end{aligned}
\]
Since $\gamma_k=\tau_k/4$, the epoch length
\eqref{eq:efg-insured-epoch-length}
satisfies
\[
N_k
=\left\lceil
\frac{4096H_i(\alpha)^2}{\rho^2\tau_k^2}
\right\rceil
\le 1+\frac{4096H_i(\alpha)^2}{\rho^2\tau_k^2}.
\]
Each iteration uses two epochs. Therefore,
\[
\begin{aligned}
T_{\bar s,i,I,a}
=2\sum_{k=0}^{J-1}N_k&\le 2K+
\frac{8192H_i(\alpha)^2}{\rho^2}
\sum_{k=0}^{J-1}\frac1{\tau_k^2}\\
&\le 2K+
\frac{2048}{1-q^2}
\frac{H_i(\alpha)^2}{\rho^2\varepsilon^2}=2K+\frac{2097152}{399}
\frac{H_i(\alpha)^2}{\rho^2\varepsilon^2},
\end{aligned}
\]
where the last equality uses
$1-q^2=399/1024$.
This proves the theorem.
\end{proof}

\begin{proof}[Proof of Theorem~\ref{thm:learn-all-efg-allx}]
If $\mathcal J=0$ or $\varepsilon\ge1/2$,
no testing rounds are required. Otherwise,
apply Theorem~\ref{thm:learn-efg-traj-pathwise}
to each of the $\mathcal J$ positive-reach
queries with failure probability
$\delta/\mathcal J$. By a union bound,
with probability of at least $1-\delta$,
\( 
\bigl|
\widehat\theta_{i,I}^{\bar s}(a)
-\theta_{i,I}^{\bar s}(a)
\bigr|\le\varepsilon
\) 
simultaneously for every queried
$(\bar s,i,I,a)$. Now, fix $(\bar s,i,I)$ with positive reach
and let
\( 
v_{\max}:=\max_{b\in A_I}
v_{i,\bar s}^{(I)}(b).
\) 
By definition,
\( 
v_{i,\bar s}^{(I)}(a)
=v_{\max}-\theta_{i,I}^{\bar s}(a).
\) 
Since the algorithm outputs
\( 
\widetilde v_{i,\bar s}^{(I)}(a)
=-\widehat\theta_{i,I}^{\bar s}(a)
+c_{i,\bar s,I},
\) 
choosing the action-independent constant
$c_{i,\bar s,I}^\star
:=c_{i,\bar s,I}-v_{\max}$
gives
\[
\begin{aligned}
&\widetilde v_{i,\bar s}^{(I)}(a)
-v_{i,\bar s}^{(I)}(a)
-c_{i,\bar s,I}^\star=
\theta_{i,I}^{\bar s}(a)
-\widehat\theta_{i,I}^{\bar s}(a).
\end{aligned}
\]
Consequently,
\[
\max_{a\in A_I}
\left|
\widetilde v_{i,\bar s}^{(I)}(a)
-v_{i,\bar s}^{(I)}(a)
-c_{i,\bar s,I}^\star
\right|
\le\varepsilon.
\]
The constant $c_{i,\bar s,I}^\star$
is used only to establish accuracy;
the algorithm does not need to know
$v_{\max}$. For the trajectory bound, each query
uses at most
\[
2K+\frac{2097152}{399}
\frac{H_i(\alpha)^2}
{\rho_{\bar s,i,I}^2\varepsilon^2}
\]
rounds by
Theorem~\ref{thm:learn-efg-traj-pathwise},
where $\alpha=\delta/(2\mathcal JK)$.
Since
$H_i(\alpha)\le H_{\max}(\alpha)$
and $\rho_{\bar s,i,I}\ge\rho_{\min}$,
summing over the $\mathcal J$ queries
yields
\[
T_{\rm tot}
\le 2\mathcal JK
+\frac{2097152}{399}
\frac{\mathcal JH_{\max}(\alpha)^2}
{\rho_{\min}^2\varepsilon^2}.
\]
Finally, each query uses at most $2K$
testing epochs, with one new
context per player per epoch. Hence
at most $2n\mathcal JK$ context identifiers
are required. The payment rules
\eqref{eq:efg-insured-opponents} and
\eqref{eq:efg-insured-payment} also give
$0\le q_j(z)\le B+1$ for every player
and terminal history. This completes
the proof.
\end{proof}

\subsection{Sufficiency of the learned
continuation values}\label{app:efg-sufficiency}

\begin{proof}[Proof of
Proposition~\ref{prop:efg-values-suffice-efce}]

We first fix player $i$, information set $I\in\mathcal I_i$,
recommended action $a\in A_I$, and a pure continuation
plan $\hat s_i\in S_i(I)$. Under the EFCE protocol, player $i$ follows the
sampled recommendations until reaching $I$ and
receiving recommendation $a$. Conditional on a
sampled profile $\bar s$, this trigger can occur
only if $\bar s_i\in S_i(I,a)$.

\vspace{6pt}
\noindent 
When the trigger occurs, player $i$ switches to
$\hat s_i$ at $I$ and subsequently, while the
opponents continue playing $\bar s_{-i}$.
The resulting behavior is represented by
$F_i^{I,\hat s_i}(\bar s_i)$.
This plan agrees with $\bar s_i$ before $I$
and uses $\hat s_i$ at $I$ and at all following
information sets.

\vspace{6pt}
\noindent
For any $\bar s_i\in S_i(I,a)$, obedience and
the trigger response induce the same terminal
distribution on histories that do not pass
through $I$. Under obedience, terminal histories
passing through $I$ must belong to $Z(I,a)$.
After triggering, the continuation may instead
reach any terminal history in $Z(I)$.
Since $\bar s\sim\mu$ is sampled independently
of Nature, expanding the corresponding expected
payoffs over terminal histories gives
\begin{equation}\label{eq:app-efce-trigger-identity}
\begin{aligned}
&\sum_{\substack{\bar s\in\mathcal S\\
\bar s_i\in S_i(I,a)}}
\mu(\bar s)
\Bigl[
U_i(F_i^{I,\hat s_i}(\bar s_i),\bar s_{-i})
-U_i(\bar s)
\Bigr]=
\sum_{z\in Z(I)}
p^{(I,a)}_{\mu,\delta_{\hat s_i}}(z)u_i(z)
-
\sum_{z\in Z(I,a)}q_\mu(z)u_i(z),
\end{aligned}
\end{equation}
where $\delta_{\hat s_i}$ denotes the distribution
concentrated on $\hat s_i$.
This is precisely the trigger-deviation gain in
Definition~\ref{def:efce}. By linearity, the
maximum over mixed continuation plans equals
the maximum over pure continuation plans, which shows that $G_i(\mu;I,a,\hat s_i)$ can determine $\delta(\mu)$. 

\vspace{6pt}
\noindent 
Now, fix $s_{-i}$ and two pure strategies
$s_i,t_i\in S_i$.
Enumerate
$\mathcal I_i=\{I_1,\ldots,I_m\}$, where
$m=|\mathcal I_i|$, and construct
$s_i^0,\ldots,s_i^m$ by setting $s_i^0=s_i$
and
\[
s_i^k(I_k)=t_i(I_k),\qquad
s_i^k(J)=s_i^{k-1}(J)
\quad(J\ne I_k).
\]
Thus $s_i^m=t_i$. Write
$s^{k-1}:=(s_i^{k-1},s_{-i})$. By telescoping and applying
\eqref{eq:efg-local-unconditional-difference}
at each information set, we obtain
\begin{equation}\label{eq:app-efg-hybrid}
\begin{aligned}
&U_i(t_i,s_{-i})-U_i(s_i,s_{-i})=\sum_{k=1}^{m}
\rho_{i,I_k}(s^{k-1})
\Bigl[
v_{i,s^{k-1}}^{(I_k)}(t_i(I_k))
-v_{i,s^{k-1}}^{(I_k)}(s_i^{k-1}(I_k))
\Bigr].
\end{aligned}
\end{equation}
Now, let $\widehat D_i(t_i,s_i;s_{-i})$
denote the right-hand side of
\eqref{eq:app-efg-hybrid}, replacing
$v$ by $\widetilde v$ at every
positive-reach term. By assumption, for each such term there
exists an action-independent constant
$c_{i,s^{k-1},I_k}$ such that
\[
\begin{aligned}
&\Bigl|
\bigl[
\widetilde v_{i,s^{k-1}}^{(I_k)}(b)
-\widetilde v_{i,s^{k-1}}^{(I_k)}(a)
\bigr]-
\bigl[
v_{i,s^{k-1}}^{(I_k)}(b)
-v_{i,s^{k-1}}^{(I_k)}(a)
\bigr]
\Bigr|
\le 2\varepsilon
\end{aligned}
\]
for all $a,b\in A_{I_k}$.
The additive constant cancels because both
actions are evaluated at the same information
set under the same reference profile. Since $\rho_{i,I_k}(s^{k-1})\le1$,
\begin{equation}\label{eq:app-efg-hybrid-error}
\begin{aligned}
&\left|
\widehat D_i(t_i,s_i;s_{-i})
-\bigl[
U_i(t_i,s_{-i})-U_i(s_i,s_{-i})
\bigr]
\right|\le
2\varepsilon
\sum_{k=1}^{m}\rho_{i,I_k}(s^{k-1})
\le 2m\varepsilon.
\end{aligned}
\end{equation}
The bound holds for every pair of pure
strategies and every fixed opponents'
pure profile, proving
\eqref{eq:efg-pure-plan-difference-accuracy}.

\vspace{6pt}
\noindent 
Now, we proceed with the estimation of trigger-deviations. For a fixed trigger $(I,a)$ and pure
continuation plan $\hat s_i\in S_i(I)$,
define
\[ 
\begin{aligned}
&\widehat G_i(\mu;I,a,\hat s_i):=
\sum_{\substack{\bar s\in\mathcal S,
\bar s_i\in S_i(I,a)}}
\mu(\bar s)\,
\widehat D_i\bigl(
F_i^{I,\hat s_i}(\bar s_i),
\bar s_i;\bar s_{-i}
\bigr).
\end{aligned}
\] 
Combining
\eqref{eq:efg-efce-gain-from-pure-profiles}
with \eqref{eq:app-efg-hybrid-error} gives
\[
\begin{aligned}
&\left|
\widehat G_i(\mu;I,a,\hat s_i)
-G_i(\mu;I,a,\hat s_i)
\right| \le
\sum_{\substack{\bar s\in\mathcal S,
\bar s_i\in S_i(I,a)}}
\mu(\bar s)\,2m\varepsilon\le 2|\mathcal I_i|\varepsilon.
\end{aligned}
\]
By Definition~\ref{def:efce} and
\eqref{eq:app-efce-trigger-identity},
the maximum trigger-deviation gain equals
\[
\delta(\mu)=
\max_{i\in[n]}
\max_{\substack{I\in\mathcal I_i\\a\in A_I}}
\max_{\hat s_i\in S_i(I)}
G_i(\mu;I,a,\hat s_i).
\]
Similarly, define
\[
\widehat\delta(\mu)=
\max_{i\in[n]}
\max_{\substack{I\in\mathcal I_i\\a\in A_I}}
\max_{\hat s_i\in S_i(I)}
\widehat G_i(\mu;I,a,\hat s_i).
\]
For two families of real numbers indexed
by the same set, the difference between
their maxima is at most the largest
absolute difference between corresponding
terms. Therefore,
\[
\begin{aligned}
|\widehat\delta(\mu)-\delta(\mu)|
&\le
\max_{i,I,a,\hat s_i}
\left|
\widehat G_i(\mu;I,a,\hat s_i)
-G_i(\mu;I,a,\hat s_i)
\right|\le
2\varepsilon\max_{i\in[n]}|\mathcal I_i|.
\end{aligned}
\]
Finally, if $\widehat\delta(\mu)\le\eta$,
then
\( 
\delta(\mu)
\le
\widehat\delta(\mu)
+2\varepsilon\max_i|\mathcal I_i|
\le
\eta+2\varepsilon\max_i|\mathcal I_i|.
\) 
By Definition~\ref{def:efce}, this means
that $\mu$ is an
$\bigl(\eta+2\varepsilon
\max_i|\mathcal I_i|\bigr)$-EFCE.
\end{proof}

\subsection{Proof of Lower Bounds for Learning Continuation Values in EFGs}

\begin{proof}[Proof of Proposition~\ref{prop:efg-inherited-lower-bounds}]
We establish the three lower bounds separately. The first one comes from the
amount of information needed to identify the continuation-value table,
the second one from the behavior permitted by the no-regret assumption, and
the third one from information sets that are rarely reached. 

\vspace{6pt}
\noindent
We first show that
\( 
T=\Omega_\delta\big(
\frac{\mathcal J\log(1/\varepsilon)}
{\log\mathcal J}
\big)
\) 
by embedding a family of normal-form games into
extensive-form games. Consider an $n$-player game in which each player has
two actions, $A_i=\{0,1\}$. Represent the game in
extensive form by letting players act sequentially,
without observing any preceding actions. Thus each
player has a single information set $I_i$, reached
with probability one.

\vspace{6pt}
\noindent 
There are $2^n$ pure strategy profiles, and the
number of continuation-value entries is
\( 
\mathcal J=2n2^n.
\) 
Moreover, for every pure profile $\bar s$,
\( 
v_{i,\bar s}^{(I_i)}(a)
=u_i(a,\bar s_{-i}).
\) 
Thus, the continuation values coincide with the
corresponding normal-form payoffs. Fix $0<\varepsilon\le 1/16$ and define the grid
\[
\mathcal G_\varepsilon
=\left\{8\varepsilon k:
0\le k\le
\left\lfloor\frac{1}{8\varepsilon}\right\rfloor
\right\},
\qquad
M_\varepsilon:=|\mathcal G_\varepsilon|.
\]
For every player $i$ and opponents' action profile
$a_{-i}$, set $u_i(0,a_{-i})=0$ and let
$u_i(1,a_{-i})$ be an unknown element of
$\mathcal G_\varepsilon$. There are
\( 
D=n2^{n-1}=\frac{\mathcal J}{4}
\) 
such unknown entries. Hence the family contains
\( 
|\mathcal U_\varepsilon|=M_\varepsilon^D
\) 
distinct utility vectors. We first observe that an $\varepsilon$-accurate
continuation-value table identifies the unknown
utility vector exactly. Indeed, for each
$(i,a_{-i})$, consider the estimated difference
\[
\widetilde\Delta_i(a_{-i})
:=
\widetilde v_{i,(0,a_{-i})}^{(I_i)}(1)
-
\widetilde v_{i,(0,a_{-i})}^{(I_i)}(0).
\]
Since the estimation error is at most $\varepsilon$
per action, up to a common additive constant,
\[
\left|
\widetilde\Delta_i(a_{-i})
-u_i(1,a_{-i})
\right|\le 2\varepsilon.
\]
Distinct grid points are separated by
$8\varepsilon$. Therefore, rounding
$\widetilde\Delta_i(a_{-i})$ to its nearest grid
point recovers $u_i(1,a_{-i})$ exactly.
Applying this to every $(i,a_{-i})$ recovers
the entire vector $U$.

\vspace{6pt}
\noindent 
It remains to count how many utility vectors can
be distinguished from $T$ observations.
Each terminal history is a binary action profile
in $\{0,1\}^n$. Consequently, the complete
observation sequence
$Y=(Z_1,\ldots,Z_T)$ has at most $2^{nT}$
possible realizations. To account for randomization, let $R$ denote
the mediator's private random seed, independent
of $U$. Conditional on $R=r$, the decoder
$\widehat U=f(Y,r)$ is deterministic.
For a uniformly selected
$U\in\mathcal U_\varepsilon$, we have
\begin{align*}
\Pr(\widehat U=U\mid R=r)
&=\frac{1}{M_\varepsilon^D}
  \sum_{u\in\mathcal U_\varepsilon}
  \sum_y
  \Pr(Y=y\mid U=u,R=r)
  \mathbf1\{f(y,r)=u\}\le \frac{2^{nT}}{M_\varepsilon^D}.
\end{align*}
The inequality holds because each observation
sequence $y$ is decoded into only one utility
vector. It remains valid for randomized agents
and adaptive mediator decisions. Averaging over $R$ gives
\( 
\Pr(\widehat U=U)
\le 2^{nT}/ {M_\varepsilon^D}.
\) 
Therefore, a learning algorithm that succeeds
with probability at least $1-\delta$ for every
game in this family must satisfy
\( 
1-\delta
\le 2^{nT}/M_\varepsilon^D.
\) 
Taking logarithms and rearranging yields
\[
T\ge
\frac{D\log M_\varepsilon+\log(1-\delta)}
{n\log 2}.
\]
Since $D=\mathcal J/4$ and
$M_\varepsilon=\Theta(1/\varepsilon)$, for
fixed $\delta<1$ and sufficiently small
$\varepsilon$,
\( 
T=\Omega_\delta\big(
\frac{\mathcal J\log(1/\varepsilon)}{n}
\big).
\) 
Finally,
\( 
\log\mathcal J
=n\log2+\log(2n)=\Theta(n),
\) 
and hence
\begin{align} \label{quaso:1}{
T=\Omega_\delta\big(
\frac{\mathcal J\log(1/\varepsilon)}
{\log\mathcal J}
\big).
}
\end{align}

\vspace{6pt}
\noindent
Second, we show that no-regret behavior alone can prevent the mediator
from distinguishing two different utility functions for
$\Omega(C^2/\varepsilon^2)$ rounds. Consider a single-agent
game with one information set and actions $\{0,1\}$. The two
possible utility vectors are
\( 
u^0=(0,0), u^1=(0,8\varepsilon).
\) 
Set
\( 
T_0:=\left\lfloor\frac{C^2}{64\varepsilon^2}\right\rfloor.
\) 

\vspace{6pt}
\noindent
We describe two admissible response rules. Under $u^1$, the agent
plays a best response to the announced payment rule $q_t$ in every
round, using fixed tie-breaking. Under $u^0$, the agent mimics
exactly this same best response for the first $T_0$ rounds and
then plays a best response to $u^0+q_t$ thereafter. Since payments
are announced before the action is chosen, these rules produce the
same actions, and hence the same mediator observations, in the first
$T_0$ rounds under any adaptive payment and context policy.

\vspace{6pt}
\noindent
We verify that both rules satisfy the regret bound on every
prefix of each context. Under $u^1$, every chosen action maximizes the current
augmented payoff, so the realized regret on every context
subsequence is non-positive. Under $u^0$, let $a_t$ be the action
played during a mimicking round. For any comparator $b$, the
best-response property under $u^1$ implies
\[
u^1(b)+q_t(b)\le u^1(a_t)+q_t(a_t),
\]
and therefore
\[
(u^0(b)+q_t(b))-(u^0(a_t)+q_t(a_t))
=q_t(b)-q_t(a_t)\le 8\varepsilon.
\]
After round $T_0$, each comparator's per-round regret contribution
under $u^0$ is nonpositive. Thus, writing $N_c(t)$ for the number
of occurrences of context $c$ through time $t$, we have
\begin{align*}
\widehat R_{1,c}(N_c(t))
&\le 8\varepsilon\min\{N_c(t),T_0\}\le 8\varepsilon\sqrt{T_0N_c(t)}
\le C\sqrt{N_c(t)}.
\end{align*}
Both environments therefore satisfy the stipulated no-regret
condition, yet their first $T_0$ rounds are observationally
indistinguishable to the mediator.

\vspace{6pt}
\noindent
Nevertheless, their continuation-value differences are $0$ and
$8\varepsilon$, respectively. An $\varepsilon$-accurate table,
up to a common additive constant, estimates either difference
with error at most $2\varepsilon$. The corresponding intervals
$[-2\varepsilon,2\varepsilon]$ and
$[6\varepsilon,10\varepsilon]$ are disjoint. Since the mediator's
output has the same distribution in both environments, it cannot
be accurate with probability greater than $1/2$ in both.
As $\delta<1/2$, uniform success probability $1-\delta$ is
impossible by time $T_0$. Hence
\begin{equation}\label{eq:lb-regret-final}
T=\Omega\!\left(\frac{C^2}{\varepsilon^2}\right).
\end{equation}

\begin{figure}[t]
\centering
\begin{tikzpicture}[
    >=Latex,
    scale=0.95,
    every node/.style={font=\small},
    chance/.style={
        circle, draw, fill=gray!20,
        minimum size=7mm, inner sep=0pt
    },
    info/.style={
        rectangle, rounded corners, draw, fill=blue!8,
        minimum width=10mm, minimum height=6mm
    },
    terminal/.style={
        circle, draw, fill=black, inner sep=1.2pt
    },
    prob label/.style={
        midway, fill=white,
        text=red!70!black,
        font=\scriptsize\itshape,
        inner sep=1.5pt
    },
    action label/.style={
        midway, fill=blue!8, draw=blue!50,
        rounded corners=1pt,
        text=blue!70!black,
        font=\scriptsize\bfseries,
        inner sep=1.2pt
    },
    payoff label/.style={
        font=\scriptsize,
        inner sep=1pt,
        fill=white
    }
]

\node[chance] (C1) at (0,0) {$c_1$};
\node[chance] (C2) at (2.7,-1.9) {$c_2$};
\node[chance] (C3) at (5.4,-3.8) {$c_3$};

\node[chance] (CHmOne) at (8.1,-5.7) {$c_{h-1}$};

\node[info] (I1) at (-2.1,-1.4) {$I_1$};
\node[above=1pt of I1] {\scriptsize P2};

\node[info] (I2) at (0.6,-3.3) {$I_2$};
\node[above=1pt of I2] {\scriptsize P2};

\node[info] (I3) at (3.3,-5.2) {$I_3$};
\node[above=1pt of I3] {\scriptsize P2};

\node[info] (IHmOne) at (6.0,-7.1) {$I_{h-1}$};
\node[above=1pt of IHmOne] {\scriptsize P2};

\node[info] (IH) at (10.2,-7.1) {$I_h$};
\node[above=1pt of IH] {\scriptsize P1};

\draw[->] (C1) --
    node[prob label, above left] {$1/2$} (I1);

\draw[->] (C2) --
    node[prob label, above left] {$1/2$} (I2);

\draw[->] (C3) --
    node[prob label, above left] {$1/2$} (I3);

\draw[->] (CHmOne) --
    node[prob label, above left] {$1/2$} (IHmOne);

\draw[->] (C1) --
    node[prob label, above right] {$1/2$} (C2);

\draw[->] (C2) --
    node[prob label, above right] {$1/2$} (C3);

\draw[dashed,->] (C3) --
    node[
        midway,
        fill=white,
        text=black,
        inner sep=2pt
    ] {$\cdots$} (CHmOne);

\draw[->] (CHmOne) --
    node[prob label, above right] {$1/2$} (IH);

\node[terminal] (T1a) at (-2.85,-2.6) {};
\node[terminal] (T1b) at (-1.35,-2.6) {};

\draw[->] (I1) --
    node[action label, above left] {$0$} (T1a);
\draw[->] (I1) --
    node[action label, above right] {$1$} (T1b);

\node[payoff label, below=3pt of T1a] {$(0,0)$};
\node[payoff label, below=3pt of T1b] {$(0,0)$};

\node[terminal] (T2a) at (-0.15,-4.5) {};
\node[terminal] (T2b) at (1.35,-4.5) {};

\draw[->] (I2) --
    node[action label, above left] {$0$} (T2a);
\draw[->] (I2) --
    node[action label, above right] {$1$} (T2b);

\node[payoff label, below=3pt of T2a] {$(0,0)$};
\node[payoff label, below=3pt of T2b] {$(0,0)$};

\node[terminal] (T3a) at (2.55,-6.4) {};
\node[terminal] (T3b) at (4.05,-6.4) {};

\draw[->] (I3) --
    node[action label, above left] {$0$} (T3a);
\draw[->] (I3) --
    node[action label, above right] {$1$} (T3b);

\node[payoff label, below=3pt of T3a] {$(0,0)$};
\node[payoff label, below=3pt of T3b] {$(0,0)$};

\node[terminal] (THmOnea) at (5.25,-8.3) {};
\node[terminal] (THmOneb) at (6.75,-8.3) {};

\draw[->] (IHmOne) --
    node[action label, above left] {$0$} (THmOnea);
\draw[->] (IHmOne) --
    node[action label, above right] {$1$} (THmOneb);

\node[payoff label, below=3pt of THmOnea] {$(0,0)$};
\node[payoff label, below=3pt of THmOneb] {$(0,0)$};

\node[terminal] (THa) at (9.45,-8.3) {};
\node[terminal] (THb) at (10.95,-8.3) {};

\draw[->] (IH) --
    node[action label, above left] {$0$} (THa);
\draw[->] (IH) --
    node[action label, above right] {$1$} (THb);

\node[payoff label, below=3pt of THa] {$(0,0)$};
\node[payoff label, below=3pt of THb] {$(\xi,0)$};

\end{tikzpicture}

\caption{
EFG used in the reachability lower bound. Red labels indicate chance probabilities, and blue boxed
labels indicate player actions. All terminal payoffs are
$(0,0)$ except after action $1$ at $I_h$, where the payoff
is $(\xi,0)$. The two utility environments correspond to
$\xi=0$ and $\xi=1$.
}
\label{fig:reachability-lb-game}
\end{figure}
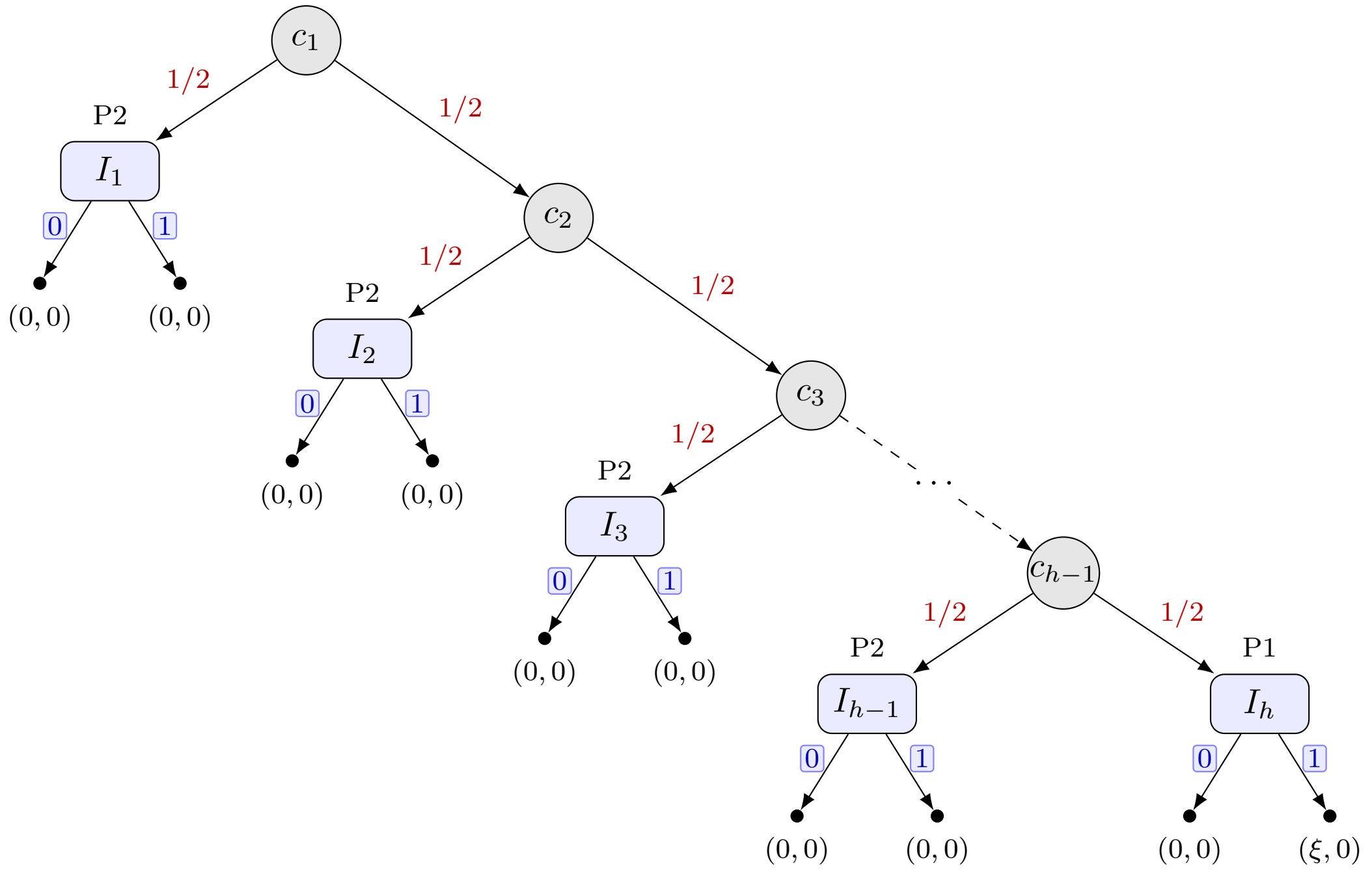

\vspace{6pt}
\noindent
The final construction captures a difficulty specific to
trajectory feedback: the mediator cannot identify a continuation
value without observing the part of the game tree on which it
depends. Fix $h\ge 2$ and consider the two-player extensive-form game
illustrated in Figure~\ref{fig:reachability-lb-game}.
The game consists of $h-1$ fair binary chance nodes
$c_1,\ldots,c_{h-1}$ arranged in a chain. At each chance
node $c_j$, one outcome leads to player~2's information
set $I_j$, while the other one continues to the next chance
node. If Nature takes the continuation branch at every
chance node, play reaches the final information set $I_h$
of player~1. At each information set, the acting player
chooses between actions $0$ and $1$, after which the
game terminates. Since every information set has two available actions,
the game has $2^h$ pure strategy profiles and
\( 
|\mathcal S|=2^h,
\mathcal J=2h2^h.
\) 
Moreover, as in
Figure~\ref{fig:reachability-lb-game}, reaching $I_h$
requires Nature to select the continuation branch
at all nodes. Therefore,
\( 
\rho_{\min}
=\Pr(I_h\text{ is reached})
=\left(\frac12\right)^{h-1}.
\) 
The game tree has only $O(h)$ histories, and every
information set has positive reach under every pure
strategy profile. Importantly, these reach probabilities
depend only on Nature's moves and cannot be influenced
by the mediator's contexts or payments.

\vspace{6pt}
\noindent
We now define two utility environments on the same game
tree. In both environments, all terminal payoffs are zero,
except possibly player~1's payoff after choosing action $1$
at the rare information set $I_h$. This payoff is $0$ in
the first environment and $1$ in the second. Thus, the
two environments differ only at the rightmost terminal
branch in Figure~\ref{fig:reachability-lb-game}.
For any reference pure profile $\bar s$, the conditional
continuation-value difference at $I_h$ satisfies
\[
v_{1,\bar s}^{(I_h)}(1)
-v_{1,\bar s}^{(I_h)}(0)
=
\begin{cases}
0, & \text{in the first environment},\\
1, & \text{in the second environment}.
\end{cases}
\]
Let $E$ denote the event that none of the first $T$
trajectories reaches $I_h$. Since Nature's chance
draws are independent across rounds,
\( 
\Pr(E)=(1-\rho_{\min})^T.
\) 
The key observation is that, conditional on $E$,
the two utility environments are indistinguishable
to the mediator. Indeed, they differ only in
player~1's payoff at $I_h$, which is never visited
on this event. Moreover, player~1 makes no decisions
before $I_h$, and we may take player~2 to follow
the same admissible learning rule in both environments,
as its utilities are identical. Consequently, the
observed terminal histories have the same conditional
distribution in both environments. Since the mediator
chooses its contexts, payments, and final estimates
based only on these observations and its own
randomization, its output also has the same conditional
distribution in both environments.

\vspace{6pt}
\noindent
However, the true continuation-value difference at
$I_h$ is $0$ in the first environment and $1$ in the
second. An $\varepsilon$-accurate continuation-value
table estimates this difference with error at most
$2\varepsilon$. Since $\varepsilon<1/4$, no estimate
can be accurate in both environments. Thus,
conditional on $E$, the sum of the two failure
probabilities is at least one. Since $E$ has the same
probability in both environments, the unconditional
failure probabilities satisfy
\[
\Pr_0(\mathrm{failure})+
\Pr_1(\mathrm{failure}) = 2\delta
\ge \Pr(E)=(1-\rho_{\min})^T.
\]
Taking logarithms and rearranging gives
\[
T\ge
\frac{\log(1/(2\delta))}
{-\log(1-\rho_{\min})}.
\]
Finally, since $h\ge2$ implies $\rho_{\min}\le1/2$,
we use $-\log(1-x)\le2x$ for $x\in(0,1/2]$ to obtain
\begin{equation}\label{eq:lb-reach-final}
T\ge
\frac{\log(1/(2\delta))}{2\rho_{\min}}
=
\Omega\left(
\frac{\log(1/(2\delta))}{\rho_{\min}}
\right).
\end{equation}
This establishes the reachability lower bound. Finally, taking the maximum of \eqref{quaso:1}, \eqref{eq:lb-regret-final}, \eqref{eq:lb-reach-final} gives the stated worst-case summary.
\end{proof}

\section{Proofs for Section \ref{sec:steering}}
\label{app:efce-steering-proofs}

\subsection{Deviation strategies and the learned gains}
\begin{lemma}
\label{lem:efce-pure-response-plan}
Fix player $i$ and a pure strategy
$\tau_i\in\mathcal D_i$.
For each $\bar s_i\in S_i$, there is a complete
original-game plan $F_i^{\tau_i}(\bar s_i)\in S_i$.
These plans satisfy, for every $\mu\in\Delta(\mathcal S)$,
\begin{equation}
u_i(\mu,\tau_i,d_{-i})-u_i(\mu,d)
=
\sum_{\bar s\in\mathcal S}\mu(\bar s)
\bigl[
U_i(F_i^{\tau_i}(\bar s_i),\bar s_{-i})
-U_i(\bar s)
\bigr].
\label{eq:app-efce-full-deviation}
\end{equation}
\end{lemma}

\begin{proof}
Condition on the sampled profile $\bar s$.
At an original-game information set $I$, perfect recall
fixes the player's previous information sets and actions.
Together with $\bar s_i$, these determine the previous
recommendations and whether advice has already stopped.
If advice has not stopped, the current recommendation
is $\bar s_i(I)$.
Thus the pure strategy $\tau_i$ prescribes a definite
action at every $I$ compatible with its induced own
history. These actions define an original-game plan;
complete incompatible information sets arbitrarily.
This construction depends on $\bar s_i$ and $\tau_i$,
but not on Nature's unobserved realization or
$\bar s_{-i}$.
With direct opponents, the terminal distribution is
therefore the distribution under
$(F_i^{\tau_i}(\bar s_i),\bar s_{-i})$.
Averaging over $\bar s$ proves
\eqref{eq:app-efce-full-deviation}.
\end{proof}

\paragraph{Constructing the estimated gains.}
Applying Proposition~\ref{prop:efg-values-suffice-efce}
with $s_i=\bar s_i$,
$t_i=F_i^{\tau_i}(\bar s_i)$, and
$s_{-i}=\bar s_{-i}$, lets 
Lemma~\ref{lem:efce-pure-response-plan} identify
the corresponding true gain with
$A_{\bar s,(i,\tau_i)}$.
Define
\begin{equation}\label{eq:efce-estimated-gain-matrix}
\widehat A_{\bar s,(i,\tau_i)}
:=
\operatorname{clip}_{[-1,1]}
\widehat D_i\bigl(
F_i^{\tau_i}(\bar s_i),\bar s_i;\bar s_{-i}
\bigr),
\qquad
\widehat A_{\bar s,0}:=0.
\end{equation}
The bound
\eqref{eq:efg-pure-plan-difference-accuracy} gives
\eqref{eq:efce-gain-matrix-accuracy}.
Clipping preserves it because every true entry is
in $[-1,1]$.

\subsection{Properties of the Lagrangian}

\begin{lemma}\label{lem:efce-lagrangian-justification}
The program \eqref{eq:efce-primal-program} satisfies
\[
u_0^\star
=
\max_{\mu\in\Delta(\mathcal S)}
\inf_{\substack{\ell\ge0\\y\in\Delta(\mathcal D)}}
\bigl[c^\top\mu-\ell\mu^\top Ay\bigr].
\]
For every fixed $\lambda\ge1$ and every $\mu$,
\[
\begin{aligned}
&\min_{\substack{\gamma\in\mathbb R_+^{\mathcal D}\\
                 \mathbf1^\top\gamma\le\lambda}}
\bigl[c^\top\mu-\mu^\top A\gamma\bigr]
=\lambda\min_{y\in\Delta(\mathcal D)}
L_\lambda(\mu,y).
\end{aligned}
\]
The game with objective $L_\lambda(\mu,y)
=
\frac{c^\top\mu}{\lambda}-\mu^\top Ay$ has a saddle
point. There is a finite $\lambda^\star$ such that,
for $\lambda>\lambda^\star$, its maximizing
\( \mu \in \Delta(\mathcal S)\) are precisely the optimal EFCEs,
and its value is $u_0^\star/\lambda$.
\end{lemma}

\begin{proof}
For every fixed $\mu\in\Delta(\mathcal S)$
and $\ell\ge0$,
\[
\min_{y\in\Delta(\mathcal D)}
\bigl[c^\top\mu-\ell\mu^\top Ay\bigr]
=
c^\top\mu-\ell\Delta_{\mathrm{eq}}(\mu).
\]
Since $\Delta_{\mathrm{eq}}(\mu)\ge0$, with equality
exactly when $A^\top\mu\le0$, taking the infimum
over $\ell\ge0$ gives
\[
\inf_{\substack{\ell\ge0\\
                 y\in\Delta(\mathcal D)}}
\bigl[c^\top\mu-\ell\mu^\top Ay\bigr]
=
\begin{cases}
c^\top\mu,
& A^\top\mu\le0,\\
-\infty,
& A^\top\mu\not\le0.
\end{cases}
\]
Maximizing over $\mu\in\Delta(\mathcal S)$
therefore recovers
\eqref{eq:efce-primal-program}, which proves the first claim.

\vspace{6pt}
\noindent
For the second claim, fix $\lambda\ge1$.
Every nonnegative vector $\gamma$ with
$\mathbf1^\top\gamma\le\lambda$ can be written
as $\gamma=\ell y$, where
$\ell=\mathbf1^\top\gamma\in[0,\lambda]$
and $y\in\Delta(\mathcal D)$.
For $\ell>0$, take $y=\gamma/\ell$;
for $\ell=0$, any $y\in\Delta(\mathcal D)$
suffices.
Conversely, every such pair $(\ell,y)$
gives an admissible $\gamma$.
Thus
\[
\begin{aligned}
\min_{\substack{\gamma\ge0\\
                 \mathbf1^\top\gamma\le\lambda}}
\bigl[c^\top\mu-\mu^\top A\gamma\bigr]
&=
\min_{0\le\ell\le\lambda}
\bigl[c^\top\mu-\ell\Delta_{\mathrm{eq}}(\mu)\bigr]\\
&\qquad=
c^\top\mu-\lambda\Delta_{\mathrm{eq}}(\mu)\\
&\qquad=
\lambda\min_{y\in\Delta(\mathcal D)}
L_\lambda(\mu,y).
\end{aligned}
\]
The minimum over $\ell$ is attained at
$\ell=\lambda$ because
$\Delta_{\mathrm{eq}}(\mu)\ge0$.
Furthermore, since $L_\lambda$ is bilinear and both domains
are finite-dimensional simplices, the finite
minimax theorem gives
\[
\max_{\mu\in\Delta(\mathcal S)}
\min_{y\in\Delta(\mathcal D)}L_\lambda(\mu,y)
=
\min_{y\in\Delta(\mathcal D)}
\max_{\mu\in\Delta(\mathcal S)}L_\lambda(\mu,y),
\]
with both extrema attained.
Hence the game has a saddle point. It remains to show that sufficiently large
$\lambda$ recovers exactly the optimal EFCEs.
The linear-programming dual of
\eqref{eq:efce-primal-program} is
\[
\min_{\substack{b\in\mathbb R\\
                 \gamma\in\mathbb R_+^{\mathcal D}}}
\bigl\{b:b\mathbf1+A\gamma\ge c\bigr\}.
\]
The primal is feasible and bounded, so
linear-programming strong duality gives an
optimal dual solution $(u_0^\star,\gamma^\star)$.
In particular,
\[
c\le u_0^\star\mathbf1+A\gamma^\star.
\]
For every $\mu\in\Delta(\mathcal S)$,
multiplying this inequality by $\mu$ gives
\begin{equation}\label{eq:efce-dual-objective-bound}
\begin{aligned}
c^\top\mu
&\le u_0^\star
   +(\gamma^\star)^\top A^\top\mu\\
&\le u_0^\star
   +\|\gamma^\star\|_1
    \Delta_{\mathrm{eq}}(\mu),
\end{aligned}
\end{equation}
where the second inequality follows from
$\gamma^\star\ge0$ and
$(A^\top\mu)_j\le\Delta_{\mathrm{eq}}(\mu)$
for every $j\in\mathcal D$. Set $\lambda^\star:=\|\gamma^\star\|_1$.
For $\lambda>\lambda^\star$,
\[
c^\top\mu-\lambda\Delta_{\mathrm{eq}}(\mu)
\le
u_0^\star
-(\lambda-\lambda^\star)
\Delta_{\mathrm{eq}}(\mu).
\]
Every infeasible law has
$\Delta_{\mathrm{eq}}(\mu)>0$ and therefore
attains a value strictly below $u_0^\star$.
For a feasible law,
$\Delta_{\mathrm{eq}}(\mu)=0$, so its value
is $c^\top\mu$, which equals $u_0^\star$
exactly when it is an optimal EFCE.
Thus
\[
\operatorname*{arg\,max}_{\mu\in\Delta(\mathcal S)}
\min_{y\in\Delta(\mathcal D)}L_\lambda(\mu,y)
=
\left\{
\mu\in\Delta(\mathcal S):
A^\top\mu\le0,\;
c^\top\mu=u_0^\star
\right\},
\]
and the game value is $u_0^\star/\lambda$.
\end{proof}

\subsection{Regret in the estimated Lagrangian}

\begin{proof}[Proof of Proposition~\ref{prop:efce-saddle-bound}]
The two regret bounds mean
\begin{align}
&\sum_{r=1}^t
\bigl[
\widehat L_r(\mu,y^{(r)})
-
\widehat L_r(\mu^{(r)},y^{(r)})
\bigr]
\le R_0(t)
&&\text{for every }\mu,
\label{eq:app-efce-max-regret}\\
&\sum_{r=1}^t
\bigl[
\widehat L_r(\mu^{(r)},y^{(r)})
-
\widehat L_r(\mu^{(r)},y)
\bigr]
\le R_{\min}(t)
&&\text{for every }y.
\label{eq:app-efce-min-regret}
\end{align}
Let $\mu^\star$ be an optimal EFCE.
Add these inequalities with $\mu=\mu^\star$.
The played payoff terms cancel out, giving
\[
\sum_{r=1}^t\widehat L_r(\mu^{(r)},y)
\ge
\sum_{r=1}^t\widehat L_r(\mu^\star,y^{(r)})
-R_0(t)-R_{\min}(t).
\]
For any probability vectors $\mu,y$,
\[
|\widehat L_r(\mu,y)-L_\lambda(\mu,y)|
\le\|\widehat A^{(r)}-A\|_\infty\le e_r.
\]
Moreover, $A^\top\mu^\star\le0$, so
$L_\lambda(\mu^\star,y^{(r)})
\ge u_0^\star/\lambda$.
Consequently, for every fixed $y$,
\[
\sum_{r=1}^tL_\lambda(\mu^{(r)},y)
\ge
\frac{t u_0^\star}{\lambda}
-R_0(t)-R_{\min}(t)-2\sum_{r=1}^te_r.
\]
Minimize the left side over $y$ and divide by $t$.
Linearity in $\mu$ yields
\begin{equation}
\frac{u_0(\bar\mu^{(t)},d)}{\lambda}
-\Delta_{\mathrm{eq}}(\bar\mu^{(t)})
\ge
\frac{u_0^\star}{\lambda}-a_t.
\label{eq:app-efce-joint-certificate}
\end{equation}
The equilibrium gap is
nonnegative by definition, which proves
\eqref{eq:efce-saddle-objective}.
Using $u_0(\bar\mu^{(t)},d)\le1$ and
$u_0^\star\ge0$ proves
\eqref{eq:efce-saddle-feasibility}.
\end{proof}

\noindent 
Now, we fix the value $\lambda\ge1$ chosen at initialization.
It is unchanged on all rounds, including every
prefix considered below.
Clipping gives
$\widehat A^{(r)}\in[-1,1]^{\mathcal S\times\mathcal D}$,
and $c\in[0,1]^{\mathcal S}$.
Hence, on every round $r$,
\[
\begin{aligned}
\frac c\lambda-\widehat A^{(r)}y^{(r)}
&\in[-1,1+\lambda^{-1}]^{\mathcal S}
\subseteq[-1,2]^{\mathcal S},\\
(\widehat A^{(r)})^\top\mu^{(r)}
&\in[-1,1]^{\mathcal D}.
\end{aligned}
\]
These ranges are independent of $\lambda$.

\begin{lemma}
\label{lem:efce-dual-refinement}
Let $\lambda^\star\ge0$ be an optimal dual multiplier for
\eqref{eq:efce-primal-program}, whose existence has been established in
Lemma~\ref{lem:efce-lagrangian-justification}.
Under the assumptions of
Proposition~\ref{prop:efce-saddle-bound}, if
$\lambda>\lambda^\star$, then
\begin{equation}\label{eq:efce-refined-feasibility}
\Delta_{\mathrm{eq}}(\bar\mu^{(t)})
\le\frac{a_t}{1-\lambda^\star/\lambda}.
\end{equation}
In particular, if $\lambda\ge2\lambda^\star$ and
$\lambda>\lambda^\star$, then
$\Delta_{\mathrm{eq}}(\bar\mu^{(t)})\le2a_t$.
\end{lemma}

\begin{proof}[Proof of Lemma~\ref{lem:efce-dual-refinement}]
By \eqref{eq:app-efce-joint-certificate},
\[
\frac{u_0(\bar\mu^{(t)},d)}{\lambda}
-\Delta_{\mathrm{eq}}(\bar\mu^{(t)})
\ge\frac{u_0^\star}{\lambda}-a_t.
\]
On the other hand, the optimal dual multiplier
$\lambda^\star$ from
Lemma~\ref{lem:efce-lagrangian-justification} satisfies
\eqref{eq:efce-dual-objective-bound}. Hence
\[
\frac{u_0(\bar\mu^{(t)},d)}{\lambda}
-\Delta_{\mathrm{eq}}(\bar\mu^{(t)})
\le\frac{u_0^\star}{\lambda}
-\left(1-\frac{\lambda^\star}{\lambda}\right)
\Delta_{\mathrm{eq}}(\bar\mu^{(t)}).
\]
Combining these two inequalities gives
\[
\left(1-\frac{\lambda^\star}{\lambda}\right)
\Delta_{\mathrm{eq}}(\bar\mu^{(t)})\le a_t.
\]
The coefficient is positive when
$\lambda>\lambda^\star$. If also
$\lambda\ge2\lambda^\star$, it is at least $1/2$.
\end{proof}

\begin{lemma}
\label{lem:efce-computed-regret}
The mediator can choose the two regret minimizers so
that
\( 
R_0(t)+R_{\min}(t)
\le C_{\mathrm{reg}}\sqrt t
\) 
for every $t\ge1$, independently
of $\lambda$.
\end{lemma}

\begin{proof}
By \eqref{eq:efce-update-vectors}, the reward
vectors of the two regret minimizers belong
to $[-1,2]^{\mathcal S}$ and
$[-1,1]^{\mathcal D}$, respectively.
Applying the standard anytime
exponential-weights \citep{cesa2006prediction} regret bound,
after rescaling the rewards to $[0,1]$,
gives
\( 
R_0(t)+R_{\min}(t)
\le
C_{\mathrm{reg}}\sqrt t,
\) 
. 
Since the reward ranges are bounded
independently of $\lambda$, so is
$C_{\mathrm{reg}}$.
\end{proof}

\subsection{Steering with changing recommendations}

For this subsection, index the steering rounds by $k$.
Fix one such round and suppress $k$ where convenient.
Let
\( 
D_i:=\Prb\{d_i[\bar s,z]=0\}, D:=\sum_iD_i.
\) 
Moreover, let $g_i$ denote the expected augmented-payoff
improvement from switching only player $i$ to $d_i$, 
\( 
g_i:=
v_i(\mu,d_i,x_{-i})-v_i(\mu,x).
\) 
The probabilities include the mediator's sampled
profile, the selected strategies' randomization,
and Nature.

\begin{lemma}
\label{lem:efce-steering-coupling}
Fix one round and condition on the pre-play history.
Assume $0\le u_i\le1$, $0<\alpha\le1$, and $P\ge2$. With payments \eqref{eq:efce-trajectory-payment}, there is an event
$E$, of probability $h$, such that
\begin{align}
Ph&\le\sum_i g_i+D,
\label{eq:app-efce-exceptional-mass}\\
\alpha D_i&\le g_i+\Delta_{\mathrm{eq},i}(\mu)+4h
\qquad\text{for every }i.
\label{eq:app-efce-one-round-steering}
\end{align}
\end{lemma}

\begin{proof}
Draw $\bar s$, a complete contingency plan, and pure response rules
implementing $x$. Use the same draws in the following plays:
\[
\begin{array}{c|c}
\text{terminal history}&\text{strategy profile}\\ \hline
Z&x\\
Z^{(j)}&(d_j,x_{-j})\\
Z_i^{\mathrm{uni}}&(x_i,d_{-i})\\
Z^d&d
\end{array}
\]
Each play has its prescribed marginal law, and coupled plays agree until a replaced player chooses a different action.
Set $W_i(z):=u_i(z)+q_i(\bar s,z)$. Define the event $E$ as 
\[
E_j:=\{d_j[\bar s,Z]=0\}
\cap\bigcup_{\ell\ne j}\{d_\ell[\bar s,Z^{(j)}]=0\},
\qquad E:=\bigcup_jE_j,\quad h:=\Pr(E).
\]
If $j$ obeys in actual play, $Z^{(j)}=Z$.
Otherwise its actual payment is zero; on $E_j$, its switched
payment is $P$. Thus,
\[
W_j(Z^{(j)})-W_j(Z)
\ge P\mathbf1_{E_j}-\mathbf1_{\{d_j[\bar s,Z]=0\}}.
\]
Taking expectations gives $P\Pr(E_j)\le g_j+D_j$.
The union bound now gives
\[
Ph=P\Pr\Bigl(\bigcup_jE_j\Bigr)
\le P\sum_j\Pr(E_j)\le\sum_jg_j+D,
\]
which is \eqref{eq:app-efce-exceptional-mass}. Now, we analyze what happens outside the event $E$.
There is at most one actual deviator: if there were two,
switching the later first-deviator to obedience would preserve
the earlier disobedience, so its event $E_j$ would occur.
For fixed $i$, the coupled identities on $E^c$ are
\[
\begin{array}{c|l}
\text{actual deviators}&\text{identities}\\ \hline
\text{none}&Z=Z^{(i)}=Z_i^{\mathrm{uni}}=Z^d\\
\text{only }i&Z=Z_i^{\mathrm{uni}},\quad Z^{(i)}=Z^d\\
\text{only }j\ne i&Z=Z^{(i)},\quad Z_i^{\mathrm{uni}}=Z^d
\end{array}
\]
For the last row, $E_j^c$ means that switching $j$ produces
all-direct play. Hence $x_i$ obeys on the direct path, so
$(x_i,d_{-i})$ produces that same path.
Consequently, the indicators of $i$'s disobedience in $Z$ and
$Z_i^{\mathrm{uni}}$ agree on $E^c$. With
$D_i^{\mathrm{uni}}:=\Pr\{d_i[\bar s,Z_i^{\mathrm{uni}}]=0\}$,
\begin{equation}
|D_i-D_i^{\mathrm{uni}}|
\le\mathbb E\left|
\mathbf1_{\{d_i[\bar s,Z]=0\}}
-\mathbf1_{\{d_i[\bar s,Z_i^{\mathrm{uni}}]=0\}}
\right|\le h.
\label{eq:app-efce-unilateral-mass}
\end{equation}
Now, to compare payoffs let
\( 
H_i:=W_i(Z^{(i)})-W_i(Z)-W_i(Z^d)+W_i(Z_i^{\mathrm{uni}}).
\) 
The table gives $H_i=0$ on $E^c$. On $E$:
\begin{itemize}
\item If $i$ obeys in actual play, $Z^{(i)}=Z$, so
$H_i\ge-(1+\alpha)\ge-2$.
\item Otherwise, $q_i(\bar s,Z)=0$,
$q_i(\bar s,Z^d)=\alpha$, and
$q_i(\bar s,Z^{(i)})\ge\alpha$.
The payment contribution to $H_i$ is nonnegative; its
utility contribution is at least $-2$.
\end{itemize}
Thus $H_i\ge-2\mathbf1_E$. Taking expectations yields
\begin{equation}
g_i\ge v_i(\mu,d)-v_i(\mu,x_i,d_{-i})-2h.
\label{eq:app-efce-terminal-comparison}
\end{equation}
Against direct opponents, $i$ earns $\alpha$ exactly when it obeys.
Therefore,
\[
v_i(\mu,d)-v_i(\mu,x_i,d_{-i})
\ge\alpha D_i^{\mathrm{uni}}-\Delta_{\mathrm{eq},i}(\mu).
\]
Combining this with \eqref{eq:app-efce-unilateral-mass} gives
\[
\alpha D_i
\le g_i+\Delta_{\mathrm{eq},i}(\mu)+(2+\alpha)h
\le g_i+\Delta_{\mathrm{eq},i}(\mu)+4h.
\]
\end{proof}

\begin{proof}[Proof of Lemma~\ref{lem:efce-changing-policy-steering}]
Apply Lemma~\ref{lem:efce-steering-coupling} in each steering round $k$.
Summing over players gives
\begin{align*}
\alpha D^{(k)}
\le\sum_i g_i^{(k)}
+\sum_i\Delta_{\mathrm{eq},i}(\mu^{(k)})+4nh_k\le\left(1+\frac{4n}{P}\right)\sum_i g_i^{(k)}
+\sum_i\Delta_{\mathrm{eq},i}(\mu^{(k)})
+\frac{4n}{P}D^{(k)}.
\end{align*}
Rearrange and sum over $k$:
\[
\left(\alpha-\frac{4n}{P}\right)\sum_{k=1}^ND^{(k)}
\le\left(1+\frac{4n}{P}\right)\sum_i\sum_{k=1}^Ng_i^{(k)}
+\sum_{k=1}^N\sum_i\Delta_{\mathrm{eq},i}(\mu^{(k)}).
\]
The same fixed response rule $d_i$ is used in every round, so
\eqref{eq:efce-steering-regret} gives
\[
\sum_{k=1}^Ng_i^{(k)}\le(P+1)R_i(N).
\]
Also,
\[
\sum_{k=1}^N\sum_i\Delta_{\mathrm{eq},i}(\mu^{(k)})
\le n\sum_{k=1}^N\Delta_{\mathrm{eq}}(\mu^{(k)})
=nN\bar\eta_N.
\]
Since $\alpha P\ge8n$,
$\alpha-4n/P\ge\alpha/2$. Division by $\alpha N/2$
proves \eqref{eq:efce-changing-policy-bound}. For payments, set $A_k:=\bigcup_i\{d_i[\bar s^{(k)},Z]=0\}$.
Then, $\Pr(A_k)\le D^{(k)}$ by the union bound, and
\[
\mathbb E[q_i^{(k)}]\le\alpha+P\Pr(A_k)
\le\alpha+PD^{(k)}.
\]
Average over $k$ to obtain result.
\end{proof}

\subsection{The accuracy available during interaction}

\begin{lemma}
\label{lem:efce-learning-calendar}
Let $\mathcal E_L(T)$ be the event that every table
completed by time $T$ satisfies the conclusion of
Theorem~\ref{thm:learn-all-efg-allx} at its own
requested accuracy.
Under \eqref{eq:efce-call-confidence} and the inequalities
\eqref{eq:efce-conditional-query-regret},
$\Prb(\mathcal E_L(T))\ge1-\delta/3$.
On this event, with parameters
\( 
m_t:=\lfloor(t-1)^q\rfloor,
\Lambda(m):=1+\log\!\left(
\frac{n\mathcal J(m+1)^2}{\delta}\right),
 a:=1/2,
\) the estimate available before round $t$ satisfies
\begin{equation}
e_t:=\|\widehat A^{(t)}-A\|_\infty
\le C_A\frac{\Lambda(m_t)^a}{\sqrt{m_t+1}}.
\label{eq:efce-calendar-explicit}
\end{equation}
\end{lemma}

\begin{proof}
The regret bounds hold.
For each testing epoch of a call requesting accuracy
$\varepsilon$, the concentration events in
Lemma~\ref{lem:certified-epoch-pathwise} fail with
probability at most $\zeta(\varepsilon)$.
A union bound over at most $M(\varepsilon)$ epochs
gives call-failure probability of at most
$M(\varepsilon)\zeta(\varepsilon)
=\delta_{\mathrm{call}}(\varepsilon)$.
The requested accuracies are successively halved,
so their sum is less than one. Consequently,
\[
\Prb(\mathcal E_L(T)^c)
\le \sum_{\text{calls}}\delta_{\mathrm{call}}(\varepsilon)
=\frac\delta3\sum_{\text{calls}}\varepsilon
\le\frac\delta3,
\]
as initial call is with $\varepsilon <\frac{1}{2}$. 

\vspace{6pt}
\noindent
Now, we upper bound the iterations made for a single accuracy level.
A call has at
most $K(\varepsilon)$ iterations even on failure paths, and
Theorem~\ref{thm:learn-all-efg-allx} gives the
trajectory bound
\( 
\mathcal B(\varepsilon)
:=
2\mathcal J K(\varepsilon)
+
\frac{2097152}{399}
\frac{\mathcal J H(\varepsilon)^2}
{\rho_{\min}^2\varepsilon^2}.
\) 
For
\( 
L(\varepsilon):=1+\log\!\left(
\frac{n\mathcal J K(\varepsilon)}{\delta\varepsilon}\right),
\) 
the fixed constants $C_i$ and
\eqref{eq:efg-epoch-constants} imply
$H(\varepsilon)\le C L(\varepsilon)^{1/2}$.
Also $K(\varepsilon)\le C L(\varepsilon)$. Consequently
\begin{equation}
\mathcal B(\varepsilon)
\le\frac{C}{\varepsilon^2}L(\varepsilon)^{2a}.
\label{eq:efce-call-budget-explicit}
\end{equation}
Suppose \(m\ge1\) learning trajectories have been completed and the current call, requesting accuracy \(\varepsilon\), is unfinished. First, suppose that at least one preceding call has been completed. Its requested accuracy was \(2\varepsilon\), so by \(\eqref{eq:efce-gain-matrix-accuracy}\), the available matrix satisfies
\(
e_t\le4I_{\max}\varepsilon.
\) 
Since all earlier calls requested accuracies \(2^k\varepsilon\), and \(L(2^k\varepsilon)\le L(\varepsilon)\), the total number of learning trajectories, including those in the current unfinished call, satisfies
\begin{align}\label{quaso:2}
m
&\le\sum_{\text{calls through current call}}
\mathcal B(\text{requested accuracy})\le\frac{CL(\varepsilon)^{2a}}{\varepsilon^2}
\sum_{k\ge0}4^{-k}
=\frac{4C}{3\varepsilon^2}L(\varepsilon)^{2a}.
\end{align}
If \(\varepsilon\le1/(m+1)\), then
\[
e_t\le\frac{4I_{\max}}{m+1}
\le4I_{\max}
\frac{\Lambda(m)^a}{\sqrt{m+1}}.
\]
Otherwise, \(1/\varepsilon<m+1\), so
\(K(\varepsilon)\le C[1+\log(m+1)]\) and
\(L(\varepsilon)\le C\Lambda(m)\).
Therefore \eqref{quaso:2} gives
\[
\begin{aligned}
e_t
&\le4I_{\max}\varepsilon\le C\frac{L(\varepsilon)^a}{\sqrt m}
\le C\frac{\Lambda(m)^a}{\sqrt{m+1}},
\end{aligned}
\]
where the last inequality uses \(m\ge1\). It remains to consider the period before the first continuation table is completed. Let \(\varepsilon_0\in(0,1/2)\) be the fixed initial requested accuracy. During this period, \(\widehat A^{(t)}=0\), so \(e_t\le1\), and the number of completed learning trajectories satisfies
\[
m\le\mathcal B(\varepsilon_0)
\le\frac{C}{\varepsilon_0^2}
L(\varepsilon_0)^{2a}.
\]
Since \(\varepsilon_0\) is fixed independently of \(T\) and \(\delta\), the definition of \(L\) implies
\( 
L(\varepsilon_0)\le C_{\varepsilon_0}\Lambda(m)
\) 
for a constant \(C_{\varepsilon_0}\) independent of \(m\) and \(\delta\). Hence
\( 
m+1\le C'_{\varepsilon_0}\Lambda(m)^{2a},
\) 
which gives
\[
e_t\le1\le
C_A\frac{\Lambda(m)^a}{\sqrt{m+1}}
\]
after enlarging \(C_A\). Finally, before round \(t\), exactly
\(m_t=\lfloor(t-1)^q\rfloor\) learning trajectories have been completed. Substituting \(m=m_t\) establishes
\(\eqref{eq:efce-calendar-explicit}\).
\end{proof}

\subsection{Gap and payment convergence guarantees}

For the proof of Theorem~\ref{thm:online-efce}, let $\mathcal F_{t^-}$ contain the full history
before play in round $t$, as well as the current mediator choices,
payment rules, and players' selected strategies, but not the
current recommendations, action samples, or Nature's moves.
Write
\begin{align*}
OG_{\mathrm{cond}}(T)
:=u_0^\star-\frac1T\sum_{t=1}^T
\sum_{z\in Z}u_0(z)\pi_{\mathrm{act}}^{(t)}[z], \quad
\mathrm{Pay}_{i,\mathrm{cond}}(T)
:=\frac1T\sum_{t=1}^T
\mathbb E\!\left[
q_i^{(t)}(\bar s^{(t)},z^{(t)})
\,\middle|\,\mathcal F_{t^-}\right].
\end{align*}
These are the quantities directly controlled by the steering
lemma; concentration will relate them to the realized quantities
of Definition~\ref{def:efce-steering-gaps}.

\begin{proof}[Proof of Theorem~\ref{thm:online-efce}]
Let $\mathcal E_L(T)$ be the event of
Lemma~\ref{lem:efce-learning-calendar}. It has
probability of at least $1-\delta/3$. We first work on
$\mathcal E_L(T)$. As $q=3/4$,  
for $m_t=\lfloor(t-1)^q\rfloor$,
\[
m_t+1\ge 2^{-3/4}t^{3/4},\qquad t\ge1.
\]
Lemma \ref{lem:efce-learning-calendar}, together with
$\Lambda(m_t)\le\Lambda_T$, gives
\[
e_t:=\|\widehat A^{(t)}-A\|_\infty
\le E_Tt^{-3/8},\qquad
E_T:=2^{3/8}C_A\Lambda_T^{1/2}.
\]
For $\mathcal J=0$, take $C_A=E_T=0$.
Using Lemma~\ref{lem:efce-computed-regret} and
$\sum_{r=1}^t r^{-b}\le t^{1-b}/(1-b)$ for
$0<b<1$, we obtain
\[
\begin{aligned}
a_t
&\le C_{\mathrm{reg}}t^{-1/2}
 +\frac{2E_T}{t}\sum_{r=1}^t r^{-3/8}\\
&\le C_{\mathrm{reg}}t^{-1/2}
 +\frac{16}{5}E_Tt^{-3/8}.
\end{aligned}
\]
Thus
\begin{equation}\label{eq:app-efce-new-average-a}
\frac1T\sum_{t=1}^T a_t
\le A_T
:=2C_{\mathrm{reg}}T^{-1/2}
 +\frac{128}{25}E_TT^{-3/8}
\le C\Lambda_T^{1/2}T^{-3/8}.
\end{equation}
Proposition~\ref{prop:efce-saddle-bound} gives,
for each $t$,
\begin{equation}\label{eq:app-efce-new-target-objective}
[u_0^\star-u_0(\bar\mu^{(t)},d)]_+
\le\lambda a_t.
\end{equation}
Now, let $\lambda^\star<\infty$ be as defined in
Lemma~\ref{lem:efce-dual-refinement}. Define
\( 
T_0:=\max\left\{64,
\left\lceil(2\lambda^\star+1)^8\right\rceil\right\}.
\) 
For $T\ge T_0$, the choice
$\lambda=T^{1/8}$ satisfies
$\lambda\ge2\lambda^\star+1>\lambda^\star$.
Lemma~\ref{lem:efce-dual-refinement} therefore
implies, simultaneously for every $t\le T$,
\begin{equation}\label{eq:app-efce-new-target-gap}
\Delta_{\mathrm{eq}}(\bar\mu^{(t)})\le2a_t.
\end{equation}
Now to denote learning rounds, let $L_T:=|\mathcal L_T|$ and $N:=T-L_T$.
If $\mathcal J>0$, then
$L_T=\lfloor T^{3/4}\rfloor$;
otherwise $L_T=0$.
For $T\ge64$, $N\ge T/2$.
By \eqref{eq:app-efce-new-target-gap} and \eqref{eq:app-efce-new-average-a},
\begin{align}\label{quaso:3}
\bar\eta_N
:=\frac1N\sum_{t\in\mathcal C_T}
\Delta_{\mathrm{eq}}(\bar\mu^{(t)})
\le\frac{2T}{N}A_T.
\end{align}
Now, we apply Lemma~\ref{lem:efce-changing-policy-steering}
with \eqref{quaso:3} and $R_i(N)=C_i^{\mathrm{st}}\sqrt N$:
\[
\begin{aligned}
\bar D_N
&:=\frac1N\sum_{k=1}^N D^{(k)}\le\frac2\alpha\left[
\left(1+\frac{4n}{P}\right)
\frac{(P+1)\sum_iC_i^{\mathrm{st}}}{\sqrt N}
+\frac{2nT}{N}A_T
\right].
\end{aligned}
\]
With $\alpha=T^{-1/8}$ and
$P=8nT^{1/8}$, the required conditions
$0<\alpha\le1$, $P\ge2$, and $\alpha P=8n$
hold. Since $N\ge T/2$, the two terms have orders
\( 
\frac{P+1}{\alpha\sqrt N}=O(T^{-1/4}),
\frac{A_T}{\alpha}
=O(\Lambda_T^{1/2}T^{-1/4}).
\) 
Consequently,
\begin{equation}\label{eq:app-efce-new-disobedience}
\bar D_N\le C\Lambda_T^{1/2}T^{-1/4}.
\end{equation}
The payment on a learning round is at most $B+1$.
By Lemma~\ref{lem:efce-changing-policy-steering},
\[
\begin{aligned}
\mathrm{Pay}_{i,\mathrm{cond}}(T)
&\le\frac NT(\alpha+P\bar D_N)
 +(B+1)\frac{L_T}{T}\le C\Lambda_T^{1/2}T^{-1/8},
\end{aligned}
\]
where $L_T/T\le T^{-1/4}$.
Couple actual and obedient plays on every steering round
using the same recommendation profiles.
The union bound and $u_0\in[0,1]$ give
\begin{align}
\|\pi(\cdot\mid\bar\mu^{(t)},x^{(t)})
-\pi(\cdot\mid\bar\mu^{(t)},d)\|_1
&\le2D^{(k)},\quad
|u_0(\bar\mu^{(t)},x^{(t)})
-u_0(\bar\mu^{(t)},d)|
\le D^{(k)}.
\label{eq:app-efce-outcome-coupling}
\end{align}
Consequently,
\[
DG(T)\le2\frac NT\bar D_N+2\frac{L_T}{T}
\le C\Lambda_T^{1/2}T^{-1/4}.
\]
Also \eqref{eq:app-efce-new-target-objective} gives
\[
\begin{aligned}
[OG_{\mathrm{cond}}(T)]_+
&\le\frac\lambda T\sum_{t\in\mathcal C_T}a_t
  +\frac NT\bar D_N+\frac{L_T}{T}\le\lambda A_T+\bar D_N+T^{-1/4}\le C\Lambda_T^{1/2}T^{-1/4},
\end{aligned}
\]
because $\lambda=T^{1/8}$ and
$A_T=O(\Lambda_T^{1/2}T^{-3/8})$.
Now, we control the conditional average payment and optimality gap bounds; the directness bound already concerns
the distributions of Definition~\ref{def:efce-steering-gaps}. Finally, by Lemma~\ref{lem:efce-realized-quantities}, on an event
$\mathcal E_R(T)$ with probability of at least $1-\delta/3$,
putting $Q_T:=\max\{B+1,P\}$ and
\( 
\xi_T:=\sqrt{\frac{2\log(6(n+1)/\delta)}{T}},
\) 
we have, simultaneously for every player,
\[
\begin{aligned}
\mathrm{Pay}_i(T)
\le\mathrm{Pay}_{i,\mathrm{cond}}(T)+Q_T\xi_T, \quad
[OG(T)]_+
\le[OG_{\mathrm{cond}}(T)]_++\xi_T.
\end{aligned}
\]
For the parameters, $Q_T=O(T^{1/8})$ and
\( 
Q_T\xi_T=O(\Lambda_T^{1/2}T^{-3/8}),
\xi_T=O(\Lambda_T^{1/2}T^{-1/2}).
\) 
Both errors are smaller than the respective of \( \mathrm{Pay}_{i,\mathrm{cond}}\) and \([OG_{\mathrm{cond}}(T)]_+ \) . Thus, for $T\ge T_0$, all three claims hold on
$\mathcal E_L(T)\cap\mathcal E_R(T)$. Finally,
\( 
\Prb(\mathcal E_L(T)\cap\mathcal E_R(T))
\ge1-\frac{2\delta}{3},
\) minding the high probability event regarding the no-regret algorithms mentioned in Remark \ref{rem:efg-regret-sampling}, provides the result with probability at least \( 1-\delta\).
\end{proof}

\begin{lemma}
\label{lem:efce-realized-quantities}
Let $Q_T:=\max\{B+1,P\}$. With probability at least
$1-\delta/3$, simultaneously for all players,
\begin{align*}
|\mathrm{Pay}_i(T)-\mathrm{Pay}_{i,\mathrm{cond}}(T)|
&\le Q_T\sqrt{\frac{2\log(6(n+1)/\delta)}{T}},
\qquad
|OG(T)-OG_{\mathrm{cond}}(T)|
\le\sqrt{\frac{2\log(6(n+1)/\delta)}{T}}.
\end{align*}
\end{lemma}

\begin{proof}
For each player $i$, set
\[
X_{i,t}:=q_i^{(t)}(\bar s^{(t)},z^{(t)})
-\mathbb E\!\left[q_i^{(t)}(\bar s^{(t)},z^{(t)})
\,\middle|\,\mathcal F_{t^-}\right],
\]
and for the mediator, set
\[
X_{0,t}:=u_0(z^{(t)})
-\sum_{z\in Z}u_0(z)\pi_{\mathrm{act}}^{(t)}[z].
\]
Conditionally on $\mathcal F_{t^-}$, the current trajectory
is sampled according to the selected strategies and mediator
choices. Hence, both sequences are martingale differences,
with $|X_{i,t}|\le Q_T$ and $|X_{0,t}|\le1$.
By two-sided Azuma--Hoeffding, each of the $n+1$
normalized sums deviates by more than its stated bound
with probability at most $\delta/(3(n+1))$.
A union bound proves the simultaneous inequalities.
\end{proof}

\end{document}